\documentclass[final,11pt]{article}
\usepackage{multirow}
\usepackage[nottoc]{tocbibind}
\usepackage{geometry}
\usepackage{subcaption}
\usepackage{placeins}
\usepackage[affil-it, auth-sc]{authblk}
\usepackage[english]{babel}
\usepackage[utf8]{inputenc}
\usepackage{graphicx}
\usepackage{amssymb}
\usepackage{amsthm}
\usepackage{amsmath}
\usepackage{amsfonts}
\usepackage{verbatim}
\usepackage{lineno}
\usepackage{wasysym}
\usepackage{todonotes}
\presetkeys{todonotes}{inline, color=blue!30, size=\small}{}
\usepackage{slashed}
\usepackage{color, soul}
\definecolor{darkred}{rgb}{0.9, 0.0, 0.0}
\definecolor{darkgreen}{rgb}{0.0, 0.5, 0.0}
\usepackage[colorlinks,bookmarks,linkcolor=darkred, citecolor=darkgreen]{hyperref}
\usepackage{eso-pic}
\usepackage[sort&compress,numbers]{natbib}
\usepackage{doi}
\usepackage{paralist,epsfig}
\usepackage{relsize}
\usepackage{nicefrac,esint}
\usepackage{float}
\usepackage{cleveref}
\usepackage{marginnote}

\newcommand{\slashpi}{\protect{\slash\hspace{-0.5em}\pi}}
\newcommand{\bk}{{\boldsymbol k}}

\allowdisplaybreaks

\def\fAt{f_{A3}}
\def\fAtbar{\bar{f}_{A3}}

\begin{document}

\title{Theory of inverse beta decay for reactor antineutrinos}

\author[1]{Oleksandr Tomalak\thanks{tomalak@itp.ac.cn}}
\affil[1]{Institute of Theoretical Physics, Chinese Academy of Sciences, Beijing 100190, P. R. China}

\author[2,3]{Qishan Liu\thanks{liuqs@ihep.ac.cn}}
\affil[2]{Institute of High Energy Physics, Chinese Academy of Sciences, Beijing 100049, China}
\affil[3]{China Center of Advanced Science and Technology, Beijing 100190, China}

\author[2]{Yu-Feng Li\thanks{liyufeng@ihep.ac.cn}}

\date{\today}

\maketitle

Inverse beta decay (IBD), $\overline{\nu}_e p \to e^+ n \left( \gamma \right)$, is the main detection channel for reactor antineutrinos in water- and hydrocarbon-based detectors. As reactor antineutrino experiments now target subpercent-level sensitivity to oscillation parameters, a precise theoretical description of IBD, including recoil, weak magnetism, nucleon structure, and radiative corrections, becomes essential. In this work, we give a detailed and precise calculation of the total and differential cross sections for radiative IBD, $\overline{\nu}_e p \to e^+ n \gamma$. We use a heavy baryon chiral perturbation theory framework, systematically incorporating electroweak, electromagnetic, and strong-interaction corrections. We derive new analytic cross-section expressions, clarify the collinear structure of radiative corrections, and provide a systematic uncertainty analysis. We also discuss phenomenological applications for reactor antineutrino experiments, e.g., JUNO, and neutron decay. Our results enable subpermille theoretical precision, supporting current and future experiments.

\newpage

\tableofcontents

\newpage

\section{Introduction}
\label{sec:introduction}

Experimental evidence for neutrino oscillations, and thus the existence of nonzero neutrino masses~\cite{Pontecorvo:1957cp,Pontecorvo:1967fh}, has been firmly established by solar~\cite{Cleveland:1998nv,Hampel:1998xg,Ahmad:2002jz,Abdurashitov:2002nt,Fukuda:2001nj,Ahmed:2003kj}, atmospheric~\cite{Fukuda:1998mi,Ashie:2004mr}, reactor~\cite{Eguchi:2002dm,Araki:2004mb,Abe:2012tg,Ahn:2012nd,An:2013uza}, and accelerator~\cite{Ahn:2002up,Michael:2006rx,Abe:2013xua} neutrino experiments. Current efforts in neutrino physics~\cite{Dentler:2018sju,Nunokawa:2007qh,Esteban:2020cvm} are focused on precision measurements of the neutrino mixing parameters~\cite{NOvA:2007rmc,T2K:2011qtm,KamLAND:2013rgu,ESSnuSB:2013dql,DUNE:2016ymp,RENO:2018dro,DayaBay:2017jkb,DayaBay:2018yms,DoubleChooz:2019qbj,NOvA:2019cyt,T2K:2019bcf,DUNE:2020ypp,DUNE:2021tad}, the determination of the neutrino mass ordering~\cite{JUNO:2015zny,DUNE:2020ypp}, and searches for charge-parity (CP) violation~\cite{Hyper-KamiokandeProto-:2015xww,T2K:2019bcf,DUNE:2020ypp}. Among these, medium-baseline reactor experiments play a critical role. They probe the structure of the lepton mixing matrix with high precision.\footnote{Currently, the neutrino oscillation parameters in the framework of the Standard Model with three massive neutrinos are determined as $\theta_{12}/^\circ = 33.76^{+0.42}_{-0.41},~\theta_{23}/^\circ = 43.29^{+0.96}_{-0.79},~\theta_{13}/^\circ = 8.62^{+0.11}_{-0.11},~\delta_\mathrm{CP}/^\circ = 212^{+26}_{-36},$ $\frac{\Delta m^2_{21}}{10^{-5}~\mathrm{eV}^2} = 7.537^{+0.094}_{-0.10},$ and $\frac{\Delta m^2_{3\ell}}{10^{-3}~\mathrm{eV}^2} = 2.511^{+0.021}_{-0.020}$ for the normal hierarchy and $\theta_{23}/^\circ = 47.90^{+0.73}_{-0.92},~\theta_{13}/^\circ = 8.65^{+0.11}_{-0.11},~\delta_\mathrm{CP}/^\circ = 274^{+22}_{-25},$ and $\frac{\Delta m^2_{3\ell}}{10^{-3}~\mathrm{eV}^2} = - 2.483^{+0.020}_{-0.020}$ for the inverted hierarchy~\cite{Esteban:2024eli}.}

The Jiangmen Underground Neutrino Observatory (JUNO)~\cite{JUNO:2015zny,JUNO:2021vlw,JUNO:2022mxj,JUNO:2022lpc,JUNO:2024jaw,JUNO:2020ijm} is an example of such a medium-baseline reactor experiment. It is designed to determine the neutrino mass hierarchy with $3\sigma$ significance by detecting antineutrinos from the Yangjiang and Taishan nuclear power plants. JUNO expects about 83 inverse beta decay (IBD) events per day with a 20-kiloton liquid scintillator detector. This yields approximately 180,000 events over six years. These statistics enable precise tests of the unitarity~\cite{Aguilar-Saavedra:2000jom,Farzan:2002ct,He:2013rba,Ellis:2020ehi} of the Pontecorvo-Maki-Nakagawa-Sakata matrix~\cite{Pontecorvo:1957qd,Maki:1962mu} at the subpercent level. JUNO aims for relative uncertainties of $0.7\%$ on the largest mixing matrix element and $0.6\%$ or better on the neutrino squared-mass differences~\cite{JUNO:2015zny}. The first JUNO measurement of reactor neutrino oscillations has already improved the accuracy of the solar oscillation parameters by a factor of about $\sim1.6$~\cite{JUNO:2025gmd,JUNO:2025fpc}. This progress has motivated broad theoretical and phenomenological research~\cite{Ding:2025dqd,Goswami:2025wla,Ding:2025dzc,Capozzi:2025ovi,Chattopadhyay:2025ccy,Petcov:2025aci,Luo:2025pqy,He:2025idv,Jiang:2025hvq,Chen:2025afg,Xing:2025xte,Ge:2025cky,Huang:2025znh,Ge:2025csr,Zhang:2025jnn,Li:2025hye,Chao:2025sao,Xing:2025bdm,Minakata:2025azk,Saad:2025dkh,Gao:2025jlw,Bora:2025xfj,Tomalak:2025okl,Tomalak:2025jtn,Borah:2025jty}. The advances in precision frontier require updates in theoretical cross sections. Such new predictions are needed to convert reactor antineutrino fluxes~\cite{Schreckenbach:1985ep,Huber:2011wv,Mueller:2011nm} into observable event rates. Achieving subpercent precision in oscillation physics also demands a precise understanding of the initial antineutrino flux and energy spectrum. These quantities, determined by complex fission processes and corresponding beta-decay spectra, introduce critical systematic uncertainties that must be controlled through refined reactor modeling, improved nuclear data, and \textit{in situ} measurements to fully exploit the statistical power of experiments like JUNO.

Inverse beta decay on a free proton,
\begin{equation} \label{eq:IBD_reaction}
	\overline{\nu}_e + p \to e^+ + n (\gamma),
\end{equation}
is the dominant reaction for reactor antineutrinos in hydrogen-rich media and was the channel of their first observation~\cite{Cowan:1956rrn}. The positron from IBD promptly deposits its kinetic energy and annihilates with an electron, yielding a prompt signal. The neutron is then captured by hydrogen or carbon nuclei with average lifetimes of about 200~$\mu$s, producing a delayed gamma signal at 2.2~MeV or 4.95~MeV, respectively. The prompt-delayed time-energy correlations enable efficient background discrimination.

Precise oscillation analysis requires accurate energy reconstruction event-by-event. JUNO aims for $\sim 3\%$ energy reconstruction precision with its liquid scintillator detector. Radiative corrections, especially real photon emission in IBD, can affect energy reconstruction in neutrino detectors. These corrections modify the correlation between measured positron and incident antineutrino energies. Current analyses do not account for these effects in event reconstruction or spectrum modeling. For precision analyses in experiments like JUNO, it is essential to account for all relevant quantum corrections to the IBD cross section. These include recoil effects, weak magnetism, nucleon structure contributions, and radiative corrections from quantum electrodynamics (QED), quantum chromodynamics (QCD), and electroweak interactions.

IBD cross sections and radiative corrections have been studied for decades. The first comprehensive treatment of QED radiative corrections was provided in Ref.~\cite{Vogel:1983hi}. This work expressed them in terms of the electromagnetic energy, the sum of the positron and photon energies, and also incorporated recoil and weak magnetism corrections. However, this work included only partially the "inner" radiative corrections from QCD and electroweak interactions, and it had a residual dependence on an ultraviolet cutoff intended to be fixed by neutron decay experiments. Subsequent refinements in Ref.~\cite{Fayans:1985uej} corrected recoil effects and presented QED corrections fully analytically, using techniques developed in Refs.~\cite{Bardin:1983yb} and~\cite{Bardin:1985fg}. This type of radiative correction can be directly applied to the positron energy spectrum within the elastic IBD kinematic range. Here, the total electromagnetic energy serves as a signature of the positron energy.

Further studies refined the theoretical framework of IBD. References~\cite{Leitner:2006ww,Ankowski:2016oyj,Giunti:2016vlh,Ivanov:2017ifp,Altarawneh:2024sxo} examined the conserved vector current hypothesis. They clarified the role of first- and second-class currents in the vector sector of the hadronic matrix element~\cite{Weinberg:1958ut}. Recoil and weak magnetism corrections were revisited in the context of supernova neutrinos in Refs.~\cite{Vogel:1999zy,Horowitz:2001xf,Vogel:2006sg,Hayes:2016qnu}. Radiative and recoil effects were also studied for primordial nucleosynthesis~\cite{Dicus:1982bz,Seckel:1993dc}.

A simplified but widely used parameterization of QED, QCD, and electroweak radiative corrections was introduced in Refs.~\cite{Towner:1998bh,Kurylov:2001av,Kurylov:2002vj,Strumia:2003zx,Hayes:2016qnu}, building on the treatment of ``inner" corrections from Ref.~\cite{Sirlin:1977sv}. This approach was adopted in Ref.~\cite{Strumia:2003zx}, which also incorporated the neutron-proton mass difference and provided a compact parameterization for IBD cross sections. The estimated uncertainties at that time were 0.4\%. Recent analyses~\cite{Ankowski:2016oyj,Ricciardi:2022pru} improved this to the 0.05-0.1\% range. They identified the axial-vector coupling constant $g_A$ and the Cabibbo-Kobayashi-Maskawa (CKM) matrix element $V_{ud}$~\cite{Cabibbo:1963yz,Kobayashi:1973fv} as dominant sources of uncertainty. QED corrections to positron angle-dependent terms were evaluated in Refs.~\cite{Fukugita:2004cq,Fukugita:2005hs} and reproduced using effective field theory methods in Ref.~\cite{Raha:2011aa}.

Despite this progress, several key aspects have remained unresolved. Most existing calculations do not clearly distinguish between the positron and electromagnetic energy spectra, neglect the change in energy range from elastic to radiative processes, and apply the static-limit approximation in radiative phase space, assuming zero neutron recoil and a fixed sum of photon and positron energies. However, this approximation works very well for precision goals above $1\permil$ with the electromagnetic energy spectrum. The radiated hard photon can significantly change the acceptance by escaping the detector or changing the kinematics of the recoil positron. Therefore, the electromagnetic energy spectrum can be applied in the analysis of IBD reactions with liquid scintillators by summing the total energy in the prompt event only neglecting the detector-specific details~\cite{Birks:1951boa,Frank:1937fk,Lin:2022htc,JUNO:2024fdc} like energy- and particle-dependent energy responses, which are known to be nonlinear~\cite{DayaBay:2019fje,JUNO:2025gmd,JUNO:2025fpc}, and bremsstrahlung-related acceptance corrections, introducing potential biases in bin-to-bin migrations, and forbidding the background rejection by the event topology. Modern reconstruction techniques do not allow efficiently distinguishing electron, positron, and photon events in liquid scintillators~\cite{Aberle:2013jba,Alonso:2014fwf,Li:2015phc,Caravaca:2016fjg,Wan:2016nhe,Liu:2018fpq,Kaptanoglu:2018sus,Theia:2019non,Psihas:2019ksa,KamLAND-Zen:2019imh,DayaBay:2019fje,Qian:2021vnh,Li:2021oos,BOREXINO:2021efb,SNO:2021xpa,PROSPECT:2021jey,Li:2022tvg,Gavrikov:2022kok,Dou:2022glt,Huang:2022zum,Lin:2022htc,Liu:2024cxo,Jiang:2024wph,Gavrikov:2024rso,Liao:2024bso,Zhang:2024okq,JUNO:2024fdc,NuDoubt:2024jax,Shi:2025zut,Takenaka:2025hgi,JUNO:2025gmd,JUNO:2025fpc}, but the progress in this direction is expected by implementing the machine learning techniques~\cite{Liu:2018fpq,Psihas:2019ksa,DayaBay:2019fje,Qian:2021vnh,Li:2021oos,Li:2022tvg,Gavrikov:2022kok,Dou:2022glt,Huang:2022zum,Lin:2022htc,Liu:2024cxo,Jiang:2024wph,Gavrikov:2024rso,Liao:2024bso,Zhang:2024okq,JUNO:2024fdc,Shi:2025zut,Takenaka:2025hgi,JUNO:2025gmd,JUNO:2025fpc} as well as transitioning to ``self-segmented" opaque detectors~\cite{Tayloe:2006ct,LiquidO:2019mxd,Buck:2019tsa,Tang:2020xoy,LiquidOConsortium:2023bqe,LiquidO:2024piw,LiquidO:2025qia} and/or water-based liquid scintillators~\cite{Yeh:2011zz,Aberle:2013jba,Alonso:2014fwf,Li:2015phc,Caravaca:2016fjg,Guo:2017nnr,Kaptanoglu:2018sus,Theia:2019non,Biller:2020uoi}. Therefore, it would be desirable to control also the positron kinematics in the radiative IBD process directly. The dedicated calibration of the energy reconstruction in liquid scintillators, based on precise calculations of radiative IBD cross sections, will improve particle identification and energy determination. Moreover, an accurate simulation of (anti)neutrino-nucleus reactions requires control over the final-state positron kinematics. For instance, the broadening of charged lepton tracks by electromagnetic interactions with the nuclear charge density can significantly change the observed electron scattering angle in experiments with large nuclei as a target~\cite{Tomalak:2022kjd,Tomalak:2023kwl,Tomalak:2024lme,Bhattacharya:2025pje,Tomalak:2026xmo}.

In this work, we address these shortcomings and present a comprehensive, high-precision evaluation of IBD cross sections that incorporates all relevant physics effects. Specifically, we
\begin{itemize}
	\item Revisit and evaluate numerically QED radiative corrections within the heavy baryon chiral perturbation theory (HBChPT) framework, using recently determined coupling constants~\cite{Cirigliano:2022hob,Cirigliano:2023fnz,Tomalak:2023xgm,Cirigliano:2024nfi} that include all short-distance contributions from scales above $\sim1$-$10~\mathrm{MeV}$;
	\item Present a detailed uncertainty budget for total and differential cross sections at the subpercent level, suitable for implementation in precision reactor antineutrino analyses;
	\item Provide analytic expressions for radiative IBD cross sections as well as update QED radiative corrections to the neutron lifetime and beta asymmetry, including real photon emission beyond the static limit, using exact three-body phase-space integrals~\cite{Lee:1964jq,Ram:1967zza};
	\item Evaluate both positron and electromagnetic energy spectra with exact phase-space boundaries, satisfying expected single-logarithm and logarithm-free collinear behavior, respectively;
	\item Present double- and triple-differential distributions suitable for high-precision simulations and data analysis in reactor and solar neutrino experiments~\cite{Abe:2010hy,Wurm:2011zn,Li:2013zyd,Giunti:2014ixa,Giunti:2015gga,Capozzi:2018dat,DARWIN:2020bnc};
	\item Quantify the effects of radiative corrections on observable spectra, reconstruction of the antineutrino energy with liquid scintillators, and discuss implications for new physics searches at leading order in nucleon recoil.
\end{itemize}
With exact three-body phase-space integrals, we violate the rigorous HBChPT power counting but properly account for enhancements near kinematic endpoints in our results.

The structure of the paper is as follows. In Section~\ref{sec:LO}, we review the kinematics of elastic IBD, introduce the HBChPT Lagrangian and relevant coupling constants, and provide the leading-order differential and total cross sections. Section~\ref{sec:corrections} evaluates recoil, weak magnetism, nucleon structure effects, and virtual radiative corrections. In Section~\ref{sec:real}, we compute real photon emission cross sections, detail the treatment of the full three-body phase space, and present new and known results for IBD cross sections. Results for total cross sections, energy spectra, and angular distributions, as well as radiative effects on the reconstruction of the antineutrino energy in liquid scintillator, are discussed in Section~\ref{sec:results}. In this section, we also provide a straightforward extension to the charged-current neutrino-neutron elastic scattering and update the radiative corrections to the neutron decay.  We present our conclusions and outline future directions in Section~\ref{sec:summary}. Supplemental material~\cite{supplement} and github.com/tomalak7/IBDxsec include a Mathematica notebook and a Python library for fast, accurate evaluation of IBD cross sections.

\section{Inverse beta decay}
\label{sec:LO}

We review the kinematics of antineutrino scattering on protons in Section~\ref{sec:kinematics}. We then describe the low-energy effective field theory for charged-current electroweak processes involving (anti)neutrinos and nucleons in Section~\ref{sec:HBChPT} and specify the low-energy coupling constants and experimental inputs. In Section~\ref{sec:cross_section}, we present the leading-order cross sections in the inverse beta decay reaction.

\subsection{Kinematics in charged-current antineutrino-proton elastic scattering}
\label{sec:kinematics}

\begin{figure}[H]
\begin{center}
	\includegraphics[scale=0.23]{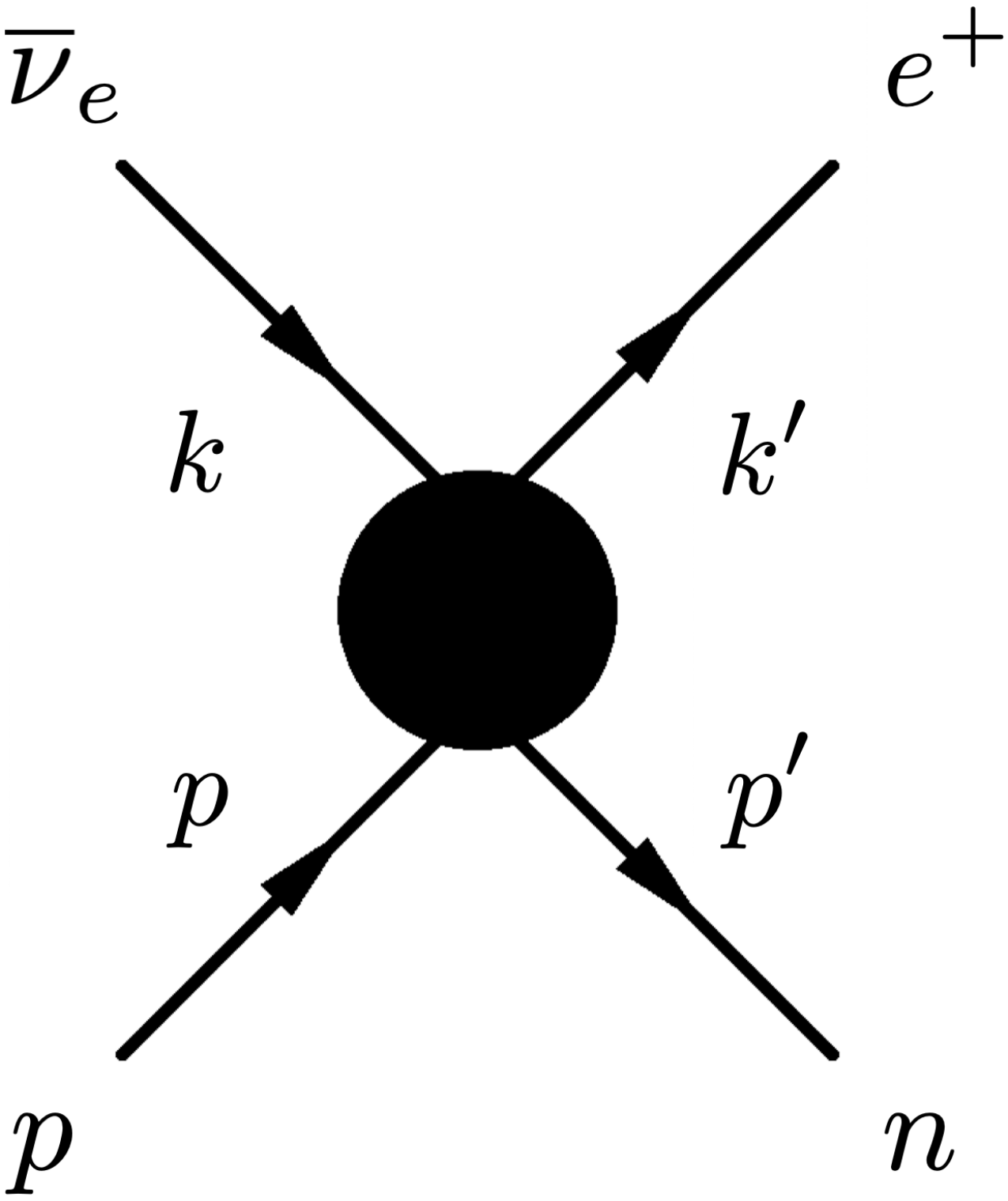}
	\caption{Kinematics in inverse beta decay. \label{fig:diagram}}
\end{center}
\end{figure}

We consider the charged-current scattering of antineutrinos on protons $\overline{\nu}_e p \to e^+ n$. For hydrocarbon or water targets, the molecular binding energy and momentum are negligible compared to the energy and momentum transferred during the scattering. Therefore, we take the initial proton to be at rest in the laboratory frame, with kinematics given by $p^\mu = (m_p,~0)$ (initial proton with $p^2=m_p^2$), $p^{\prime\mu} = (E_n,~\bk-\bk^\prime)$ (final neutron with $p'^2 = m_n^2$), $ k^\mu = (E_{\overline{\nu}_e},~\bk)$ (initial antineutrino), and $k^{\prime\mu} = (E_e,~\bk^\prime)$ (final positron with $k'^2 = m_e^2$) (see Fig.~\ref{fig:diagram}). The antineutrino mass is negligible compared to the masses of the other particles and typical beam energies and is thus ignored in all expressions. We denote the momentum transfer on the nucleon line as $q^\mu = p^{\prime\mu} - p^\mu$.

This two-to-two elastic scattering process is described by two independent kinematic variables. We define the invariant squared momentum transfer as
\begin{equation} \label{eq:squared_momentum_transfer}
	Q^2 = - q^2 = -\left(p^\prime - p \right)^2 = m^2_p - m^2_n + 2 m_p \left( E_{\overline{\nu}_e} - E_e \right),
\end{equation}
and the squared energy in the center-of-mass reference frame as
\begin{equation} \label{eq:squared_cmf_energy}
	s = \left(p + k \right)^2 = m^2_p + 2 m_p E_{\overline{\nu}_e}.
\end{equation}
The IBD process occurs only if the antineutrino energy exceeds the threshold required to produce a positron and compensate for the neutron-proton mass difference. At the level of invariants, $s = m_p^2 + 2 m_p E_{\overline{\nu}_e} \ge \left(m_n + m_e \right)^2$, which translates to the threshold energy $E_{\overline{\nu}_e}^\mathrm{thr}$: $E_{\overline{\nu}_e} \ge E_{\overline{\nu}_e}^\mathrm{thr} = \frac{\left( m_n + m_e \right)^2}{2 m_p} - \frac{m_p}{2}$, with $E_{\overline{\nu}_e}^\mathrm{thr} \approx 1.806066~\mathrm{MeV}$. Expanding in powers of $1/m_n$, we obtain
\begin{equation} \label{eq:threshold_energy_1_over_m}
	E_{\overline{\nu}_e}^\mathrm{thr} = \left( E_0 + m_e \right) \left( 1 + \frac{E_0}{m_n} \right) + \mathcal{O} \left( \frac{1}{m^2_n} \right),
\end{equation}
with $E_0 = \frac{ m^2_n + m^2_e - m^2_p}{2 m_n} \approx 1.292581~\mathrm{MeV}$. Note that $E_0 \approx \Delta m = m_n - m_p$, aligning with conventions in $\beta$-decay literature. For muon and tau antineutrinos, the thresholds are much higher, $E_{\overline{\nu}_\mu}^\mathrm{thr} \approx 113~\mathrm{MeV}$ and $E_{\overline{\nu}_\tau}^\mathrm{thr} \approx3.46~\mathrm{GeV}$, respectively. The positron energy $E_e$ is restricted to a narrow range $E_e^\mathrm{min} \le E_e \le E_e^\mathrm{max}$,
\begin{align}
	E_e^\mathrm{min} &= \frac{ \left( s - m^2_n + m^2_e \right) \left( m_p + E_{\overline{\nu}_e} \right) - E_{\overline{\nu}_e} \sqrt{\Sigma \left( s, m^2_n, m^2_e \right)} }{2 s}, \label{eq:positron_energy_min} \\
	E_e^\mathrm{max} &= \frac{ \left( s - m^2_n + m^2_e \right) \left( m_p + E_{\overline{\nu}_e} \right) + E_{\overline{\nu}_e} \sqrt{\Sigma \left( s, m^2_n, m^2_e \right)} }{2 s}, \label{eq:positron_energy_max}
\end{align}
where $\Sigma \left( s, m^2_n, m^2_e \right) = \left( s - \left( m_n + m_e \right)^2 \right) \left( s - \left( m_n - m_e \right)^2 \right)$ is the kinematic triangle function. The leading terms in $1/m_n$ expansion for the boundaries in the positron energy are given by
\begin{align}
	E_e^\mathrm{min} &= E_{\overline{\nu}_e} - E_0 - \frac{\left( E_{\overline{\nu}_e} - E_0 \right)^2 - m^2_e + E_{\overline{\nu}_e} \left( E_0 + \sqrt{ \left( E_{\overline{\nu}_e} - E_0 \right)^2 - m^2_e} \right)}{m_n} + \mathcal{O} \left( \frac{1}{m^2_n} \right), \label{eq:positron_energy_min_1_over_m} \\
	E_e^\mathrm{max} &= E_{\overline{\nu}_e} - E_0 - \frac{\left( E_{\overline{\nu}_e} - E_0 \right)^2 - m^2_e + E_{\overline{\nu}_e} \left( E_0 - \sqrt{ \left( E_{\overline{\nu}_e} - E_0 \right)^2 - m^2_e} \right)}{m_n} + \mathcal{O} \left( \frac{1}{m^2_n} \right). \label{eq:positron_energy_max_1_over_m}
\end{align}

The antineutrino energy $E_{\overline{\nu}_e}$ can be expressed in terms of the recoil positron energy $E_e$ and scattering angle $\Theta_e$ as
\begin{equation} \label{eq:antineutrino_energy_1_over_m}
	E_{\overline{\nu}_e} = \frac{1}{2} \frac{ m^2_n - m^2_p + 2 m_p E_e - m^2_e}{m_p - E_e + \sqrt{E_e^2 - m^2_e} \cos \Theta_e},
\end{equation}
and its $1/m_n$ expansion is
\begin{equation} \label{eq:antineutrino_energy}
	E_{\overline{\nu}_e} = E_e + E_0 - \frac{E_e E_0 + m^2_e - \left(E_e + E_0 \right) \left( E_e + E_0 - \sqrt{E_e^2 - m^2_e} \cos \Theta_e \right) }{m_n} + \mathcal{O} \left( \frac{1}{m^2_n} \right).
\end{equation}

In the absence of radiation, the angle $\Theta_e$ between the positron and the antineutrino beam direction is
\begin{equation} \label{eq:scattering_angle}
	\cos \Theta_e = \frac{ 2 \left( E_{\overline{\nu}_e} E_e + m_p E_e - m_p E_{\overline{\nu}_e} \right) +m^2_n - m^2_p - m^2_e}{2 E_{\overline{\nu}_e} \sqrt{E^2_e - m^2_e}},
\end{equation}
without any kinematic restrictions on the positron scattering angle, while the neutron always goes in the forward direction. These kinematic relations also apply to muon and tau antineutrinos, with appropriate substitutions for lepton masses and energies. The $1/m_n$ expansion of $\cos \Theta_e$ is given by
\begin{equation} \label{eq:scattering_angle_1_over_m}
	\cos \Theta_e = \frac{ m_n \left( E_e + E_0 - E_{\overline{\nu}_e} \right) }{E_{\overline{\nu}_e} \sqrt{E^2_e - m^2_e}} - \frac{m^2_e + E_e E_0 - \left(E_e + E_0 \right) E_{\overline{\nu}_e} }{E_{\overline{\nu}_e} \sqrt{E^2_e - m^2_e}} + \frac{E_0}{2 m_n} \frac{ E^2_0 - m^2_e}{E_{\overline{\nu}_e} \sqrt{E^2_e - m^2_e}} + \mathcal{O} \left( \frac{1}{m^2_n} \right).
\end{equation}

\subsection{Heavy baryon chiral perturbation theory}
\label{sec:HBChPT}

Inverse beta decay induced by reactor antineutrinos is a low-energy process characterized by scales well below the pion mass and typical hadronic energies. Specifically, the neutron-proton mass difference $\Delta m = m_n - m_p \approx 1.3~\mathrm{MeV}$, the electron mass $m_e \approx 511~\mathrm{keV}$, and the reactor antineutrino energies $\lesssim 10~\mathrm{MeV}$ are all much smaller than the chiral symmetry breaking scale and nucleon mass. This hierarchy enables a systematic description of strong-interaction dynamics using effective field theories. Treating nucleons as heavy static fields, we adopt heavy baryon chiral perturbation theory (HBChPT) to analyze IBD.

In this framework, nucleon mass effects are incorporated as recoil corrections characterized by the small parameter $\epsilon_{\rm recoil} = \frac{q_{\rm ext}}{m_N}$, where $m_N \approx m_p \approx m_n$ denotes the nucleon mass, and $q_{\rm ext}$ is the characteristic momentum transfer in IBD. By integrating out pion degrees of freedom, the low-energy interactions are encoded in the pionless effective field theory (EFT) Lagrangian~\cite{Ando:2004rk,Falkowski:2021vdg,Cirigliano:2022hob},
\begin{equation} \label{eq:Lagrangian_at_leading_order}
	\mathcal L_{\slashpi} = - \sqrt{2} G_F V^\star_{ud} \overline{\overline{\nu}}_e \gamma_\rho \mathrm{P}_\mathrm{L} \overline{e} \overline{N}_v \left( g_V v^\rho - 2 g_A S^\rho \right) \tau^- N_v + \mathcal{O} \left( \frac{\alpha}{\pi}, \epsilon_{\rm recoil}, \epsilon_{\slashpi}, \epsilon_\chi \right) + \mathrm{h.c.},
\end{equation}
with the scale-independent Fermi coupling constant $G_F = 1.1663787(6)\times 10^{-5}~\mathrm{GeV}^{-2}$~\cite{Fermi:1934hr,Feynman:1958ty,vanRitbergen:1999fi,MuLan:2012sih} and the matrix element of the CKM quark mixing matrix $V_{ud}$~\cite{Cabibbo:1963yz,Kobayashi:1973fv,Hardy:2020qwl,ParticleDataGroup:2020ssz}. $\overline{e}$ and $\overline{\nu}_e$ are the positron and antineutrino fields, respectively, $\mathrm{P}_\mathrm{L} = \frac{1-\gamma_5}{2}$ is the projection operator onto left-handed fermions, $N_v = \left( p,~n \right)^T$ denotes the heavy-nucleon field doublet with the field operators for proton and neutron $p$ and $n$, respectively, $v^\rho$ is the nucleon velocity, and $S^\rho = \left( 0, \frac{\vec\sigma}{2} \right)$ denotes the nucleon spin. $\vec{\sigma}$ are the Pauli matrices acting on spin indices, while $\vec{\tau}$ act in the isospin space, satisfying $ [\tau^a, \tau^b] = 2 i \varepsilon^{a b c} \tau^c$, $ \{\tau^a, \tau^b \} = 2 \delta^{a b}$, and $\tau^\pm = \frac{1}{2} \left( \mathrm{\tau}^1 \pm i \mathrm{\tau}^2 \right)$. Besides recoil expansion in $\epsilon_{\rm recoil}$, the scale-dependent nucleon vector $g_V$ and axial-vector $g_A$ coupling constants in Eq.~(\ref{eq:Lagrangian_at_leading_order}) have an expansion in the electromagnetic coupling constant $\alpha$, radiative pion contributions $\epsilon_{\slashpi} = \frac{q_{\rm ext}}{m_\pi}$, and HBChPT chiral corrections $\epsilon_\chi = \frac{m_\pi}{\Lambda_\chi}$ with the pion mass $m_\pi$ and the scale $\Lambda_\chi = 4 \pi F_\pi \approx 1~\mathrm{GeV}$, where $F_\pi$ is the pion decay constant.

Advantageously, we can work only with nucleon, positron, antineutrino, and photon fields in this effective field theory. All chiral corrections are captured inside the low-energy coupling constants $g_V$ and $g_A$, while radiative pion contributions in low-energy theory, which do not have large logarithm enhancements, are suppressed as $\frac{\alpha}{\pi} \epsilon_{\slashpi} \lesssim 1.7 \times 10^{-4}$.\footnote{$\mathcal{O} \left( \frac{\alpha}{\pi} \epsilon_{\slashpi}\right)$ corrections to the nucleon magnetic moment and tensor interaction are evaluated in Ref.~\cite{Cirigliano:2022hob}. The effects from these corrections on IBD cross sections are well below the dominant error sources from the axial-vector coupling constant $g_A$ and the CKM matrix element $V_{ud}$.} Recoil corrections to loops in QED expansion are suppressed as $\frac{\alpha}{\pi} \epsilon_{\rm recoil} \lesssim 2.5 \times 10^{-5}$. Hence, the dominant corrections to the leading order in $1/m_n$ result arise from QED, recoil, and nucleon structure effects.

The coupling constants $g_V$ and $g_A$ include short-distance electroweak contributions from heavy particles ($W$, $Z$, Higgs, top quark), as well as heavy quarks ($b$, $c$), and nonperturbative hadronic physics. These have been recently determined via a top-down systematically improvable EFT matching procedure~\cite{Tomalak:2023xgm,Cirigliano:2023fnz,Cirigliano:2024nfi}, which integrates out electroweak-scale physics and matches the Standard Model onto the low-energy effective field theory (LEFT)~\cite{Hill:2019xqk} to describe weak processes through effective four-fermion interactions down to scales of hadronic physics. Nonperturbative matching between LEFT and HBChPT~\cite{Jenkins:1990jv,Cirigliano:2022hob} at hadronic scales then determines the low-energy coupling constants, which are evolved down to the electron mass scale. In this work, we adopt the vector coupling constant at the electron-mass renormalization scale $g_V = 1.02499(13)$~\cite{Marciano:2005ec,Seng:2018qru,Seng:2018yzq,Czarnecki:2019mwq,Hayen:2020cxh,Shiells:2020fqp,Feng:2020zdc,Cirigliano:2023fnz,Ma:2023kfr,Moretti:2025qxt,Crosas:2025xyv,Cao:2025zxs}. For the axial-vector-to-vector coupling constant ratio at low energy $\lambda \equiv g_A / g_V$, we take the PDG quote $\lambda^{\mathrm{I}} = 1.2754(13)$~\cite{Beck:2019xye,ParticleDataGroup:2024cfk,Beringer:2024ady,Workman:2022ynf} for evaluating central values and estimating errors. We also consider the most precise measurement of the beta asymmetry in polarized neutron decay by PERKEO-III $\lambda^{\mathrm{II}} = 1.27641(45)_\mathrm{stat}(33)_\mathrm{sys}$~\cite{Markisch:2018ndu,Dubbers:2018kgh} for optimistic error analysis. To avoid connection to discrepancies in the axial-vector coupling constant and potential biases related to the neutron lifetime puzzle~\cite{Arzumanov:2000ma,Dewey:2003hc,Nico:2004ie,Serebrov:2004zf,Paul:2009md,Pichlmaier:2010zz,Arzumanov:2012zz,Steyerl:2012zz,Yue:2013qrc,Ezhov:2014tna,Arzumanov:2015tea,Serebrov:2017bzo,Pattie:2017vsj,Wietfeldt:2018upi,Czarnecki:2018okw,Castelvecchi:2021hec,UCNt:2021pcg,Gardner:2023wyl,Fuwa:2024cdf}, we take the matrix element $V_{ud} = 0.97348(31)$ from the $g_A$-independent analysis~\cite{Zimmer:2000nti,aSPECT:2008vas,Markisch:2018ndu,Dubbers:2018kgh,Beck:2019xye,ParticleDataGroup:2024cfk,Beringer:2024ady,Workman:2022ynf} of superallowed nuclear $\beta$ decays~\cite{Hardy:2020qwl}, incorporating recent updates to short-distance radiative corrections~\cite{Cirigliano:2023fnz}.\footnote{For both $g_V$ and $\lambda$, we account for further refinements of the phase-space integration from Section~\ref{sec:neutron_decay} of this paper.} This ensures that the dominant uncertainties in the IBD cross section stem solely from the axial-vector coupling constant $g_A$ and the CKM matrix element $V_{ud}$.

\subsection{Leading-order cross sections}
\label{sec:cross_section}

The leading-order differential cross section for the IBD process can be expressed in terms of the spin-averaged leading-order (LO) matrix element $T_\mathrm{LO}$,
\begin{equation} \label{eq:leading_order_amplitude} 
	T_\mathrm{LO} = - \sqrt{2} G_F V^\star_{ud} \overline{\overline{\nu}}_e \gamma_\rho \mathrm{P}_\mathrm{L} \overline{e} \overline{N}_v \left( g_V v^\rho - 2 g_A S^\rho \right) \tau^- N_v,
\end{equation}
where the notation follows Section~\ref{sec:HBChPT}. Using a standard flux normalization, the corresponding differential cross section is
\begin{equation} \label{eq:LO_in_terms_of_the amplitude}
	\frac{\mathrm{d}\sigma_{\rm LO}}{\mathrm{d} Q^2} = \frac{1}{128 \pi m^2_p E_{\overline{\nu}_e}^2} \sum \limits_{\mathrm{spins}} |T_\mathrm{LO} |^2,
\end{equation}
with $Q^2$ being the squared momentum transfer and $E_{\overline{\nu}_e}$ being the incident antineutrino energy. Evaluating the spin sums, the differential cross section can be compactly written as
\begin{equation} \label{eq:LO_static_diff}
	\frac{\mathrm{d}\sigma_{\rm LO}}{\mathrm{d} Q^2} = \frac{{G}_{\rm F}^2 |V_{ud}|^2}{2\pi} \bigg[ \left( 1 - \frac{E_0}{E_{\overline{\nu}_e}} \right) \left( g_V^2 + g_A^2 \right) - \frac{Q^2+m^2_e}{4 E_{\overline{\nu}_e}^2} \left( g_V^2 - g_A^2 \right) \bigg].
\end{equation}
Notably, this cross section is nonzero for all physically allowed kinematics.

The positron angular distribution and the energy spectrum can be easily obtained as\footnote{To avoid extra definitions for recoil corrections, we specify the proton mass $m_p$ instead of the nucleon mass $m_N$. Previous derivations~\cite{LlewellynSmith:1971uhs,Vogel:1983hi,Fayans:1985uej,Lopez:1997ki,Vogel:2006sg,Tomalak:2021hec,Tomalak:2022xup,Altarawneh:2024sxo} do not include $E_0/E_{\overline{\nu}_e}$ contribution to leading order. References~\cite{Vogel:1983hi,Fayans:1985uej,Seckel:1993dc,Lopez:1997ki,Esposito:1998rc,Vogel:1999zy,Strumia:2003zx,Raha:2011aa,Ankowski:2016oyj,Hayes:2016qnu,Ricciardi:2022pru,Altarawneh:2024sxo} consider $m_n-m_p$ corrections.}
\begin{align}
	\frac{\mathrm{d} \sigma_{\rm LO}}{\mathrm{d} \cos \Theta_e} &= \frac{E_{\overline{\nu}_e} \sqrt{E_e^2 - m_e^2}}{m_p + E_{\overline{\nu}_e} - \frac{E_e E_{\overline{\nu}_e}}{\sqrt{E_e^2 - m_e^2}} \cos \Theta_e} 2 m_p \frac{\mathrm{d} \sigma_{\rm LO}}{\mathrm{d} Q^2}, \label{eq:LO_static_diff_angle} \\
	\frac{\mathrm{d} \sigma_{\rm LO}}{\mathrm{d} \cos \Theta_e} &= \frac{E_{\overline{\nu}_e} \sqrt{E_e^2 - m_e^2}}{m_p + E_{\overline{\nu}_e} - \frac{E_e E_{\overline{\nu}_e}}{\sqrt{E_e^2 - m_e^2}} \cos \Theta_e} \frac{m_p}{\pi} {G}_{\rm F}^2 |V_{ud}|^2 \nonumber \\
&\bigg[ \left( 1 - \frac{E_0}{E_{\overline{\nu}_e}} - \frac{E_e}{2 E_{\overline{\nu}_e}} \left( 1 - \beta \cos \Theta_e \right) \right) g_V^2 + \left( 1 - \frac{E_0}{E_{\overline{\nu}_e}} + \frac{E_e}{2 E_{\overline{\nu}_e}} \left( 1 - \beta \cos \Theta_e \right) \right) g_A^2 \bigg], \label{eq:LO_static_diff_angle_full} \\
	\frac{\mathrm{d} \sigma_{\rm LO}}{\mathrm{d} E_e} &= 2 m_p \frac{\mathrm{d} \sigma_{\rm LO}}{\mathrm{d} Q^2}, \label{eq:LO_static_diff_energy} 
\end{align}
where $\Theta_e$ is the positron scattering angle, and $\beta = \sqrt{1 - \frac{m_e^2}{E_e^2}}$ is the positron velocity. Expanding to leading order in the large nucleon mass, the angular distribution simplifies to the well-known form
\begin{equation} \label{eq:LO_static_diff_angle_full_1_over_m}
	\frac{\mathrm{d} \sigma_{\rm LO}}{\mathrm{d} \cos \Theta_e} = \frac{ E_e \sqrt{E_e^2 - m_e^2} }{2 \pi} {G}_{\rm F}^2 |V_{ud}|^2 \bigg[ \left( 1 + \beta \cos \Theta_e \right) g_V^2 + \left( 3 - \beta \cos \Theta_e \right) g_A^2 \bigg] + \mathcal{O} \left( \frac{1}{m_n} \right).
\end{equation}

The total cross section at leading order is given by
\begin{equation} \label{eq:LO_static_tot}
	\sigma_{\rm LO} = \frac{{G}_{\rm F}^2 |V_{ud}|^2}{\pi} \left( E_{\overline{\nu}_e} - E_0 \right) \frac{\sqrt{\Sigma \left( s, m^2_n, m^2_e \right)}}{2m_n} \left( g_V^2 + 3 g_A^2 \right).
\end{equation}
To avoid infinite relative recoil corrections and potential threshold problems similar to the HBChPT calculation of the nucleon vector form factors~\cite{Bernard:1996cc,Becher:1999he,Gegelia:1999gf,Becher:2001hv,Fuchs:2003qc,Kaiser:2003qp}, as it was also pointed in Refs.~\cite{Fayans:1985uej,Vogel:1999zy} and fully included analytically in Ref.~\cite{Fayans:1985uej}, we keep the full kinematic dependence of the triangle function $\Sigma \left( s, m^2_n, m^2_e \right)$, which vanishes at the IBD threshold. The leading term of the total cross section in $1/m_n$ expansion reads
\begin{equation} \label{eq:LO_static_tot_1_over_m}
	\sigma_{\rm LO} = \frac{{G}_{\rm F}^2 |V_{ud}|^2}{\pi} \left( E_{\overline{\nu}_e} - E_0 \right) \sqrt{\left( E_{\overline{\nu}_e} - E_0 \right)^2 - m_e^2} \left( g_V^2 + 3 g_A^2 \right) + \mathcal{O} \left( \frac{1}{m_n} \right).
\end{equation}

At leading order, the total cross section can be expressed in terms of the neutron lifetime $\tau_n$ as~\cite{Vogel:1983hi,Fayans:1985uej,Esposito:1998rc,Vogel:1999zy,Bernabeu:2003rx,Formaggio:2013kya,Hayes:2016qnu}
\begin{equation} \label{eq:LO_tot_neutron_lifetime}
	\sigma_{\rm LO} = \frac{\pi^2}{ m_n \tau_n} \frac{ \left( E_{\overline{\nu}_e} - E_0 \right) \sqrt{\Sigma \left( s, m^2_n, m^2_e \right)}}{ f_0m_e^5},
\end{equation}
where the phase-space integral in neutron decay $f_0$ is
\begin{equation} \label{eq:neutron_lifetime_phase_space_LO}
	f_0 =\frac{2x_0^4-9x_0^2-8}{60} \sqrt{x_0^2-1}+ \frac{x_0}{4} \ln \left(x_0+\sqrt{x_0^2-1} \right),
\end{equation}
with $x_0 = \frac{E_0}{m_e}$ and $f_0( x_0 ) = 1.62989$. Similarly, the differential cross section can be expressed in terms of $\tau_n$, and the axial-vector-to-vector coupling constant ratio $\lambda = g_A / g_V$ as
\begin{equation} \label{eq:LO_diff_neutron_lifetime}
	\frac{\mathrm{d}\sigma_{\rm LO}}{\mathrm{d} Q^2} = \frac{\pi^2}{ \tau_n} \frac{1}{ f_0m_e^5} \frac{\left( 1 - \frac{E_0}{E_{\overline{\nu}_e}} \right) \left( 1 + \lambda^2 \right) - \frac{Q^2+m^2_e}{4 E_{\overline{\nu}_e}^2} \left( 1 - \lambda^2 \right)}{ 1 + 3 \lambda^2 }.
\end{equation}

Incorporating QED, recoil, weak magnetism, and phase-space corrections in the neutron decay can be effectively achieved by replacing the neutron lifetime in Eqs.~(\ref{eq:LO_tot_neutron_lifetime}) and~(\ref{eq:LO_diff_neutron_lifetime}) as
\begin{equation} \label{eq:LO_neutron_lifetime_with_corrections}
	\frac{1}{\tau_n} \to \frac{1}{1 + \Delta_{\rm TOT} - \left( g^2_V - 1\right)} \frac{1}{\tau_n},
\end{equation}
with the relative correction $\Delta_{\rm TOT} - \left( g^2_V - 1\right) = 2.700(7) \%$~\cite{Cirigliano:2023fnz,VanderGriend:2025mdc}, which is dominated by the Coulomb enhancement.

\section{Higher-order corrections}
\label{sec:corrections}

In this section, we discuss higher-order contributions to the inverse beta decay process. We begin with recoil effects in Section~\ref{sec:recoil}, which include kinematic and structure-dependent corrections due to the finite nucleon mass and the neutron-proton mass difference. In Section~\ref{sec:weak_magnetism}, we consider the leading contribution from the nucleon isovector-vector magnetic moment, commonly referred to as weak magnetism. The corrections arising from strong interaction dynamics via the nucleon form factors are presented in Section~\ref{sec:form_factors}. Finally, in Section~\ref{sec:virtual_QED}, we discuss QED vertex corrections arising from virtual photon exchange.

\subsection{Recoil corrections}
\label{sec:recoil}

In this section, we specify the recoil corrections to unpolarized differential and total cross sections in Eqs.~(\ref{eq:LO_static_diff}) and~(\ref{eq:LO_static_tot}), denoted as $\mathrm{d} \sigma_\mathrm{recoil}$ and $\sigma_\mathrm{recoil}$, respectively.

The correction to the differential cross section arises from recoil effects and the neutron-proton mass difference within the nucleon matrix element. It takes the form
\begin{equation} \label{eq:recoil_differential}
	\frac{\mathrm{d}\sigma_\mathrm{recoil}}{\mathrm{d} Q^2} = \frac{{G}_{\rm F}^2 |V_{ud}|^2}{2\pi} \frac{E_0}{m_n} \bigg[ - \frac{E_0}{E_{\overline{\nu}_e}} \left( g_V^2 + g_A^2 \right) - \frac{Q^2+m^2_e}{4 E_{\overline{\nu}_e}^2} \left( g_V^2 - g_A^2 \right) + \left( \frac{Q^2+m^2_e}{4 E_{\overline{\nu}_e}^2} - \frac{Q^2}{2 E_0 E_{\overline{\nu}_e}}\right) \left( g_V - g_A \right)^2 \bigg].
\end{equation}
The total cross-section correction also accounts for kinematic modifications of the integration limits and is given by
\begin{align}
	\sigma_\mathrm{recoil} &= - 2 \frac{{G}_{\rm F}^2 |V_{ud}|^2}{\pi} \sqrt{\left( E_{\overline{\nu}_e} - E_0 \right)^2 - m_e^2} \nonumber \\
&\left[ \left( \frac{ \left(E_{\overline{\nu}_e} - E_0 \right)^2}{m_n} + \frac{E_0 E_{\overline{\nu}_e}}{2 m_n} \right) \left( g_V^2 + 3 g_A^2 \right) - \left( \frac{\left(2 E_{\overline{\nu}_e} - E_0 \right) \left( E_{\overline{\nu}_e} - E_0 \right)}{m_n} - \frac{m^2_e}{m_n} \right) g_A \left( g_V - g_A \right) \right]. \label{eq:recoil_total}
\end{align}

The recoil correction to the differential unpolarized IBD cross section is negative and can reach $1.5\%$ at $E_{\overline{\nu}_e} = 5~\mathrm{MeV}$ and $3.1\%$ at $E_{\overline{\nu}_e} = 10~\mathrm{MeV}$. This suppression becomes more pronounced near the IBD threshold, reaching up to $3.2\%$. The corresponding impacts on the total unpolarized IBD cross section are $1.9\%$ at $E_{\overline{\nu}_e} = 5~\mathrm{MeV}$ and $4.0\%$ at $E_{\overline{\nu}_e} = 10~\mathrm{MeV}$, with the effect continuing to grow with the antineutrino energy. Theoretical uncertainties in the recoil correction are negligible for practical purposes, remaining below $10^{-5}$ for the total and below $10^{-6}$ for the differential cross section.

\subsection{Weak magnetism}
\label{sec:weak_magnetism}

The leading weak magnetism contributions arise from the nucleon magnetic moments, encoded in the $\mathcal{O}(p^2)$ pion-nucleon interaction Lagrangians~\cite{Gasser:1987rb,Krause:1990xc,Ecker:1995rk,Bernard:1995dp,Muller:1999ww}.\footnote{We define the weak magnetism correction to be proportional to the anomalous magnetic moment, similar to Refs.~\cite{Vogel:1983hi} and~\cite{Seckel:1993dc}.} These contributions correct the unpolarized differential and total cross sections in Eqs.~(\ref{eq:LO_static_diff}) and~(\ref{eq:LO_static_tot}) and are denoted as $\mathrm{d}\sigma_\mathrm{wm}$ and $\sigma_\mathrm{wm}$, respectively.

The corrections to the differential and total cross sections are given by
\begin{align}
	\frac{\mathrm{d}\sigma_\mathrm{wm}}{\mathrm{d} Q^2} &= \frac{{G}_{\rm F}^2 |V_{ud}|^2}{2\pi} \left( \frac{E_0}{m_n} \frac{Q^2+m^2_e}{2E_{\overline{\nu}_e}^2} - \frac{Q^2}{m_n E_{\overline{\nu}_e} } \right) g_A \left( \mu_p - \mu_n - 1 \right), \label{eq:weak_magnetism_differential} \\
	\sigma_\mathrm{wm}&= 2 \frac{{G}_{\rm F}^2 |V_{ud}|^2}{\pi} \sqrt{\left( E_{\overline{\nu}_e} - E_0 \right)^2 - m_e^2} \left( \frac{E_0 \left( E_{\overline{\nu}_e} - E_0 \right)}{m_n} - \frac{Q^2_0}{m_n} \right) g_A \left( \mu_p - \mu_n - 1 \right), \label{eq:weak_magnetism_total}
\end{align}
with the proton $\mu_p$ and neutron $\mu_n$ magnetic moments entering the isovector-vector combination under the assumption of isospin symmetry.

In contrast to neutron beta decay, where the weak magnetism corrections to the total decay rate vanish at leading order, the IBD process receives a nonzero leading-order correction from the weak magnetism.

Numerically, the weak magnetism correction to the differential unpolarized IBD cross section is negative and increases with the antineutrino energy. It reaches approximately $2.6\%$ at $E_{\overline{\nu}_e} = 5~\mathrm{MeV}$ and $5.6\%$ at $E_{\overline{\nu}_e} = 10~\mathrm{MeV}$. The total cross section is similarly affected, with corrections of $1.4\%$ and $3.1\%$ at these energies, respectively. The parametric uncertainty associated with the weak magnetism correction is below $10^{-5}$ for both total and differential cross sections and is therefore negligible for phenomenological applications.

\subsection{Corrections from nucleon radii}
\label{sec:form_factors}

The coupling constants in Eq.~(\ref{eq:Lagrangian_at_leading_order}) serve as the normalization of the nucleon form factors and incorporate all chiral corrections. However, strong interactions also induce a nontrivial dependence on the momentum transfer, $Q^2$, along the nucleon line in IBD. The leading-order effect in the low-momentum transfer expansion is governed by the nucleon mean-squared radii. Specifically, the vector and axial-vector coupling constants are modified as
\begin{align}
	g_V \to g_V \left( 1 - \frac{r^2_V Q^2}{6} + \mathcal{O} \left( \frac{Q^4}{\Lambda^4_\chi} \right)\right), \label{eq:vector_radii_expansion} \\
	g_A \to g_A \left( 1 - \frac{r^2_A Q^2}{6} + \mathcal{O} \left( \frac{Q^4}{\Lambda^4_\chi} \right)\right), \label{eq:axial_vector_radii_expansion}
\end{align}
with the isovector-vector $r_V$ and axial-vector $r_A$ radii of the nucleon. These radii cannot be predicted within chiral perturbation theory alone, as they depend on low-energy constants specific to nucleon structure~\cite{Bernard:1998gv,Bernard:2001rs,Schindler:2006it,Yao:2017fym}. Therefore, we take them from fits to the electron-proton and electron-deuteron scattering and muonic hydrogen spectroscopy data in Ref.~\cite{Borah:2020gte} for the isovector-vector radius, $r^2_V = 0.578\left(2\right)~\mathrm{fm}^2$,\footnote{We neglect percent-level isospin-breaking corrections in the relation between neutral- and charged-current processes that are below the current and near-future errors from the nucleon axial-vector radius.} and from the weighted average of fits to neutrino-deuterium bubble-chamber data~\cite{Meyer:2016oeg}, muon capture on protons~\cite{Hill:2017wgb}, and antineutrino scattering off hydrogen in plastic scintillator targets~\cite{MINERvA:2023avz} for the axial-vector radius, $r^2_A = 0.48\left(14\right)~\mathrm{fm}^2$. The resulting nucleon radii corrections to unpolarized cross sections in Eqs.~(\ref{eq:LO_static_diff}) and (\ref{eq:LO_static_tot}), $\mathrm{d} \sigma_\mathrm{FF}$ and $\sigma_\mathrm{FF}$, respectively, are given by
\begin{align}
	\frac{\mathrm{d}\sigma_\mathrm{FF}}{\mathrm{d} Q^2} &= -\frac{{G}_{\rm F}^2 |V_{ud}|^2}{2\pi} \bigg[ \left( 1 - \frac{E_0}{E_{\overline{\nu}_e}} \right) \left( g_V^2 r^2_V + g_A^2 r^2_A \right) - \frac{Q^2+m^2_e}{4E_{\overline{\nu}_e}^2} \left(g_V^2 r^2_V - g_A^2 r^2_A \right) \bigg] \frac{Q^2}{3}, \label{eq:radii_differential} \\
	\sigma_\mathrm{FF} &= - 2 \frac{{G}_{\rm F}^2 |V_{ud}|^2}{\pi} \sqrt{\left( E_{\overline{\nu}_e} - E_0 \right)^2 - m_e^2} \nonumber \\
 &\left[ \frac{\left( E_{\overline{\nu}_e} - E_0 \right) Q^2_0}{3} \left( g_V^2 r^2_V + g_A^2 r^2_A - \frac{2}{3} \left(g_V^2 r^2_V - g_A^2 r^2_A \right) \right) + \frac{\left( E_{\overline{\nu}_e} + E_0 \right) m^2_e}{18} \left( g_V^2 r^2_V - g_A^2 r^2_A \right) \right], \label{eq:radii_total}
\end{align}
with $Q^2_0 = 2 E_{\overline{\nu}_e} \left( E_{\overline{\nu}_e} - E_0 \right) - m^2_e$.

These nucleon radii corrections are negative and relatively small. For the differential unpolarized IBD cross section, they reach at most $0.029(14)\%$ at $E_{\overline{\nu}_e} = 5~\mathrm{MeV}$ and $0.14(6)\%$ at $E_{\overline{\nu}_e} = 10~\mathrm{MeV}$. The corresponding total cross-section corrections are $0.016(6)\%$ and $0.07(3)\%$, respectively. As expected, the impact of nucleon radii grows with increasing antineutrino energy. With future neutrino experiments like DUNE~\cite{DUNE:2020ypp,DUNE:2022aul}, the corresponding form-factor uncertainties in IBD cross sections can be reduced to a negligible relative size below $10^{-4}$ after improving the precision of the nucleon axial-vector radius~\cite{Petti:2023abz}.

\subsection{Virtual QED corrections}
\label{sec:virtual_QED}

\begin{figure}[H]
\begin{center}
	\includegraphics[scale=0.25]{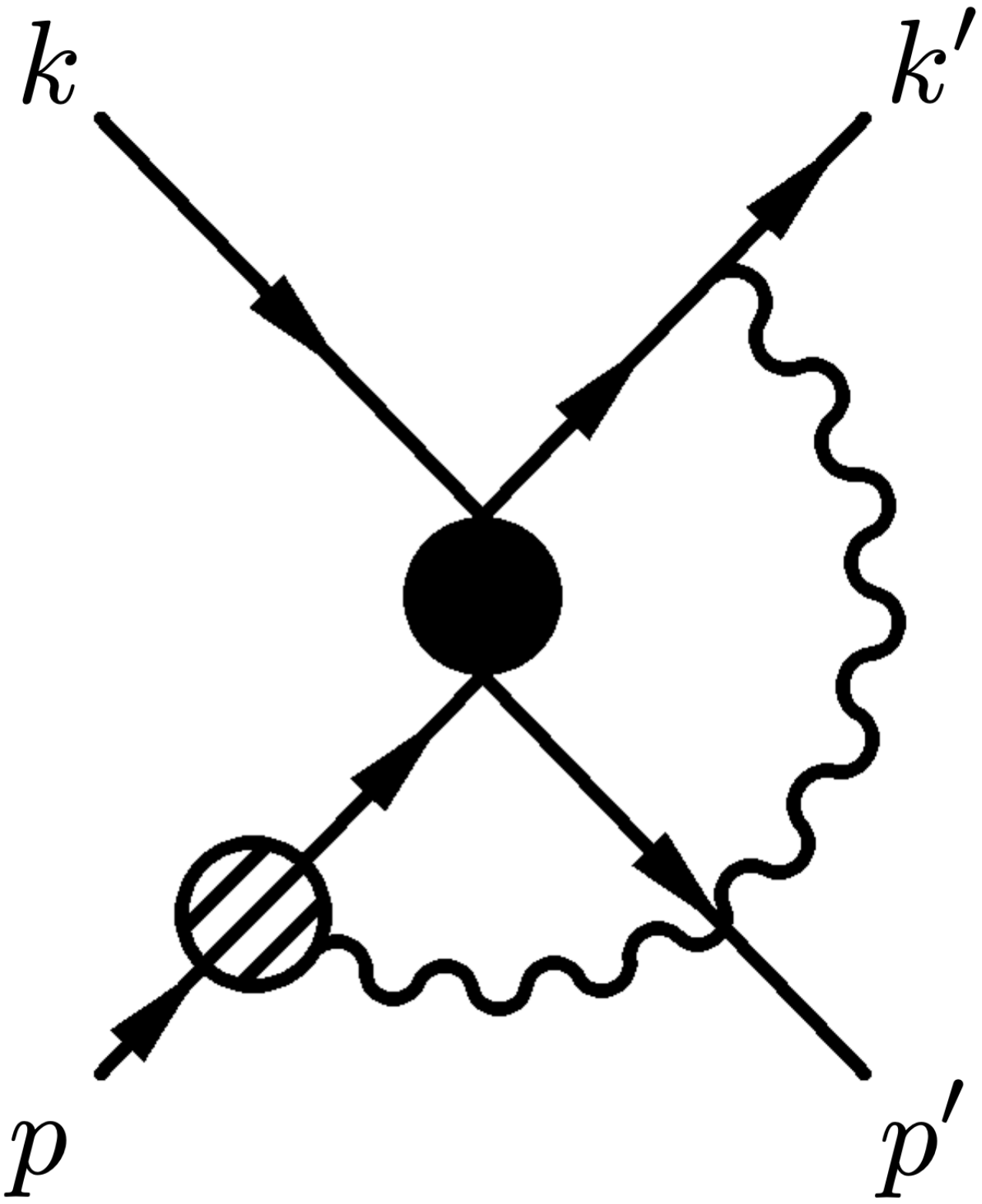}
	\caption{One-loop virtual QED diagram contributing to the IBD process. \label{fig:one_loop_QED}}
\end{center}
\end{figure}

We analyze the one-loop virtual QED corrections to charged-current antineutrino-proton elastic scattering, $\overline{\nu}_e p \to e^+ n$. Within the effective field theory framework, only the single triangle diagram in Fig.~\ref{fig:one_loop_QED} contributes. We also include the standard external field renormalization factors for the heavy proton and positron lines.

The one-loop vertex correction to the matrix element is computed in the $\overline{\mathrm{MS}}_\chi$ renormalization scheme~\cite{Gasser:1983yg,Cirigliano:2023fnz,Cirigliano:2024nfi} in dimensional regularization with $d = 4 - 2 \varepsilon$ dimensions of spacetime to regularize the ultraviolet divergence. The leading-order amplitude, $T_\mathrm{LO}$, receives a correction $\delta T_\mathrm{LO}$,\footnote{For numerical evaluations in this paper, we take the electromagnetic coupling constant $\alpha$ in the $\overline{\mathrm{MS}}_\chi$ renormalization scheme at the scale $\mu_\chi = m_e$, which is related to its value in the Thomson limit $\alpha_0$ as $\frac{1}{\alpha} = \frac{1}{\alpha_0} + \frac{1}{3 \pi}$~\cite{Cirigliano:2023fnz}.}
\begin{equation} \label{eq:next_to_leading_order_amplitude}
	\delta T_\mathrm{LO} = \sqrt{2} G_F V^\star_{ud} e^2 \int \frac{\mathrm{d}^d L}{ \left( 2 \pi \right)^d} \overline{\overline{\nu}}_e \gamma_\rho \mathrm{P}_\mathrm{L} \frac{ \slashed{k}^\prime+ \slashed{L} + m_e}{\left( k^\prime + L \right)^2-m_e^2} \gamma_\mu \overline{e} \mathrm{\Pi}^{\mu \nu} \left( L \right) \overline{N}_v \left( g_V v^\rho - 2 g_A S^\rho \right) \tau^- \frac{v_\nu}{v \cdot L} N_v,
\end{equation}
where $\slashed{k} \equiv k_\mu \gamma^\mu$ for any four-vector $k$, and the photon propagator $\mathrm{\Pi}^{\mu \nu}$ in momentum space is
\begin{equation} \label{eq:photon_propagator}
	\mathrm{\Pi}^{\mu \nu} \left( L \right) = \frac{i}{L^2 - \lambda_\gamma^2} \left[ - g^{\mu \nu} + \left( 1 - \xi_\gamma \right) \frac{L^{\mu} L^{\nu}}{L^2 - a \xi_\gamma \lambda_\gamma^2 }\right].
\end{equation}
Here, $\lambda_\gamma$ is the photon mass acting as an infrared (IR) regulator, $\xi_\gamma$ is the gauge-fixing parameter, and $a$ is an arbitrary constant.

The field renormalization factors for the external positron and proton, $Z_\ell$ and $Z_h$, respectively, are given by
\begin{align}
	Z_\ell &= 1 - \frac{\alpha}{4\pi} \frac{\xi_\gamma}{\varepsilon}- \frac{\alpha}{4\pi} \left( \ln \frac{\mu^2}{m_e^2} + 2\ln \frac{\lambda_\gamma^2}{m_e^2} + 4 \right) + \frac{\alpha}{4\pi} \left( 1 - \xi_\gamma \right) \left( \ln \frac{\mu^2}{\lambda_\gamma^2} + 1 + \frac{a \xi_\gamma \ln a \xi_\gamma }{1 - a \xi_\gamma } \right), \label{eq:Z_positron} \\
	Z_h &= 1 + \frac{\alpha}{2 \pi} \left( \frac{1}{\varepsilon} + \ln \frac{\mu^2}{\lambda_\gamma^2} \right)+ \frac{\alpha}{4 \pi} \left( 1 - \xi_\gamma \right) \left( \frac{1}{\varepsilon} + \ln \frac{\mu^2}{\lambda_\gamma^2} + 1 + \frac{a \xi_\gamma \ln a \xi_\gamma }{1 - a \xi_\gamma } \right). \label{eq:Z_proton}
\end{align}

We omit Lorentz structures that vanish upon contraction with the lepton current in the massless neutrino limit, as they do not contribute. The resulting correction to the scattering amplitude can be written as
\begin{equation} \label{eq:f1f2}
	\left( \sqrt{Z_h Z_\ell } - 1 \right) T_\mathrm{LO} + \delta T_\mathrm{LO} = \frac{\alpha}{\pi} \left( f_1 T_\mathrm{LO} + f_2 T_v \right),
\end{equation}
where the new tensor structure $T_v$ is
\begin{equation} \label{eq:vertex_structure}
	T_v = \sqrt{2} G_F V^\star_{ud} \overline{\overline{\nu}}_e \gamma_\rho \mathrm{P}_\mathrm{L} \slashed{v} \overline{e} \overline{N}_v \left( g_V v^\rho - 2 g_A S^\rho \right) \tau^- N_v,
\end{equation}
and the renormalized gauge-invariant QED form factors $f_1$ and $f_2$ are
\begin{align}
	f_1 \left(\beta \right) &= \frac{3}{8} \left( \ln \frac{\mu^2}{m_e^2} - 1 \right) + \frac{1}{2} \left(1 - \frac{1}{2 \beta} \ln \frac{1 + \beta}{1-\beta} \right) \ln \frac{m_e^2}{\lambda_\gamma^2} \nonumber \\
&+ \frac{1}{4\beta} \left( \mathrm{Li}_2 \frac{\beta +1}{2\beta} - \mathrm{Li}_2 \frac{\beta-1}{2\beta} + \left( 1 - \ln \frac{2\beta}{1-\beta} \right) \ln \frac{1 + \beta}{1-\beta} +\frac{1}{2}\ln^2 \frac{1 + \beta}{1-\beta} - \frac{\pi^2}{2}+ 2 \pi^2 \delta_C \right), \label{eq:F1_FF} \\
	f_2 \left(\beta \right) &= \frac{\rho}{4 \beta} \ln \frac{1+\beta}{1-\beta}, \label{eq:F2_FF}
\end{align}
where the positron velocity $\beta$ and parameter $\rho$ are defined as
\begin{equation} \label{eq:beta_introduced}
	\beta = \sqrt{1 - \frac{m_e^2}{E_e^2}}, \qquad \rho = \sqrt{1-\beta^2} = \frac{m_e}{E_e}.
\end{equation}
The structure of virtual QED corrections matches those in the neutron beta decay when evaluated in the neutron rest frame~\cite{Ando:2004rk,Cirigliano:2022hob,Cirigliano:2023fnz}. In IBD, there are no Coulomb corrections and $\delta_C = 0$. In the neutron beta decay and charged-current neutrino-neutron elastic scattering, the leading-order Coulomb correction is determined by $\delta_C = 1$. To include the $\frac{\pi}{\beta}$-enhanced higher-order Coulomb corrections and to properly resum them in the nonperturbative regime when $\beta \lesssim \alpha$, one can introduce the nonrelativistic Fermi function, as it is described in Refs.~\cite{Wilkinson:1982hu} and~\cite{Cirigliano:2023fnz}, cf. also Refs.~\cite{Gamow:1928zz,Sommerfeld:1931qaf,Fermi:1934hr,Konopinski:1935zz,Morita:1963zz,Wilson:1968pwx,Halpern:1968zz,Halpern:1970it,Hoang:1997sj,Hoang:1997ui,Czarnecki:1997vz,Beneke:1999qg,Hoferichter:2009gn,Matsuzaki:2012qb,Matsuzaki:2013twa,Hill:2023acw}.

The vertex correction to the unpolarized cross section can be expressed as a sum of factorizable and nonfactorizable parts,
\begin{equation} \label{eq:virtual_correction} 
	\mathrm{d} \sigma_{v} = \frac{\alpha}{\pi} \delta_v \mathrm{d} \sigma_{\mathrm{LO}} + \mathrm{d} \sigma_{v, \mathrm{NF}},
\end{equation}
where the factorizable correction $\delta_v$ is
\begin{equation} \label{eq:vector_form_factor_correction}
	\delta_v = 2 f_1,
\end{equation}
and the nonfactorizable component $\mathrm{d} \sigma_{v, \mathrm{NF}}$ reads
\begin{equation} \label{eq:virtual_correction_nf2}
	\frac{\mathrm{d} \sigma_{v, \mathrm{NF}}}{\mathrm{d} Q^2} = \frac{\alpha}{\pi} f_2 \frac{{G}_{\rm F}^2 |V_{ud}|^2}{2\pi} \frac{m_e}{E_{\overline{\nu}_e} } \left( g^2_V + 3 g^2_A \right).
\end{equation}
In the massless positron limit, the Pauli form factor vanishes, $f_2 \left(\beta \right) \to 0$, and the correction becomes exactly factorizable.

The relative contribution to the differential and total cross sections from the form factor $f_2$ is positive and below $0.23\%$ value by magnitude for the correction at the threshold. The size of this contribution decreases as the antineutrino energy increases away from the production threshold.

\section{Real photon emission}
\label{sec:real}

Oscillation analyses of reactor antineutrino data depend critically on the accuracy of the antineutrino energy reconstructed from the final-state positron kinematics. The emission of real photons can impair this reconstruction, as the energy response of liquid scintillators to photons and positrons is strongly dependent on both the particle type and its energy~\cite{JUNO:2025gmd}. Therefore, a detailed understanding, both direct and indirect, of the kinematics of the positron and photon in radiative IBD is essential. In this section, we evaluate the bremsstrahlung process in inverse beta decay, which involves the emission of a single real photon. We begin in Section~\ref{sec:bremsstrahlung_general} by presenting the general formalism and basic expressions for radiative IBD working with matrix elements at leading order in nucleon recoil. In Section~\ref{sec:soft_photons}, we focus on the emission of soft photons with energies below a cutoff $\varepsilon_\gamma$. Next, we analyze various observables related to real photon emission. In Sections~\ref{sec:3xsec_positron_energy_positron_angle_photon_energy} and~\ref{sec:3xsec_neutron_energy_neutron_angle_photon_energy}, we present triple-differential cross sections with respect to the positron energy, positron angle, and photon energy, as well as to the neutron energy, neutron angle, and photon energy, respectively. Although measurements of neutron recoil kinematics in liquid scintillators are not very feasible, we study neutron kinematic distributions because they reflect electromagnetic energy dependencies and can be evaluated analytically. By integrating over the photon energy, we obtain double-differential distributions in the positron energy and angle in Section~\ref{sec:2xsec_positron_energy_positron_angle} and the neutron energy and angle in Section~\ref{sec:2xsec_neutron_energy_neutron_angle}. Further integration over the positron angle in Section~\ref{sec:1xsec_positron_energy} yields the positron energy spectrum. We then distinguish between the positron energy spectrum associated with the elastic IBD process in Section~\ref{sec:above} and the contribution arising solely from radiative IBD in Section~\ref{sec:below}, with positron energies below the elastic IBD kinematics. The neutron energy spectrum, discussed in Section~\ref{sec:1xsec_electromagnetic_energy}, is shown to coincide with the spectrum of total electromagnetic energy. Finally, we also present the electromagnetic energy spectrum within the static-limit approximation for the phase-space integration in the radiative IBD in Section~\ref{sec:static_limit}.

\subsection{One-photon bremsstrahlung}
\label{sec:bremsstrahlung_general}

\begin{figure}[H]
\begin{center}
\includegraphics[scale=.35]{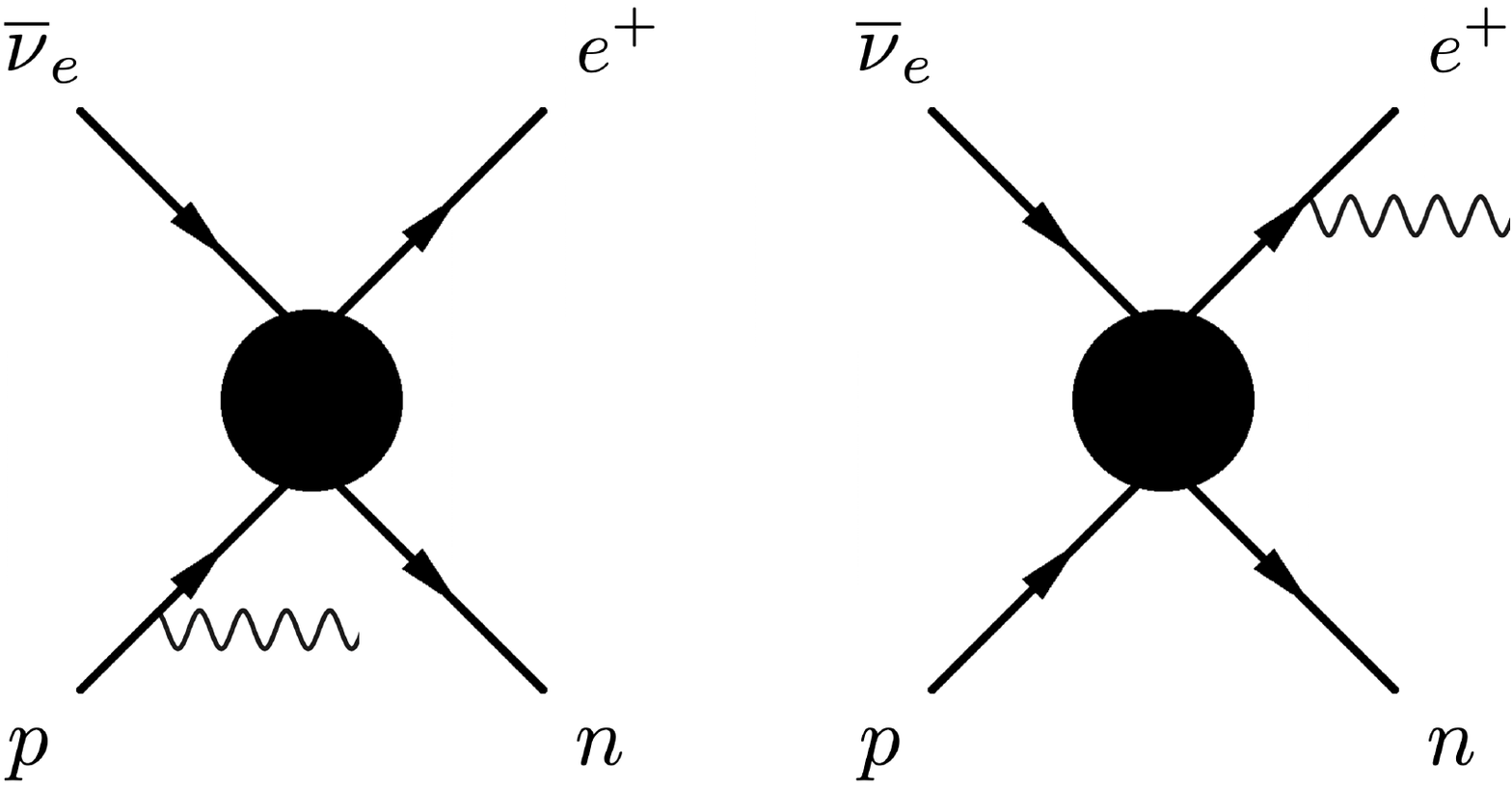}
\end{center}
	\caption{One-photon bremsstrahlung diagrams for $\overline{\nu}_e + p \rightarrow e^+ + n + \gamma$. \label{fig:bremsstrahlung_graphs}}
\end{figure}

We begin by analyzing bremsstrahlung processes in antineutrino-proton scattering that involve the emission of a single photon. At leading order in nucleon recoil, the relevant Feynman diagrams are shown in Fig.~\ref{fig:bremsstrahlung_graphs}. The total amplitude is the sum of two contributions: photon emission from the initial-state proton $T_i^{1\gamma}$ and from the final-state positron $T_f^{1\gamma}$,
\begin{equation} \label{eq:bremsstrahlung_amplitude}
	T^{1\gamma} = T_{i}^{1\gamma} + T_{f}^{1\gamma}.
\end{equation}

The amplitude for photon emission from the initial proton $T_{i}^{1\gamma}$ is obtained by attaching a photon to the leading-order amplitude in Eq.~(\ref{eq:leading_order_amplitude}) as
\begin{equation} \label{eq:initial_radiation}
	T_{i}^{1\gamma} = - \sqrt{2} e G_F V^\star_{ud} \overline{\overline{\nu}}_e \gamma_\rho \mathrm{P}_\mathrm{L} \overline{e} \overline{N}_v \left( g_V v^\rho - 2 g_A S^\rho \right) \tau^- \frac{v \cdot \varepsilon^\star}{v \cdot k_\gamma} N_v,
\end{equation}
where $k_\gamma$ and $\varepsilon^\star$ are the photon four-momentum and polarization vector, respectively. Similarly, the amplitude for photon emission from the outgoing positron $T_{f}^{1\gamma}$ is
\begin{equation} \label{eq:final_radiation}
	T_{f}^{1\gamma} = \sqrt{2} e G_F V^\star_{ud} \overline{\overline{\nu}}_e \gamma_\rho \mathrm{P}_\mathrm{L} \frac{ \slashed{k}^\prime + \slashed{k}_\gamma + m_e}{\left( k^\prime + k_\gamma \right)^2 - m^2_e} \slashed{\varepsilon}^\star \overline{e} \overline{N}_v \left( g_V v^\rho - 2 g_A S^\rho \right) \tau^- N_v.
\end{equation}

The corresponding bremsstrahlung cross section can be expressed in terms of the spin-averaged matrix element $\sum \limits_{\mathrm{spins}}|T^{1\gamma} |^2$ as
\begin{equation} \label{eq:bremsstrahlung_in_terms_of_the_amplitude}
	\mathrm{d} \sigma^{1\gamma}_{\rm LO} = \frac{1}{4 m_p E_{\overline{\nu}_e}} \frac{1}{\left( 2 \pi \right)^5} \frac{1}{2} \sum \limits_{\mathrm{spins}}|T^{1\gamma} |^2 \delta^4 \left( k + p - k^\prime -p^\prime - k_\gamma \right) \frac{\mathrm{d}^3 \vec{k}_\gamma}{ 2 E_\gamma} \frac{\mathrm{d}^3 \vec{k}^\prime}{2 E_e} \frac{\mathrm{d}^3 \vec{p}^{~\prime}}{2 E_n}.
\end{equation}
At leading order in $1/m_n$, the spin-averaged squared matrix element takes the form~\cite{Raha:2011aa,Fukugita:2004cq}
\begin{align}
	\sum \limits_{\mathrm{spins}}|T^{1\gamma} |^2 &= - e^2 \left( \frac{1}{E^2_\gamma} + \frac{m_e^2}{\left( k^\prime \cdot k_\gamma \right)^2} - \frac{2 E_e}{E_\gamma \left( k^\prime \cdot k_\gamma \right)}\right) \sum \limits_{\mathrm{spins}} |T_\mathrm{LO} |^2 \nonumber \\
&+ 64 e^2 {G}_{\rm F}^2 |V_{ud}|^2 m^2_n \left[ c^{1\gamma}_+ \left( g^2_V + g^2_A \right) + c^{1\gamma}_- \left( g^2_V - g^2_A \right) \right], \label{eq:bremsstrahlung_squared_matrix_element}
\end{align}
where the leading-order amplitude without radiation $T_\mathrm{LO}$ is expressed in terms of Lorentz contractions of momenta in the radiative IBD process, and the structure-independent prefactors $c^{1\gamma}_\pm$ are given by
\begin{align}
	c^{1\gamma}_+ &= - 2 \frac{E_{\overline{\nu}_e}}{E_\gamma} \left( 1 - \frac{2 E_e E_\gamma}{k^\prime \cdot k_\gamma} - \frac{E^2_\gamma}{k^\prime \cdot k_\gamma} \left( 1 - \frac{m^2_e}{k^\prime \cdot k_\gamma} \right) \right), \label{eq:c_plus} \\
	c^{1\gamma}_- &= \frac{E_{\overline{\nu}_e}}{E_\gamma} - \frac{k \cdot k_\gamma}{k^\prime \cdot k_\gamma} \left(\frac{E_e}{E_\gamma} - \frac{m^2_e}{k^\prime \cdot k_\gamma} \right) - \frac{k \cdot k_\gamma + k \cdot k^\prime}{k^\prime \cdot k_\gamma}, \label{eq:c_minus}
\end{align}
which coincides with the expression in the radiative neutron decay~\cite{Sirlin:1967zza,Shann:1971fz,Garcia:1978bq,Gluck:1992tg,Ando:2004rk,Bernard:2004cm,Gudkov:2005bu,Gardner:2012rp} and charged-current neutrino-neutron elastic scattering, under the no-recoil approximation for the final-state proton $p \cdot k_\gamma = m_p E_\gamma$ with the energy $E_\gamma$ in the rest frame of the neutron.

\subsection{Soft-photon bremsstrahlung}
\label{sec:soft_photons}

The bremsstrahlung process in inverse beta decay includes a universal IR contribution arising from soft-photon emission when the emitted photon is too low in energy to be experimentally distinguished from the nonradiative process. This soft-photon region corresponds to kinematics where the elastic and radiative processes become indistinguishable, and must therefore be included to obtain physically meaningful, i.e., IR-finite, cross sections. Soft-photon corrections affect differential cross sections with respect to any kinematic variable except the photon energy, for which the spectrum is evaluated above a chosen minimum value. For a single soft photon with energy $E_\gamma \leq \varepsilon_\gamma$, where $\varepsilon_\gamma \ll m_e, E_0, E_{\overline{\nu}_e}$ acts as an energy cutoff, the soft-photon bremsstrahlung amplitude $T^{1\gamma}_\mathrm{soft}$ factorizes as
\begin{equation} \label{eq:soft_photon_amplitude}
	{T}^{1\gamma}_\mathrm{soft} = - e \left[ \frac{k^\prime \cdot \varepsilon^\star}{k^\prime \cdot k_\gamma} - \frac{v \cdot \varepsilon^\star}{v \cdot k_\gamma} \right] T_\mathrm{LO},
\end{equation} 
where $T_\mathrm{LO}$ is the leading-order amplitude for the elastic IBD. This factorization reflects the universal and model-independent nature of soft-photon emission. The corresponding contribution to the bremsstrahlung cross section $\mathrm{d} \sigma^{1\gamma}_\mathrm{soft}$ is given by
\begin{equation} \label{eq:soft_photon_cross_section}
	\mathrm{d} \sigma^{1\gamma}_\mathrm{soft} = \frac{\alpha}{\pi} \delta_s \mathrm{d} \sigma_{\mathrm{LO}},
\end{equation}
with the soft-photon correction factor $\delta_s$~\cite{Lee:1964jq,Aoki:1980ix,Sarantakos:1982bp,Passera:2000ug,Tomalak:2019ibg,Tomalak:2021hec,Tomalak:2022xup},
\begin{equation} \label{eq:soft_photon_correction}
	\delta_s = \frac{1}{ \beta}\left( \mathrm{Li}_2 \frac{1-\beta}{1+\beta} - \frac{\pi^2}{6} \right)- \left( 1 - \frac{1}{2 \beta} \ln \frac{1+\beta}{1-\beta} \right)\ln \frac{4 \varepsilon_\gamma^2}{\lambda_\gamma^2} + \frac{1}{2 \beta} \ln \frac{1+\beta}{1-\beta}\left( 1 + \ln \frac{\rho \left(1+ \beta \right) }{4 \beta^2} \right)+1.
\end{equation}
Here, $\beta$ is the velocity defined as
\begin{equation} \label{eq:beta_for_radiation}
	\beta = \sqrt{1 - \frac{m^2}{\overline{E}^2}},
\end{equation}
and $\rho = \sqrt{1 - \beta^2}$. The quantity $\overline{E}$ denotes the relevant energy, either the positron energy $E_e$ or the total electromagnetic energy $E_{\rm EM} = E_e + E_\gamma$, depending on the observable under consideration. As expected from the Kinoshita-Lee-Nauenberg theorem~\cite{Bloch:1937pw,Nakanishi:1958ur,Kinoshita:1962ur,Lee:1964is}, the infrared divergence associated with soft-photon emission cancels exactly when combined with the virtual correction. Consequently, the sum $\delta_s + \delta_v$ is finite and independent of the fictitious photon mass $\lambda_\gamma$ used as an IR regulator.

The radiation of multiple soft photons can be accounted for by factorizing the cross section into the soft and hard~\cite{Tomalak:2019ibg} or soft, collinear, and hard contributions~\cite{Tomalak:2021hec,Tomalak:2022xup} and exploiting our exact results from Section~\ref{sec:corrections} for the hard function.

\subsection{Triple-differential distribution in positron energy, positron angle, and photon energy}
\label{sec:3xsec_positron_energy_positron_angle_photon_energy}

We compute the bremsstrahlung cross section for photons with energy $E_\gamma \ge \varepsilon_\gamma$ using the integration technique introduced in Ref.~\cite{Ram:1967zza}. To describe the angular distribution of the positron in the presence of radiation, we define the four-vector $l$,
\begin{equation} \label{eq:vector_l_positron_spectrum}
	l = p + k - k^\prime = \left( l_0,~\vec{f} \right),
\end{equation}
with components in the laboratory frame,
\begin{align}
	l_0 &= m_p + E_{\overline{\nu}_e} - E_e, \label{eq:vector_l0_positron_spectrum} \\
	f^2 &= |\vec{f}|^2 = E^2_{\overline{\nu}_e} + \beta^2 E_e^2 - 2 \beta E_{\overline{\nu}_e} E_e \cos \theta_e, \label{eq:vector_f_positron_spectrum}
\end{align}
where the positron scattering angle $\theta_e$ in the radiative process differs from the scattering angle $\Theta_e$ defined in the elastic case [cf. Eq.~(\ref{eq:scattering_angle})].

After performing the relevant phase-space integrations, the triple-differential cross section in the positron energy, positron angle, and photon energy can be expressed as
\begin{equation} \label{eq:radiation_3d}
	\mathrm{d} \sigma^{1\gamma} = \frac{\alpha}{\pi} {G}_{\rm F}^2 |V_{ud}|^2 \left[ c^{3d}_+ \left( g^2_V + g^2_A \right) + c^{3d}_- \left( g^2_V - g^2_A \right) \right] {\cal{D}}_m,
\end{equation}
where the structure-independent coefficients $c^{3d}_\pm$ are given by
\begin{align}
	c^{3d}_+ &= - \frac{\rho^2 f^4 m^2_e \sigma E_\gamma}{d^{3/2}} \left( E_e + E_\gamma \right) + \frac{f^2 m_e}{\sqrt{d} E_e} \left( \left( E_e + E_\gamma \right)^2 + E^2_e \right) - \frac{E_e + E_\gamma}{2 E_\gamma}, \label{eq:c_plus_3d} \\
	c^{3d}_- &= \left( \frac{\rho^2 f^4 m^2_e \sigma E_\gamma}{2 d^{3/2} E_{\overline{\nu}_e}} - \frac{f^2 m_e}{2 \sqrt{d} E_{\overline{\nu}_e}} \frac{ E_e + E_\gamma}{E_e} \right) \left( m^2_e - m_n E_0 + m_p \left( E_{\overline{\nu}_e} - E_e - E_\gamma \right) \right) + \frac{E_{\overline{\nu}_e} - E_e - E_\gamma}{4 E_{\overline{\nu}_e}} \nonumber \\
&+ \frac{f^2 m_e }{4 \sqrt{d} E_{\overline{\nu}_e} E_e} \left( \left( \left( E_{\overline{\nu}_e} - E_e \right)^2 - f^2 \right) E_e + m^2_e \left( 2 E_\gamma - E_e \right) \right) - \frac{ \left( E_{\overline{\nu}_e} - E_e \right)^2 - f^2 - m^2_e}{8 E_{\overline{\nu}_e} E_\gamma}, \label{eq:c_minus_3d}
\end{align}
and the auxiliary kinematic functions are defined as
\begin{align}
	\sigma &= \rho \left( E^2_{\overline{\nu}_e} - f^2 - E^2_e +m^2_e \right) \left(l^2 - m^2_n - 2 E_{\gamma } l_0 \right) + 4 E_{\gamma } m_e f^2, \label{eq:sigma} \\
	d &= \beta^2 m_e^2 E^2_{\overline{\nu}_e} \left( \left( l^2 - m^2_n \right) \left( l^2 - m^2_n - 4 l_0 E_{\gamma } \right)+ 4 E_{\gamma }^2 l^2\right) \sin^2 \theta_{e} + \frac{\sigma^2}{4}, \label{eq:d}
\end{align}
with the phase-space factor ${\cal{D}}_m$,
\begin{equation} \label{eq:measure_3D_positron}
	{\cal{D}}_m = \frac{m_p}{\pi} \frac{\mathrm{d} E_\gamma}{E_\gamma} \frac{\mathrm{d} f}{E_{\overline{\nu}_e}} \mathrm{d} E_e.
\end{equation}

The presence of a photon with energy $E_\gamma \ge \varepsilon_\gamma$ increases the reaction threshold to $E_{\overline{\nu}_e}^\mathrm{\gamma, thr}$,
\begin{equation} \label{eq:threshold_energy_radiative}
	E_{\overline{\nu}_e}^\mathrm{\gamma, thr} = \frac{\left( m_n + m_e \right)^2}{2 m_p \left( 1 - \frac{2 \varepsilon_\gamma}{m_p} \right)} - \frac{m_p}{2} \ge E_{\overline{\nu}_e}^\mathrm{thr}.
\end{equation}
The physical phase space for radiative events with hard photons includes both the following region inherent to the elastic process for arbitrary antineutrino energy above the threshold $E_{\overline{\nu}_e} \ge E_{\overline{\nu}_e}^\mathrm{\gamma, thr}$,
\begin{align}
	E_e^\mathrm{min} &\le E_e \le E_e^\mathrm{max}, \label{eq:radiative_positron_energy_range_from_elastic} \\
	E_e^\mathrm{min} &= \frac{ \left( s \left( 1 - \frac{2 \varepsilon_\gamma}{m_p} \right) - m^2_n + m^2_e \right) \left( m_p + E_{\overline{\nu}_e} - \varepsilon_\gamma \right) - \left( E_{\overline{\nu}_e} + \varepsilon_\gamma \right) \sqrt{\Sigma \left( s \left( 1 - \frac{2 \varepsilon_\gamma}{m_p} \right), m^2_n, m^2_e \right)} }{2 s \left( 1 - \frac{2 \varepsilon_\gamma}{m_p} \right)}, \label{eq:radiative_positron_energy_min_from_elastic} \\
	E_e^\mathrm{max} &= \frac{ \left( s \left( 1 - \frac{2 \varepsilon_\gamma}{m_p} \right) - m^2_n + m^2_e \right) \left( m_p + E_{\overline{\nu}_e} - \varepsilon_\gamma \right) + \left( E_{\overline{\nu}_e} + \varepsilon_\gamma \right) \sqrt{\Sigma \left( s \left( 1 - \frac{2 \varepsilon_\gamma}{m_p} \right), m^2_n, m^2_e \right)} }{2 s \left( 1 - \frac{2 \varepsilon_\gamma}{m_p} \right)}, \label{eq:radiative2_positron_energy_max_from_elastic} \\
	E_{\overline{\nu}_e} - \beta E_e &\le f \le \sqrt{ \left( l_0 - \varepsilon_\gamma \right)^2 - m^2_n} - \varepsilon_\gamma, \label{eq:radiative_f_from_elastic} \\
	 \frac{ l^2 - m^2_n }{2 \left( l_0 +f \right)} &\le E_\gamma \le \frac{ l^2 - m^2_n }{2\left( l_0 -f \right)}, \label{eq:radiative_Egamma}
\end{align}
and an additional region not accessible in the elastic process and without constraints on the variable $f$,
\begin{align}
	m_e &\le E_e \le E_e^\mathrm{min}, \label{eq:radiative_positron_energy_range_new} \\
	E_{\overline{\nu}_e} - \beta E_e &\le f \le E_{\overline{\nu}_e} + \beta E_e, \label{eq:radiative_f_new} \\
	\frac{ l^2 - m^2_n }{2 \left( l_0 +f \right)} &\le E_\gamma \le \frac{ l^2 - m^2_n }{2\left( l_0 -f \right)}, \label{eq:radiative_Egamma_new}
\end{align}
which opens for $E_{\overline{\nu}_e} \ge E_{\overline{\nu}_e}^\mathrm{\gamma} = \frac{1}{2} \left( \frac{m^2_n}{m_p - m_e - 2 \varepsilon_\gamma} + m_e - m_p \right) > E_{\overline{\nu}_e}^\mathrm{\gamma, thr}$. For small cutoff values $\varepsilon_\gamma$, we find numerically $E_{\overline{\nu}_e}^\mathrm{\gamma} \approx 1.806067~\mathrm{MeV}$ and $E_{\overline{\nu}_e}^\mathrm{\gamma, thr} \approx E_{\overline{\nu}_e}^\mathrm{thr} \approx 1.806066~\mathrm{MeV}$.

For practical applications, we also present the integration regions after change of variables $f$ and $E_\gamma$. The region of Eqs.~(\ref{eq:radiative_f_from_elastic}) and~(\ref{eq:radiative_Egamma}) corresponds to
\begin{align}
	- E_{\overline{\nu}_e} + \frac{\left(m_p + 2 E_{\overline{\nu}_e} \right) \left( m_p + 2 E_{\overline{\nu}_e} - 2 E_e \right) - m^2_n + m^2_e}{2 \left( m_p + 2 E_{\overline{\nu}_e} - \sqrt{E^2_e - m^2_e} - E_e \right)} &\le E_\gamma \le E_{\overline{\nu}_e} + \frac{m_p \left( m_p - 2 E_e \right) - m^2_n + m^2_e}{2 \left( m_p + \sqrt{E^2_e - m^2_e} - E_e \right)}, \label{eq:radiative_Egamma_inverse1} \\
	E_{\overline{\nu}_e} - \beta E_e &\le f \le \sqrt{ \left( l_0 - E_\gamma \right)^2 - m^2_n} + E_\gamma, \label{eq:radiative_f_inverse1}  \\
	\left( 1 + \frac{2 \left(  \sqrt{ \left( l_0 - \varepsilon_\gamma \right)^2 - m^2_n} - \varepsilon_\gamma \right) }{l_0 - \left(  \sqrt{ \left( l_0 - \varepsilon_\gamma \right)^2 - m^2_n} - \varepsilon_\gamma \right) }\right) \varepsilon_\gamma  &\le E_\gamma, \label{eq:radiative_Egamma_inverse21} \\
	- E_{\overline{\nu}_e} + \frac{\left(m_p + 2 E_{\overline{\nu}_e} \right) \left( m_p + 2 E_{\overline{\nu}_e} - 2 E_e \right) - m^2_n + m^2_e}{2 \left( m_p + 2 E_{\overline{\nu}_e} - \sqrt{E^2_e - m^2_e} - E_e \right)} &\ge E_\gamma, \label{eq:radiative_Egamma_inverse22} \\
	\sqrt{ \left( l_0 - E_\gamma \right)^2 - m^2_n} - E_\gamma&\le f \le \sqrt{ \left( l_0 - E_\gamma \right)^2 - m^2_n} + E_\gamma, \label{eq:radiative_f_inverse2}  \\
	\varepsilon_\gamma &\le E_\gamma \le \left( 1 + \frac{2 \left(  \sqrt{ \left( l_0 - \varepsilon_\gamma \right)^2 - m^2_n} - \varepsilon_\gamma \right) }{l_0 - \left(  \sqrt{ \left( l_0 - \varepsilon_\gamma \right)^2 - m^2_n} - \varepsilon_\gamma \right) }\right) \varepsilon_\gamma, \label{eq:radiative_Egamma_inverse3} \\
	 \sqrt{ \left( l_0 - E_\gamma \right)^2 - m^2_n} - E_\gamma &\le f \le \sqrt{ \left( l_0 - \varepsilon_\gamma \right)^2 - m^2_n} - \varepsilon_\gamma. \label{eq:radiative_f_inverse3}
\end{align}
While the region of Eqs.~(\ref{eq:radiative_f_new}) and~(\ref{eq:radiative_Egamma_new}) corresponds to
\begin{align}
    E_e  &\le E_e^0, \label{eq:radiative_Egamma_new_inverse_Eerange1} \\
	E_{\overline{\nu}_e}  + \frac{m_p \left( m_p - 2 E_e \right) - m^2_n + m^2_e}{2 \left( m_p - \sqrt{E^2_e - m^2_e} - E_e \right)} &\le E_\gamma \le E_{\overline{\nu}_e} + \frac{m_p \left( m_p - 2 E_e \right) - m^2_n + m^2_e}{2 \left( m_p + \sqrt{E^2_e - m^2_e} - E_e \right)}, \label{eq:radiative_Egamma_new_inverse1} \\
	E_{\overline{\nu}_e} - \beta E_e &\le f \le \sqrt{ \left( l_0 - E_\gamma \right)^2 - m^2_n} + E_\gamma, \label{eq:radiative_f_new2_inverse1}  \\
	- E_{\overline{\nu}_e} + \frac{\left(m_p + 2 E_{\overline{\nu}_e} \right) \left( m_p + 2 E_{\overline{\nu}_e} - 2 E_e \right) - m^2_n + m^2_e}{2 \left( m_p + 2 E_{\overline{\nu}_e} - \sqrt{E^2_e - m^2_e} - E_e \right)} &\le E_\gamma \le E_{\overline{\nu}_e} + \frac{m_p \left( m_p - 2 E_e \right) - m^2_n + m^2_e}{2 \left( m_p - \sqrt{E^2_e - m^2_e} - E_e \right)}, \label{eq:radiative_Egamma_new_inverse2} \\
	E_{\overline{\nu}_e} - \beta E_e &\le f \le E_{\overline{\nu}_e} + \beta E_e, \label{eq:radiative_f_new_inverse2}  \\
	- E_{\overline{\nu}_e} + \frac{\left(m_p + 2 E_{\overline{\nu}_e} \right) \left( m_p + 2 E_{\overline{\nu}_e} - 2 E_e \right) - m^2_n + m^2_e}{2 \left( m_p + 2 E_{\overline{\nu}_e} + \sqrt{E^2_e - m^2_e} - E_e \right)} &\le E_\gamma, \label{eq:radiative_Egamma_new_inverse31} \\
	- E_{\overline{\nu}_e} + \frac{\left(m_p + 2 E_{\overline{\nu}_e} \right) \left( m_p + 2 E_{\overline{\nu}_e} - 2 E_e \right) - m^2_n + m^2_e}{2 \left( m_p + 2 E_{\overline{\nu}_e} - \sqrt{E^2_e - m^2_e} - E_e \right)} &\ge E_\gamma, \label{eq:radiative_Egamma_new_inverse32} \\
	 \sqrt{ \left( l_0 - E_\gamma \right)^2 - m^2_n} - E_\gamma&\le f \le  E_{\overline{\nu}_e} + \beta E_e,\label{eq:radiative_f_new_inverse3}
\end{align}
and
\begin{align}
    E_e  &\ge E_e^0, \label{eq:radiative_Egamma_new_inverse_Ee2range1} \\
	 - E_{\overline{\nu}_e} + \frac{\left(m_p + 2 E_{\overline{\nu}_e} \right) \left( m_p + 2 E_{\overline{\nu}_e} - 2 E_e \right) - m^2_n + m^2_e}{2 \left( m_p + 2 E_{\overline{\nu}_e} - \sqrt{E^2_e - m^2_e} - E_e \right)} &\le E_\gamma \le E_{\overline{\nu}_e} + \frac{m_p \left( m_p - 2 E_e \right) - m^2_n + m^2_e}{2 \left( m_p + \sqrt{E^2_e - m^2_e} - E_e \right)}, \label{eq:radiative_Egamma_new2_inverse1} \\
	E_{\overline{\nu}_e} - \beta E_e &\le f \le \sqrt{ \left( l_0 - E_\gamma \right)^2 - m^2_n} + E_\gamma, \label{eq:radiative_f_new_inverse1}  \\
	E_{\overline{\nu}_e}  + \frac{m_p \left( m_p - 2 E_e \right) - m^2_n + m^2_e}{2 \left( m_p - \sqrt{E^2_e - m^2_e} - E_e \right)} &\le E_\gamma, \label{eq:radiative_Egamma_new2_inverse21} \\
	- E_{\overline{\nu}_e} + \frac{\left(m_p + 2 E_{\overline{\nu}_e} \right) \left( m_p + 2 E_{\overline{\nu}_e} - 2 E_e \right) - m^2_n + m^2_e}{2 \left( m_p + 2 E_{\overline{\nu}_e} - \sqrt{E^2_e - m^2_e} - E_e \right)} &\ge E_\gamma, \label{eq:radiative_Egamma_new2_inverse22} \\
	\sqrt{ \left( l_0 - E_\gamma \right)^2 - m^2_n} - E_\gamma &\le f \le \sqrt{ \left( l_0 - E_\gamma \right)^2 - m^2_n} + E_\gamma, \label{eq:radiative_f_new2_inverse2}  \\
	- E_{\overline{\nu}_e} + \frac{\left(m_p + 2 E_{\overline{\nu}_e} \right) \left( m_p + 2 E_{\overline{\nu}_e} - 2 E_e \right) - m^2_n + m^2_e}{2 \left( m_p + 2 E_{\overline{\nu}_e} + \sqrt{E^2_e - m^2_e} - E_e \right)} &\le E_\gamma \le  E_{\overline{\nu}_e} + \frac{m_p \left( m_p - 2 E_e \right) - m^2_n + m^2_e}{2 \left( m_p - \sqrt{E^2_e - m^2_e} - E_e \right)} , \label{eq:radiative_Egamma_new2_inverse3} \\
	 \sqrt{ \left( l_0 - E_\gamma \right)^2 - m^2_n} - E_\gamma&\le f \le  E_{\overline{\nu}_e} + \beta E_e,\label{eq:radiative_f_new2_inverse3}
\end{align}
with $E_e^0 = \frac{m_p + E_{\overline{\nu}_e}}{2} \left( 1 + \frac{m^2_e}{m_p \left( m_p + 2 E_{\overline{\nu}_e} \right) - m^2_n }\right) - \frac{\sqrt{m_n^2 + E^2_{\overline{\nu}_e}}}{2} \left( 1 + \frac{m^2_e}{m_p \left( m_p + 2 E_{\overline{\nu}_e} \right) - m^2_n }\right)$.

We retain the dependence on the unphysical regulator $\varepsilon_\gamma$, which is essential for the accurate evaluation of the positron energy spectrum in Section~\ref{sec:1xsec_positron_energy} and for computing the averaged positron scattering angle in Section~\ref{sec:averaged_angle}. The phase-space region defined by Eqs.~(\ref{eq:radiative_positron_energy_range_from_elastic})-(\ref{eq:radiative_Egamma}) corresponds to region I in Section~\ref{sec:above}.

In the radiative IBD process, the photon energy range is narrow. The photon energy $E_\gamma$ can be estimated in the no-recoil approximation as
\begin{equation}
	E_\gamma = E_{\overline{\nu}_e} - E_e - E_0 - \frac{E_e^2 - m^2_e + E_{\overline{\nu}_e}  E_0}{m_n} + \mathcal{O} \left( \frac{E_{\overline{\nu}_e} \sqrt{ E_e^2 - m^2_e} }{2 m_n} \right), \label{eq:photon_energy_from_positron_energy}
\end{equation}
with an accuracy only of the order of the elastic IBD energy range in Eqs.~(\ref{eq:positron_energy_min})-(\ref{eq:positron_energy_max_1_over_m}), i.e., below $100~\mathrm{keV}$ for the whole range of reactor antineutrino energies.

Finally, to obtain the distribution with respect to the neutron energy $E_n$ instead of the photon energy $E_\gamma$, we perform the substitution $E_\gamma \to l_0 - E_n$ in Eqs.~(\ref{eq:c_plus_3d}),~(\ref{eq:c_minus_3d}), and~(\ref{eq:measure_3D_positron}), and replace $\mathrm{d} E_\gamma \to \mathrm{d} E_n$ in Eq.~(\ref{eq:measure_3D_positron}), accordingly.

\subsection{Triple-differential distribution in neutron energy, neutron angle, and photon energy}
\label{sec:3xsec_neutron_energy_neutron_angle_photon_energy}

We evaluate the bremsstrahlung cross section with respect to the neutron energy, equivalently to the sum of positron and photon energies, considering the final neutron energy spectrum instead of the positron energy spectrum~\cite{Ram:1967zza}. For the fixed antineutrino energy, the total electromagnetic energy, defined as the sum of the positron and photon energies, $E_\mathrm{EM}$: $E_\mathrm{EM} = E_e + E_\gamma = m_p + E_{\overline{\nu}_e} - E_n$, is fully determined by the neutron recoil energy. The corresponding distributions can be obtained by expressing $E_n$ in terms of $E_\mathrm{EM}$ and changing the measure as $\mathrm{d} E_n = - \mathrm{d} E_\mathrm{EM}$. To study angular distributions of the neutron, we introduce the four-vector $\tilde{l}$,
\begin{equation} \label{eq:vector_l_electromagnetic_spectrum}
	\tilde{l} = k + p - p^\prime = \left( \tilde{l}_0,~\vec{\tilde{f}} \right),
\end{equation}
where the components in the laboratory frame are
\begin{align}
	\tilde{l}_0 &= E_\mathrm{EM}, \label{eq:vector_l0_electromagnetic_spectrum} \\
	\tilde{f}^2 &= |\vec{\tilde{f}}|^2= E_{\overline{\nu}_e}^2 + E_n^2 - m^2_n - 2 \beta_n E_{\overline{\nu}_e} E_n \cos \theta_n, \label{eq:vector_f_electromagnetic_spectrum}
\end{align}
with the neutron velocity $\beta_n = \sqrt{ 1 - \frac{m^2_n}{E^2_n}}$.

The triple-differential distribution in the neutron energy, neutron angle, and photon energy is obtained by modifying the kinematic coefficients $c^{3d}_i$ in Eq.~(\ref{eq:radiation_3d}) to their neutron-based counterparts $\tilde{c}^{3d}_i$ as $c^{3d}_i \to \tilde{c}^{3d}_i$, with the explicit expressions
\begin{align}
	\tilde{c}^{3d}_+ &= \frac{2\tilde{l}_0^2 + E^2_\gamma}{\tilde{l}^2-m^2_e} - \frac{2 E_\gamma \tilde{l}_0 \tilde{l}^2}{\left( \tilde{l}^2 - m^2_e \right)^2} - \frac{\tilde{l}_0}{2 E_\gamma}, \label{eq:c_plus_3d_neutron} \\
	\tilde{c}^{3d}_- &= \frac{E^2_n - E^2_{\overline{\nu}_e} - m^2_n}{8 \tilde{f}^2} \left( \frac{\left( \tilde{l}^2_0 - m^2_e \right) \left( 3 \tilde{l}^2_0 - m^2_e \right)}{ \left( \tilde{l}^2 - m^2_e \right)^2 } \frac{E_\gamma}{E_{\overline{\nu}_e}} - 2 \frac{\tilde{l}_0}{E_{\overline{\nu}_e}} + \frac{\tilde{l}^2_0 + 2 \tilde{f}^2 - m^2_e}{2 E_{\overline{\nu}_e} E_\gamma}\right) - \frac{\tilde{f}^2}{16 E_{\overline{\nu}_e} E_\gamma} - \frac{\tilde{l}_0}{4 E_{\overline{\nu}_e}} \nonumber \\
	& + \frac{m^2_e}{\tilde{l}^2 - m^2_e} \left(\frac{1}{2} \frac{E^2_\gamma}{\tilde{f}^2} + \frac{E_\gamma}{4 E_{\overline{\nu}_e}} \right) + 3 \left( 1 + \frac{ \tilde{l}^2_0 - \frac{m^2_e}{3}}{\tilde{l}^2 - m^2_e}\right) \frac{\tilde{l}_0 \left( 2 E_{\overline{\nu}_e} - \tilde{l}_0 \right) + m^2_e}{\tilde{l}^2 - m^2_e} \frac{E_\gamma}{8 E_{\overline{\nu}_e}} - \frac{\tilde{l}_0 - 2 E_{\overline{\nu}_e}}{4 E_{\overline{\nu}_e}} \frac{E^2_\gamma}{\tilde{f}^2} \nonumber \\
	&+ \frac{m_n E_0 - m_p E_n + m^2_p - m^2_e}{\tilde{l}^2 - m^2_e} \left( \frac{\tilde{l}_0}{2 E_{\overline{\nu}_e}} \frac{E^2_\gamma}{\tilde{f}^2} + \frac{E_\gamma}{4 E_{\overline{\nu}_e}} \frac{3 \tilde{l}^2_0 - 5 m^2_e}{\tilde{l}^2 - m^2_e} - \frac{E_\gamma}{4 E_{\overline{\nu}_e}} + \frac{\tilde{l}_0}{E_{\overline{\nu}_e}} \right) + \frac{E_\gamma}{8 E_{\overline{\nu}_e}} \nonumber \\
	&+ \frac{ 2 \tilde{l}_0 E_{\overline{\nu}_e} - \left( \tilde{l}_0 - E_{\overline{\nu}_e} \right) \left( \tilde{l}_0 - m_p \right) + m_n E_0}{8 E_{\overline{\nu}_e} E_\gamma}. \label{eq:c_minus_3d_neutron}
\end{align}
The corresponding phase-space measure ${\cal{D}}_m$ in Eq.~(\ref{eq:measure_3D_positron}) becomes
\begin{equation} \label{eq:measure_3D_neutron}
	\tilde{\cal{D}}_m = \frac{m_p}{\pi} \frac{ \mathrm{d} E_\gamma}{E_\gamma } \frac{\mathrm{d} \tilde{f}}{E_{\overline{\nu}_e}} \mathrm{d} E_n.
\end{equation}

The physically allowed integration region for hard-photon radiation $E_\gamma \geq \varepsilon_\gamma$ consists of the following region inherent to the elastic process:
\begin{align}
	E_n^\mathrm{min} &\le E_n \le E_n^\mathrm{max}, \label{eq:radiative_neutron_energy_range_from_elastic} \\
	E_n^\mathrm{min} &= \frac{ \left( s \left( 1 - \frac{2 \varepsilon_\gamma}{m_p} \right) - m^2_e + m^2_n \right) \left( m_p + E_{\overline{\nu}_e} - \varepsilon_\gamma \right) - \left( E_{\overline{\nu}_e} + \varepsilon_\gamma \right) \sqrt{\Sigma \left( s \left( 1 - \frac{2 \varepsilon_\gamma}{m_p} \right), m^2_n, m^2_e \right)} }{2 s \left( 1 - \frac{2 \varepsilon_\gamma}{m_p} \right)}, \label{eq:radiative_neutron_energy_min_from_elastic} \\
	E_n^\mathrm{max} &= \frac{ \left( s \left( 1 - \frac{2 \varepsilon_\gamma}{m_p} \right) - m^2_e + m^2_n \right) \left( m_p + E_{\overline{\nu}_e} - \varepsilon_\gamma \right) + \left( E_{\overline{\nu}_e} + \varepsilon_\gamma \right) \sqrt{\Sigma \left( s \left( 1 - \frac{2 \varepsilon_\gamma}{m_p} \right), m^2_n, m^2_e \right)} }{2 s \left( 1 - \frac{2 \varepsilon_\gamma}{m_p} \right)}, \label{eq:radiative2_neutron_energy_max_from_elastic} \\
	|E_{\overline{\nu}_e} - \beta_n E_n| &\le f \le \sqrt{ \left( l_0 - \varepsilon_\gamma \right)^2 - m^2_e} - \varepsilon_\gamma, \label{eq:radiative_f_from_elastic_neutron} \\
	 \frac{ l^2 - m^2_e }{2 \left( l_0 +f \right)} &\le E_\gamma \le \frac{ l^2 - m^2_e }{2\left( l_0 -f \right)}. \label{eq:radiative_Egamma_neutron}
\end{align}
This integration region corresponds to the region I in Section~\ref{sec:1xsec_electromagnetic_energy}. We retain the full dependence on the infrared cutoff parameter $\varepsilon_\gamma$, essential for constructing the complete electromagnetic energy spectrum discussed in Section~\ref{sec:1xsec_electromagnetic_energy}.

Notably, just as IBD exhibits two distinct regions in the positron energy and one in the electromagnetic energy, elastic (anti)neutrino-electron scattering features one region in the electron energy and two in the electromagnetic energy. This analogy arises due to the large mass hierarchy between the positron and nucleon, similar to the neutrino-electron mass difference in elastic (anti)neutrino-electron scattering~\cite{Sarantakos:1982bp,Bardin:1983yb,Bardin:1985fg,Tomalak:2019ibg}.

Finally, to obtain the distribution in the positron energy $E_e$ instead of the photon energy $E_\gamma$, we can substitute $E_\gamma \to \tilde{l}_0 - E_e$ in Eqs.~(\ref{eq:c_plus_3d_neutron})-(\ref{eq:c_minus_3d_neutron}) and replace $\mathrm{d} E_\gamma$ with $\mathrm{d} E_e$ in Eq.~(\ref{eq:measure_3D_neutron}).

\subsection{Double-differential distribution in positron energy and positron angle}
\label{sec:2xsec_positron_energy_positron_angle}

By integrating the triple-differential cross section in Eq.~(\ref{eq:radiation_3d}) over the photon energy $E_\gamma$, we obtain the double-differential cross section with respect to the recoil positron energy and its scattering angle. This result can be obtained by the following substitution in the kinematic structure of Eq.~(\ref{eq:radiation_3d}),
\begin{align}
	c^{3d}_i {\cal{D}}_m &\to \frac{m_p}{\pi} \left( a_i + b_i \ln \frac{1+\beta}{1-\beta} + c_i \ln \frac{l_0+f}{l_0-f} + d_i \ln \frac{l_0 - \beta f \cos \delta - \sqrt{g}}{l_0 - \beta f \cos \delta + \sqrt{g}} \right) \frac{\mathrm{d} f}{E_{\overline{\nu}_e}} \mathrm{d} E_e \nonumber \\
&- 2 \left( 1 - \frac{1}{2 \beta} \ln \frac{1+\beta}{1-\beta} \right) c^\mathrm{LO}_i \mathrm{d} E_e \frac{ \mathrm{d} f^2}{l^2 - m^2_n}, \label{eq:radiation_2d_positron_energy_positron_angle}
\end{align}
where $g$ is defined as $g = \left( f \cos \delta - \beta l_0 \right)^2 + \rho^2 f^2 \sin^2 \delta$, and the angle $\delta$ between vectors $\vec{f}$ and $\vec{k}^\prime$ is defined as
\begin{equation}
	\cos \delta = \frac{E^2_{\overline{\nu}_e} - \beta^2 E_e^2 - f^2 }{2 \beta E_e f}. \label{eq:angle_delta_positron}
\end{equation}
The kinematic factors $c^\mathrm{LO}_\pm$ in Eq.~(\ref{eq:radiation_2d_positron_energy_positron_angle}) represent the leading-order (LO) kinematic factors from the $2 \to 2$ elastic scattering process in Eq.~(\ref{eq:LO_static_diff}),
\begin{align}
	c^\mathrm{LO}_+ &= \frac{m_p}{\pi} \frac{E_e}{E_{\overline{\nu}_e}}, \label{eq:c_plus_LO}\\
	c^\mathrm{LO}_- &= \frac{m_p}{\pi} \frac{ \left( E_{\overline{\nu}_e} - E_e \right)^2 - f^2 - m^2_e}{4 E^2_{\overline{\nu}_e}}. \label{eq:c_minus_LO}
\end{align}

The coefficients $a_i,~b_i,~c_i,$ and $d_i$ are given by
\begin{align}
a_+ &= - \beta \frac{f}{\sqrt{g}} \frac{\beta l_0 - f \cos \delta}{\sqrt{g}} + \frac{f}{E_e} \frac{f}{\sqrt{g}} \frac{f - \beta l_0 \cos \delta}{\sqrt{g}} \frac{l^2 - m^2_n}{2 l^2}, \\
b_+ &= 0, \\
c_+ &= - \frac{1}{2}, \\
d_+ &= - \frac{f}{\sqrt{g}} - \frac{\beta}{4} \frac{f}{E_e} \frac{\beta l_0 - f \cos \delta}{\sqrt{g}} \frac{l^2 - m^2_n}{g} + \rho^2 \frac{f}{E_e} \frac{f - \beta l_0 \cos \delta}{2 \sqrt{g}} \frac{m^2_e}{g}, \\
a_- &= \left( \frac{1}{2} - \beta \frac{m_p \left( \beta l_0 - f \cos \delta \right)}{2 g} - \frac{l^2 - m^2_n}{2 l^2} + \frac{f m_p}{g} \frac{f - \beta l_0 \cos \delta}{E_e} \frac{l^2 - m^2_n}{2 l^2} \right) \frac{f}{E_{\overline{\nu}_e}}, \\
b_- &= - \frac{1}{4 \beta} \frac{f}{E_{\overline{\nu}_e}}, \\
c_- &= \frac{E_{\overline{\nu}_e} - E_e}{4 E_{\overline{\nu}_e}}, \\
d_- &= \frac{m_p}{2 E_{\overline{\nu}_e}} d_+ + \frac{f}{\sqrt{g}} \frac{m_n E_0 - m_p E_{\overline{\nu}_e}}{4 E_{\overline{\nu}_e} E_e}.
\end{align}

\subsection{Double-differential distribution in neutron energy and neutron angle}
\label{sec:2xsec_neutron_energy_neutron_angle}

To obtain the double-differential cross section with respect to the neutron energy $E_n$ and its scattering angle, we integrate the triple-differential expression in Eq.~(\ref{eq:radiation_3d}), with the neutron-specific coefficients from Eqs.~(\ref{eq:c_plus_3d_neutron}) and~(\ref{eq:c_minus_3d_neutron}) and the phase-space factor $\tilde{\cal{D}}_m$ from Eq.~(\ref{eq:measure_3D_neutron}), over the photon energy $E_\gamma$. This integration results in the following substitution in Eq.~(\ref{eq:radiation_3d}):
\begin{equation} \label{eq:radiation_2d_neutron_energy_neutron_angle}
	c^{3d}_i {\cal{D}}_m \to \frac{m_p}{\pi} \left( \tilde{a}_i + \tilde{c}_i \ln \frac{\tilde{l}_0+\tilde{f}}{\tilde{l}_0-\tilde{f}} \right) \frac{\mathrm{d} \tilde{f}}{E_{\overline{\nu}_e}} \mathrm{d} E_n, 
\end{equation}
where the coefficients $\tilde{a}_i$ and $\tilde{c}_i$ describe the bremsstrahlung contributions to the neutron kinematic distribution and are given by
\begin{align}
	\tilde{a}_+ &= - \left( \frac{m^2_e}{\tilde{l}^2} \left( 1 + \frac{1}{2} \frac{m^2_e}{\tilde{l}^2 - m^2_e} \right) + \frac{7}{2} \frac{\tilde{l}^2}{\tilde{l}^2 - m^2_e} \right) \frac{\tilde{f} \tilde{l}_0}{\tilde{l}^2}, \\
	\tilde{c}_+ &= \frac{2 \tilde{l}^2_0}{\tilde{l}^2 - m^2_e}, \\
	\tilde{a}_- &= - \frac{\tilde{l}_0}{\tilde{f}} + \frac{\tilde{l}^2 + 2 m_p E_n - m^2_n - m^2_p}{8 \tilde{f} E_{\overline{\nu}_e}} \frac{3 \tilde{f}^2 \left( \tilde{l}^2 + m^2_e \right)^2 + 4 \left( \tilde{l}^2_0 \tilde{l}^2 + m^2_e \tilde{f}^2 \right) \left( \tilde{l}^2_0 - m^2_e \right) - 4 \tilde{l}^2_0 \tilde{f}^2 \left( \tilde{l}^2 + m^2_e \right)}{ \left( \tilde{l}^2 \right)^2 \left( \tilde{l}^2 - m^2_e \right)}, \\
	\tilde{c}_- &= \left( \frac{1}{4} \frac{\tilde{f}^2}{\tilde{l}^2 - m^2_e} + \frac{1}{2} \frac{ \tilde{l}_0 \left( m_p - E_{\overline{\nu}_e} \right) + m_n E_0 - m_p E_{\overline{\nu}_e} - m^2_e}{\tilde{l}^2 - m^2_e} - \frac{E^2_n - E^2_{\overline{\nu}_e} - m^2_n}{4 \tilde{f}^2} \frac{ \tilde{l}^2_0 - m^2_e }{\tilde{l}^2 - m^2_e } \right) \frac{\tilde{l}_0}{E_{\overline{\nu}_e}}.
\end{align}

\subsection{Positron energy spectrum}
\label{sec:1xsec_positron_energy}

In addition to the soft-photon correction, the factorizable terms from Eq.~(\ref{eq:soft_photon_amplitude}) contribute to the positron energy spectrum in the regime of hard photon emission with $ E_\gamma \ge \varepsilon_\gamma$. The notations and integration regions used in the spectrum evaluation were introduced in Section~\ref{sec:3xsec_positron_energy_positron_angle_photon_energy}. Above the minimum positron energy allowed by elastic inverse beta decay kinematics, $E_e \ge E_e^\mathrm{min}$, we directly apply the integration techniques developed in Ref. \cite{Ram:1967zza} to compute the positron energy spectrum. Below this threshold, the photon energy is bounded from below as $E_\gamma \ge E_e^\mathrm{min} - E_e$, and there is no corresponding elastic process and no contribution from soft photons. We consider these kinematic regions separately in Sections~\ref{sec:above} and~\ref{sec:below}.

\subsubsection{Above positron lowest energy in elastic IBD}
\label{sec:above}

In the region where the positron energy is above the minimum value allowed in elastic inverse beta decay, $E_e \ge E_e^\mathrm{min}$, radiative corrections modify leading-order IBD cross section. The soft-photon region with $E_\gamma \le \varepsilon_\gamma$ contributes according to Eqs.~(\ref{eq:soft_photon_cross_section}) and~(\ref{eq:soft_photon_correction}). For the radiation of hard photons with $E_\gamma \ge \varepsilon_\gamma$ from Eq.~(\ref{eq:soft_photon_amplitude}), we divide the phase space into two regions:
\begin{itemize}
	\item \textbf{Region I:} $l^2 - m_n^2 \ge 2\varepsilon_\gamma \left( l_0 + f \right)$. Here, the photon phase-space integration is unrestricted.
	\item \textbf{Region II:} $l^2 - m_n^2 \le 2\varepsilon_\gamma \left( l_0 + f \right)$, where the photon angle is constrained as
\end{itemize}
\begin{equation} \label{eq:photon_angle_gamma}
	\cos \gamma \ge \frac{1}{f} \left( l_0 - \frac{l^2 - m^2_n}{2 \varepsilon_\gamma} \right),
\end{equation}
where $\gamma$ is the angle between the vectors $\vec{f}$ and $\vec{k}_\gamma$. This region corresponds to elastic kinematics. The bremsstrahlung correction from region II is given by
\begin{align}
	\delta_\mathrm{II} &= \frac{1 - \beta \beta_n \cos \delta}{2\sqrt{\left( 1 - \beta \beta_n \cos \delta \right)^2 - \rho^2 \rho^2_n}} \ln \frac{1 - \beta \beta_n \cos \delta + \sqrt{\left( 1 - \beta \beta_n \cos \delta \right)^2 - \rho^2 \rho^2_n}}{1 - \beta \beta_n \cos \delta- \sqrt{\left( 1 - \beta \beta_n \cos \delta \right)^2 - \rho^2 \rho^2_n}} - 1 \nonumber \\
&+ \frac{1}{2 \beta_n} \ln \frac{1 + \beta_n}{1 - \beta_n} - \frac{1}{2 \beta} \ln \frac{1 + \beta}{1 - \beta} + \frac{1}{\beta} \ln \frac{2 \beta \left( 1 + \beta_n \right)}{\rho \beta_n \left( 1 + \cos \delta \right)} \ln \frac{1 + \beta}{1 - \beta} - \ln \frac{1 + \beta_n}{1 - \beta_n} \nonumber \\
& + \sum \limits_{\sigma_1, \sigma_2 = \pm 1} \frac{\sigma_1}{\beta} \mathrm{Li}_2 \left[ \frac{\left( 1 + \sigma_1 \beta \right) \beta_n \left( 1 - \cos \delta \right) }{\beta - \beta_n \cos \delta + \sigma_2 \sqrt{\left( 1 - \beta \beta_n \cos \delta \right)^2 - \rho^2 \rho^2_n}} \right], \label{eq:delta_2_positron}
\end{align}
with $\rho_n = \frac{m_n}{l_0}$ and $\beta_n = \sqrt{1- \rho^2_n}$. The angle $\delta$ is defined in Eq.~(\ref{eq:angle_delta_positron}).

The bremsstrahlung contribution from region I, $\mathrm{d} \sigma^{1 \gamma}_\mathrm{I}$, cancels the $\ln\varepsilon_\gamma$ divergence arising in the soft-photon correction. It consists of a factorizable part $\delta_\mathrm{I}$ and a nonfactorizable remainder $\mathrm{d} \sigma^{1\gamma}_{\mathrm{I}, \mathrm{NF}}$,
\begin{equation} \label{eq:region_1_differential}
	\mathrm{d} \sigma^{1\gamma}_\mathrm{I} = \frac{\alpha}{\pi} \delta_\mathrm{I} \mathrm{d} \sigma_{\mathrm{LO}} + \mathrm{d} \sigma^{1\gamma}_{\mathrm{I}, \mathrm{NF}}.
\end{equation}
The factorizable correction $\delta_\mathrm{I}$ originates from factorizable terms in Eq.~(\ref{eq:soft_photon_amplitude}) and is given by
\begin{equation} \label{eq:delta_1_positron_above}
	\delta_\mathrm{I} = 2 \left( 1 - \frac{1}{2 \beta} \ln \frac{1+\beta}{1-\beta} \right) \ln \frac{2 \left( l_0 + \sqrt{l_0^2 - m^2_n} \right) \varepsilon_\gamma}{l^2_0 - \left( E_{\overline{\nu}_e} - \beta E_e \right)^2 - m^2_n}.
\end{equation}

The contributions from regions I and II are combined later with the remaining piece from terms beyond Eq.~(\ref{eq:soft_photon_amplitude}) into the contribution from hard photons $\mathrm{d} \sigma^{1\gamma}_\mathrm{LO}$. Integrating the positron angle and positron energy distribution over the variable $f$, equivalent to the positron scattering angle $\theta_e$, we obtain the total hard-photon contribution in the following form:
\begin{equation} \label{eq:radiative_positron_energy_spectrum_above}
	\mathrm{d} \sigma^{1\gamma}_\mathrm{LO} = \frac{\alpha}{\pi} \left( \delta_\mathrm{I} + \delta_\mathrm{II} + \delta_{\gamma} \right)\mathrm{d} \sigma_{\mathrm{LO}} + \mathrm{d} \sigma^{1\gamma}_{\mathrm{NF}}.
\end{equation}
In the first term, the cross section of the elastic process is expressed as a function of the positron energy. Another factorizable correction $\delta_\gamma$ is given by
\begin{equation} \label{eq:radiative_factorizable_positron_energy_spectrum_above}
	\delta_\gamma = \frac{1}{2} \mathrm{Li}_2 \frac{s \sqrt{1-\beta}}{m_p m_e \sqrt{1+\beta}} + \frac{1}{2} \mathrm{Li}_2 \frac{m_p \sqrt{1+\beta}}{m_e \sqrt{1-\beta}} - \frac{1}{2} \sum \limits_{\sigma_1 = \pm 1} \mathrm{Li}_2\left[ \frac{\overline{\Delta} + m^2_e + \frac{\sigma_1}{2} \sqrt{\Sigma}}{m^2_e} \right],
\end{equation}
with $\overline{\Delta} = m_p E_{\overline{\nu}_e} - m_n E_0$ and $\Sigma = \Sigma \left( s, m^2_n, m^2_e \right)$. The nonfactorizable contribution $\mathrm{d} \sigma^{1\gamma}_{\mathrm{NF}}$ to the positron energy spectrum is given by the following substitutions in Eq.~(\ref{eq:radiation_3d}):
\begin{equation} \label{eq:radiative_NF_positron_energy_spectrum_above}
	c^{3d}_i {\cal{D}}_m \to \frac{m_p}{\pi} \left( A_i + B_i \ln \frac{1+\beta}{1-\beta} + C_i \ln \frac{l^2_0 - \left( E_{\overline{\nu}_e} - \beta E_e \right)^2}{l^2_0 - \left( E_{\overline{\nu}_e} + \beta E_e \right)^2} + \frac{D_i}{2} l_1 - \frac{E_i}{2} l_2 + F_i \delta_\gamma \right) \mathrm{d} E_\mathrm{e},
\end{equation}
where the coefficients $A_\pm,~B_\pm,~C_\pm,~D_\pm,~E_\pm$, and $F_\pm$ can be expressed as
\begin{align}
	A_+&= 0, \\
	B_+&= 0, \\
	C_+&= 0, \\
	D_+&= 1, \\
	E_+&= 1, \\
	F_+&= 0, \\
	A_-&= \frac{l^2_0 - m^2_n - \left( E_{\overline{\nu}_e} - \beta E_e \right)^2}{8 E^2_{\overline{\nu}_e}} + \left( 1 - \beta \right) \frac{E_e}{E_{\overline{\nu}_e}} - \frac{Q^2 + m^2_e}{2 E^2_{\overline{\nu}_e}}, \\
	B_-&= \frac{\beta - 2}{4} \frac{E_e}{E_{\overline{\nu}_e}} + \frac{Q^2 + m^2_e}{2 \beta E^2_{\overline{\nu}_e}} - \frac{s - 2 E_e \left( l_0 + E_e \right) - m^2_n + m^2_e}{8 \beta E^2_{\overline{\nu}_e}}, \\
	C_-&= \frac{m^2_n}{8 E^2_{\overline{\nu}_e}}, \\
	D_-&= \frac{E_e - E_{\overline{\nu}_e}}{2 E_{\overline{\nu}_e}}, \\
	E_-&= \frac{m_p}{2 E_{\overline{\nu}_e}}, \\
	F_-&= \frac{m^2_e}{2 E^2_{\overline{\nu}_e}}. \label{eq:radiative_NF_positron_energy_spectrum_coefficients_above}
\end{align}
The combinations of logarithms $l_1$ and $l_2$ are given by
\begin{align}
	l_1&= \frac{\sqrt{l^2_0 - m^2_n}}{E_{\overline{\nu}_e}} \ln \frac{l_0 - \sqrt{l^2_0 - m^2_n}}{l_0 + \sqrt{l^2_0 - m^2_n}} + \frac{E_{\overline{\nu}_e} - \beta E_e}{E_{\overline{\nu}_e}} \ln \frac{l_0 + \left( E_{\overline{\nu}_e} - \beta E_e \right)}{l_0 - \left( E_{\overline{\nu}_e} - \beta E_e \right)} + \frac{l_0}{E_{\overline{\nu}_e}} \ln \frac{l^2_0 - \left( E_{\overline{\nu}_e} - \beta E_e \right)^2}{m^2_n}, \\
	l_2&= \frac{m^2_n}{s-m^2_e} \frac{l_0}{E_{\overline{\nu}_e}} \ln \frac{s + m^2_e - 2 E_e \left( l_0 + E_e - \beta E_{\overline{\nu}_e} \right)}{m^2_n} + \frac{\overline{\Delta}}{s-m^2_e} \frac{l_0 + E_e}{E_{\overline{\nu}_e}} \ln \frac{\Sigma - 4 \left( \overline{\Delta} + m^2_e \right)^2}{4 m^4_e} \nonumber \\
&- \frac{l^2_0 - s - m^2_e}{s-m^2_e} \frac{E_e}{E_{\overline{\nu}_e}} \ln \frac{\Sigma - 4 \left( \overline{\Delta} + m^2_e \right)^2}{4 m^2_e s} + \frac{s - m^2_n - m^2_e}{s-m^2_e} \left[ \frac{s - m^2_e}{s} \ln \left( 1 - \frac{s}{m_e m_p} \frac{\sqrt{1-\beta}}{\sqrt{1+\beta}}\right) \right. \nonumber \\
&\left. + \frac{m^2_p + m^2_e - 2 m_p E_e}{2 m_p E_{\overline{\nu}_e}} \ln \left( 1 - \frac{s}{m_e m_p} \frac{\sqrt{1-\beta}}{\sqrt{1+\beta}} \right) \left( 1 - \frac{m_p}{m_e} \frac{\sqrt{1+\beta}}{\sqrt{1-\beta}} \right) - \frac{E_e}{2 E_{\overline{\nu}_e}} \ln \frac{\Sigma - 4 \left( \overline{\Delta} + m^2_e \right)^2}{4 m^4_e} \right. \nonumber \\
&\left. + \left( \frac{s - m^2_e}{2 m_e \sqrt{s}} \frac{E_e}{E_{\overline{\nu}_e}} - \frac{m_e}{\sqrt{s}} \frac{l_0}{2 E_{\overline{\nu}_e}} \right) \ln \frac{\left( \sqrt{\Sigma} - 2 m_e \sqrt{s} \right)^2 - 4 \left( \overline{\Delta} + m^2_e \right)^2}{\left( \sqrt{\Sigma} + 2 m_e \sqrt{s} \right)^2 - 4 \left( \overline{\Delta} + m^2_e \right)^2} + \frac{s - m^2_e}{s} \frac{l_0 + E_e}{E_{\overline{\nu}_e}} \ln \frac{\sqrt{\Sigma} - 2 \overline{\Delta}}{2 m^2_e} \right. \nonumber\\
&\left. + \frac{l_0}{2 E_{\overline{\nu}_e}} \ln \frac{16 m^{10}_e}{s \left( \sqrt{\Sigma} - 2 \overline{\Delta} \right)^4} - \left( 4 \frac{\overline{\Delta} + m^2_e}{\sqrt{\Sigma} + 2 \left( \overline{\Delta} + m^2_e \right)} \frac{l_0 + E_e}{E_{\overline{\nu}_e}} + 2 \frac{\left( \overline{\Delta} + m^2_e \right) \left( 2 \overline{\Delta} + m^2_e \right) + \left( \overline{\Delta} - m^2_e \right) s}{\sqrt{\Sigma}\left(\sqrt{\Sigma} + 2 \left( \overline{\Delta} + m^2_e \right) \right)} \frac{l_0}{E_{\overline{\nu}_e}} \right. \right. \nonumber \\
&\left. \left. + \frac{s - m^2_e}{\sqrt{\Sigma} + 2 \left( \overline{\Delta} + m^2_e \right)} \frac{ \sqrt{\Sigma} - 2 \overline{\Delta}}{\sqrt{\Sigma}} \frac{l_0 + E_e}{E_{\overline{\nu}_e}} - \frac{s + m^2_e + 2 \overline{\Delta}}{\sqrt{\Sigma} + 2 \left( \overline{\Delta} + m^2_e \right)} \frac{l_0 + 2 E_e}{E_{\overline{\nu}_e}} \right) \ln \frac{\sqrt{\Sigma} + 2 \overline{\Delta}}{\sqrt{\Sigma} - 2 \overline{\Delta}}\right]. \label{eq:radiative_NF_positron_energy_spectrum_logarithms_above}
\end{align}

Consequently, the complete positron energy spectrum in IBD with reactor antineutrinos at subpermille precision is given by
\begin{equation} \label{eq:positron_energy_spectrum_above}
	\mathrm{d} \sigma_\mathrm{NLO} + \mathrm{d} \sigma^{1 \gamma}_\mathrm{LO} \to \left[ 1 + \frac{\alpha}{\pi} \left( \delta_v + \delta_s + \delta_\mathrm{I} + \delta_\mathrm{II} + \delta_\gamma \right) \right] \mathrm{d} \sigma_{\mathrm{LO}} + \mathrm{d} \sigma_{v, \mathrm{NF}} + \mathrm{d} \sigma^{1\gamma}_\mathrm{NF} + \mathrm{d} \sigma_\mathrm{recoil} + \mathrm{d} \sigma_\mathrm{wm} + \mathrm{d} \sigma_\mathrm{FF},
\end{equation}
and does not depend on the unphysical parameters $\varepsilon_\gamma$ and $\lambda_\gamma$.

As in elastic (anti)neutrino-electron scattering and other QED and QCD processes, the full radiative correction to the cross section is free from Sudakov double logarithms~\cite{Sudakov:1954sw,Yennie:1961ad}. This cancellation becomes evident in the high-energy limit $\beta \to 1$, where the individual contributions behave as
\begin{equation} \label{eq:sudakov}
	\delta_v \underset{\beta \to 1}{\sim} -\frac{1}{4} \ln^2 \left( 1- \beta \right), \qquad \delta_s \underset{\beta \to 1}{\sim} -\frac{1}{4} \ln^2 \left( 1- \beta \right), \qquad \delta_\mathrm{I} \underset{\beta \to 1}{\sim} \frac{1}{2} \ln^2 \left( 1- \beta \right).
\end{equation}
The sum of these contributions ensures the cancellation of the double-logarithmic terms, leaving only single logarithms in the positron mass. This collinear behavior is consistent with findings in other charged-lepton spectra arising from (anti)neutrino-induced processes~\cite{Dicus:1982bz,Tomalak:2019ibg,Tomalak:2021lif,Tomalak:2022uwv,Cirigliano:2023fnz} and in high-energy (anti)neutrino-nucleon scattering~\cite{Tomalak:2021hec,Tomalak:2022xup}.

At the production threshold in exactly forward kinematics, when the outgoing neutron is at rest with $E_n \to m_n$, the positron energy spectrum simplifies. In this limit, it reduces to the sum of the leading-order result and the nonfactorizable vertex correction, as well as recoil, weak magnetism, and nucleon radii contributions,
\begin{equation} \label{eq:threshold_result_positron}
	\mathrm{d} \sigma_\mathrm{NLO} + \mathrm{d} \sigma^{1 \gamma}_\mathrm{LO} \underset{E_e \to m_e}{\longrightarrow} \mathrm{d} \sigma_\mathrm{NLO} \to \mathrm{d} \sigma_\mathrm{LO} + \mathrm{d} \sigma_{v, \mathrm{NF}} + \mathrm{d} \sigma_\mathrm{recoil} + \mathrm{d} \sigma_\mathrm{wm} + \mathrm{d} \sigma_\mathrm{FF}.
\end{equation}
This result represents a universal threshold behavior for the positron and electromagnetic energy spectra.

Near the endpoint of the elastic IBD kinematics, just below the maximum positron energy, the positron energy spectrum exhibits the logarithmic enhancement governed by infrared dynamics,
\begin{equation} \label{eq:logarithm_near_end_point_positron}
	\frac{\mathrm{d} \sigma^{1 \gamma}_\mathrm{LO}}{\mathrm{d} \sigma_\mathrm{LO}} \approx \frac{\alpha}{\pi} 2 \left( 1 - \frac{1}{2 \beta} \ln \frac{1+\beta}{1-\beta} \right) \ln \frac{\left( l_0 + \sqrt{l_0^2 - m^2_n} \right) m_e}{l^2_0 - \left( E_{\overline{\nu}_e} - \beta E_e \right)^2 - m^2_n},
\end{equation}
as determined by the infrared logarithms in Eqs.~(\ref{eq:F1_FF}) and~(\ref{eq:soft_photon_correction}).

\subsubsection{Below positron lowest energy in elastic IBD}
\label{sec:below}

In the region where the positron energy is below the minimum value allowed in elastic inverse beta decay, $E_e \le E_e^\mathrm{min}$, the elastic channel is kinematically forbidden. Consequently, the photon must carry sufficient energy to satisfy energy conservation, such that $E_\gamma > E_e^\mathrm{min} - E_e$. This lower bound on the photon energy ensures that no infrared divergence arises in this kinematic regime, eliminating the need for IR regularization. The positron energy spectrum below the energy $E_e^\mathrm{min}$ can be decomposed into factorizable and nonfactorizable contributions,
\begin{equation} \label{eq:positron_energy_spectrum_below}
	\mathrm{d} \sigma^{1\gamma}_\mathrm{LO} = \frac{\alpha}{\pi} \left( \delta_\mathrm{I} + \delta_{\gamma} \right) \mathrm{d} \sigma_{\mathrm{LO}} + \mathrm{d} \sigma^{1\gamma}_\mathrm{NF}.
\end{equation}
The factorizable term $\delta_\mathrm{I}$ arises from IR-sensitive contributions to the soft-photon bremsstrahlung amplitude in Eq.~(\ref{eq:soft_photon_amplitude}),
\begin{equation} \label{eq:delta_1_positron_below}
	\delta_\mathrm{I} = 2 \left( 1 - \frac{1}{2 \beta} \ln \frac{1+\beta}{1-\beta} \right) \ln \frac{l^2_0 - \left( E_{\overline{\nu}_e} + \beta E_e \right)^2 - m^2_n}{l^2_0 - \left( E_{\overline{\nu}_e} - \beta E_e \right)^2 - m^2_n}.
\end{equation}
The remaining factorizable contribution $\delta_\gamma$ is given by
\begin{equation} \label{eq:radiative_factorizable_positron_energy_spectrum_below}
	\delta_\gamma = \frac{1}{2} \mathrm{Li}_2 \frac{s \sqrt{1-\beta}}{m_p m_e \sqrt{1+\beta}} - \frac{1}{2} \mathrm{Li}_2 \frac{s \sqrt{1+\beta}}{m_p m_e \sqrt{1-\beta}} + \frac{1}{2} \mathrm{Li}_2 \frac{m_p \sqrt{1+\beta}}{m_e \sqrt{1-\beta}} - \frac{1}{2} \mathrm{Li}_2 \frac{m_p \sqrt{1-\beta}}{m_e \sqrt{1+\beta}}.
\end{equation}
The nonfactorizable contribution $\mathrm{d} \sigma^{1\gamma}_{\mathrm{NF}}$ to the positron energy spectrum is obtained by the following substitutions in Eq.~(\ref{eq:radiation_3d}):
\begin{equation} \label{eq:radiative_NF_positron_energy_spectrum_below}
	c^{3d}_i {\cal{D}}_m \to \frac{m_p}{\pi} \left( A_i + B_i \ln \frac{1+\beta}{1-\beta} + C_i \ln \frac{l^2_0 - \left( E_{\overline{\nu}_e} - \beta_n E_n \right)^2}{l^2_0 - \left( E_{\overline{\nu}_e} + \beta_n E_n \right)^2} + \frac{D_i}{2} l_1 - \frac{E_i}{2} l_2 + F_i \delta_\gamma \right) \mathrm{d} E_\mathrm{e},
\end{equation}
where the coefficients $A_\pm,~B_\pm,~C_\pm,~D_\pm,~E_\pm$, and $F_\pm$ can be expressed as
\begin{align}
	A_+&= 0, \\
	B_+&= 0, \\
	C_+&= 0, \\
	D_+&= 1, \\
	E_+&= 1, \\
	F_+&= 0, \\
	A_-&= - \frac{3}{2} \frac{\beta E_e}{E_{\overline{\nu}_e}}, \\
	B_-&= \frac{3}{2} \frac{E_e}{E_{\overline{\nu}_e}}, \\
	C_-&= \frac{m^2_n}{8 E^2_{\overline{\nu}_e}}, \\
	D_-&= \frac{E_e - E_{\overline{\nu}_e}}{2 E_{\overline{\nu}_e}}, \\
	E_-&= \frac{m_p}{2 E_{\overline{\nu}_e}}, \\
	F_-&= \frac{m^2_e}{2 E^2_{\overline{\nu}_e}}. \label{eq:radiative_NF_positron_energy_spectrum_coefficients_below}
\end{align}
The logarithmic terms $l_1$ and $l_2$ are defined as
\begin{align}
	l_1&= \frac{E_{\overline{\nu}_e} + \beta E_e}{E_{\overline{\nu}_e}} \ln \frac{l_0 - E_{\overline{\nu}_e} - \beta E_e}{l_0 + E_{\overline{\nu}_e} + \beta E_e} - \frac{E_{\overline{\nu}_e} - \beta E_e}{E_{\overline{\nu}_e}} \ln \frac{l_0 - E_{\overline{\nu}_e} + \beta E_e}{l_0 + E_{\overline{\nu}_e} - \beta E_e} + \frac{l_0}{E_{\overline{\nu}_e}} \ln \frac{l_0^2 - \left( E_{\overline{\nu}_e} - \beta E_e \right)^2}{l_0^2 - \left(E_{\overline{\nu}_e} + \beta E_e \right)^2}, \\
	l_2&= \frac{m^2_n}{s-m^2_e} \frac{l_0}{E_{\overline{\nu}_e}} \ln \frac{s + m^2_e - 2 E_e \left( l_0 + E_e - \beta E_{\overline{\nu}_e} \right)}{s + m^2_e - 2 E_e \left( l_0 + E_e + \beta E_{\overline{\nu}_e} \right)} + \frac{s - m^2_n - m^2_e}{s} \ln \frac{1 - \frac{s}{m_e m_p} \frac{\sqrt{1-\beta}}{\sqrt{1+\beta}}}{1 - \frac{s}{m_e m_p} \frac{\sqrt{1+\beta}}{\sqrt{1-\beta}}} \nonumber \\
&+ \frac{s - m^2_n - m^2_e}{s-m^2_e} \frac{m^2_p + m^2_e - 2 m_p E_e}{2 m_p E_{\overline{\nu}_e}} \ln \frac{1 - \frac{s}{m_e m_p} \frac{\sqrt{1-\beta}}{\sqrt{1+\beta}}}{1 - \frac{s}{m_e m_p} \frac{\sqrt{1+\beta}}{\sqrt{1-\beta}}} \frac{1 - \frac{m_p}{m_e} \frac{\sqrt{1+\beta}}{\sqrt{1-\beta}}}{1 - \frac{m_p}{m_e} \frac{\sqrt{1+\beta}}{\sqrt{1-\beta}}}. \label{eq:radiative_NF_positron_energy_spectrum_logarithms_below}
\end{align}

As expected, the positron energy spectrum vanishes in the $\beta \to 0$ limit and remains finite near the kinematic threshold $E_e \le E_e^\mathrm{min}$. In this region, the spectrum exhibits a logarithmic enhancement in the positron mass, consistent with general features of radiative charged-lepton spectra. The positron energy spectrum has the following logarithmic behavior just below the lowest positron energy in the elastic IBD kinematics:
\begin{equation} \label{eq:logarithm_near_end_point_positron_below}
	\frac{\mathrm{d} \sigma^{1 \gamma}_\mathrm{LO}}{\mathrm{d} \sigma_\mathrm{LO}} \approx \frac{\alpha}{\pi} 2 \left( 1 - \frac{1}{2 \beta} \ln \frac{1+\beta}{1-\beta} \right) \ln \frac{l^2_0 - \left( E_{\overline{\nu}_e} + \beta E_e \right)^2 - m^2_n}{l^2_0 - \left( E_{\overline{\nu}_e} - \beta E_e \right)^2 - m^2_n},
\end{equation}
in accordance with the infrared logarithms identified in Eqs.~(\ref{eq:F1_FF}) and~(\ref{eq:soft_photon_correction}).

\subsection{Electromagnetic and neutron energy spectra}
\label{sec:1xsec_electromagnetic_energy}

To compute the electromagnetic energy spectrum, we adopt the notations and integration domains introduced in Section~\ref{sec:3xsec_neutron_energy_neutron_angle_photon_energy}. The relevant phase space of hard photons with $E_\gamma \ge \varepsilon_\gamma$ is split into two regions, following the approach in Section~\ref{sec:above},
\begin{itemize}
	\item \textbf{Region I:} $\tilde{l}^2 - m_e^2 \ge 2\varepsilon_\gamma \left( \tilde{l}_0 + \tilde{f} \right)$. Here, the photon phase-space integration is unrestricted.
	\item \textbf{Region II:} $\tilde{l}^2 - m_e^2 \le 2\varepsilon_\gamma \left( \tilde{l}_0 - \tilde{f} \right)$, where the photon angle is constrained.
\end{itemize}
The allowed emission angle $\tilde{\gamma}$ between $\vec{\tilde{f}}$ and the photon momentum $\vec{k}_\gamma$ must satisfy
\begin{equation} \label{eq:photon_angle_gamma_tilde}
	\cos \tilde{\gamma} \ge \frac{1}{\tilde{f}} \left( \tilde{l}_0 - \frac{\tilde{l}^2-m_e^2}{2 \varepsilon_\gamma} \right).
\end{equation}
The correction factor from region II, denoted $\delta_\mathrm{II}$, matches that of elastic (anti)neutrino-electron scattering,
\begin{equation} \label{eq:delta_2_electromagnetic}
	\delta_\mathrm{II} = - \left( 1 - \frac{1}{2 \beta} \ln \frac{1+\beta}{1-\beta} \right) \ln \frac{1+\beta}{1-\beta}.
\end{equation}
Here $\beta$ depends on the total electromagnetic energy [cf. Eq.~(\ref{eq:beta_for_radiation})].

As for the positron energy spectrum, the bremsstrahlung contribution from region I can be separated into factorizable and nonfactorizable parts. The factorizable correction $\delta_\mathrm{I}$ is obtained from the factorizable terms in the soft-photon bremsstrahlung amplitude in Eq.~(\ref{eq:soft_photon_amplitude}),
\begin{equation} \label{eq:delta_1_electromagnetic}
	\delta_\mathrm{I} = 2 \left( 1 - \frac{1}{2 \beta} \ln \frac{1+\beta}{1-\beta} \right) \ln \frac{2 m_e \varepsilon_\gamma}{E^2_\mathrm{EM} - \left( E_{\overline{\nu}_e} - \beta_n E_n \right)^2 - m^2_e}.
\end{equation}
Combining the remaining nonfactorizable piece from terms beyond Eq.~(\ref{eq:soft_photon_amplitude}) with the contribution from regions I and II gives the full hard photon correction $\mathrm{d} \sigma^{1\gamma}_{\mathrm{LO}}$,
\begin{equation} \label{eq:radiative_electromagnetic_energy_spectrum}
	\mathrm{d} \sigma^{1\gamma}_{\mathrm{LO}} = \frac{\alpha}{\pi} \left( \delta_\mathrm{I} + \delta_\mathrm{II} + \delta_{\gamma} \right)\mathrm{d} \sigma_{\mathrm{LO}} + \mathrm{d} \sigma^{1\gamma}_{\mathrm{NF}},
\end{equation}
where $\mathrm{d} \sigma_\mathrm{LO}$ is the elastic cross section expressed in terms of the neutron energy and $\delta_\gamma$,
\begin{align}
	\delta_\gamma &= \sum \limits_{\sigma_1 = \pm 1} \frac{\sigma_1}{\beta} \mathrm{Li}_2 \left[ \frac{2 E_\mathrm{EM} }{m_e} \left( \frac{ 1 + \beta }{ 1 - \beta } \right)^{\frac{\sigma_1}{2}} - 1 \right] - \sum \limits_{\sigma_1, \sigma_2 = \pm 1} \frac{\sigma_1}{\beta} \mathrm{Li}_2 \left[ \frac{E_\mathrm{EM} + \sigma_2 \left( E_{\overline{\nu}_e} - \beta_n E_n \right)}{m_e} \left( \frac{ 1 + \beta }{ 1 - \beta } \right)^{\frac{\sigma_1}{2}} \right]\nonumber \\
	&- \frac{1}{\beta} \ln \left[ \frac{E^2_\mathrm{EM} - \left( E_{\overline{\nu}_e} - \beta_n E_n \right)^2}{4m^2_e} \frac{E^2_\mathrm{EM} - \left( E_{\overline{\nu}_e} - \beta_n E_n \right)^2 - m^2_e}{E^2_\mathrm{EM} - m^2_e} \left( \frac{ 1 + \beta }{ 1 - \beta } \right)^{\frac{1}{2}} \right] \ln \frac{ 1 + \beta }{ 1 - \beta } \nonumber \\
&- \frac{1}{\beta} \ln \frac{E_\mathrm{EM} + E_{\overline{\nu}_e} - \beta_n E_n}{E_\mathrm{EM} - E_{\overline{\nu}_e} + \beta_n E_n} \ln \frac{\beta E_\mathrm{EM} + E_{\overline{\nu}_e} - \beta_n E_n}{\beta E_\mathrm{EM} - E_{\overline{\nu}_e} + \beta_n E_n}. \label{eq:radiative_factorizable_electromagnetic_energy_spectrum}
\end{align}
The nonfactorizable contribution $\mathrm{d} \sigma^{1\gamma}_{\mathrm{NF}}$ to the electromagnetic energy spectrum in Eq.~(\ref{eq:radiative_electromagnetic_energy_spectrum}) is given by the following substitutions in Eq.~(\ref{eq:radiation_3d}):
\begin{equation} \label{eq:radiative_NF_electromagnetic_energy_spectrum}
	c^{3d}_i {\cal{D}}_m \to \frac{m_p}{\pi} \left( \overline{a}_i + \overline{b}_i \ln \frac{1+\beta}{1-\beta} + \overline{c}_i \ln \frac{E^2_\mathrm{EM} - \left( E_{\overline{\nu}_e} - \beta_n E_n \right)^2}{m^2_e} + \overline{d}_i \ln \frac{E_\mathrm{EM} + E_{\overline{\nu}_e} - \beta_n E_n}{E_\mathrm{EM} - E_{\overline{\nu}_e} + \beta_n E_n} \right) \mathrm{d} E_\mathrm{EM},
\end{equation}
where the coefficients $\overline{a}_\pm,~\overline{b}_\pm,~\overline{c}_\pm$, and $\overline{d}_\pm$ can be expressed as
\begin{align}
	\overline{a}_+&= - \frac{E_\mathrm{EM}}{4 E_{\overline{\nu}_e}} \left( 1 - \frac{m^2_e}{E^2_\mathrm{EM} - \left( E_{\overline{\nu}_e} - \beta_n E_n \right)^2 } \right), \\
	\overline{b}_+&= \frac{E_\mathrm{EM}}{E_{\overline{\nu}_e}}, \\
	\overline{c}_+&= \frac{E_\mathrm{EM}}{4 E_{\overline{\nu}_e}}, \\
	\overline{d}_+&= 0, \\
	\overline{a}_-&= \left( 1 - \frac{m^2_e}{E^2_\mathrm{EM} - \left( E_{\overline{\nu}_e} - \beta_n E_n \right)^2 } \right) \frac{E^2_n - m^2_n - E^2_\mathrm{EM} - \left(E_{\overline{\nu}_e} - E_\mathrm{EM} \right)^2 + \left(E_{\overline{\nu}_e} - \beta_n E_n\right)^2}{16 E^2_{\overline{\nu}_e}} \nonumber \\
&+ \frac{E^2_\mathrm{EM} - \left( E_{\overline{\nu}_e} - \beta_n E_n \right)^2 - m^2_e}{4 E^2_{\overline{\nu}_e}}, \\
	\overline{b}_-&= - \frac{E_\mathrm{EM}}{2 E_{\overline{\nu}_e}} + \frac{1-\beta}{\beta} \frac{E^2_n - m^2_n - E^2_{\overline{\nu}_e} - E^2_\mathrm{EM} + m^2_e}{4 E^2_{\overline{\nu}_e}}, \\
	\overline{c}_-&= - \frac{E^2_n - m^2_n - E^2_{\overline{\nu}_e} - E^2_\mathrm{EM} }{4 E^2_{\overline{\nu}_e}} - \frac{\overline{\Delta} + m_p E_\mathrm{EM}}{8 E^2_{\overline{\nu}_e}}, \\
	\overline{d}_-&= \frac{E_\mathrm{EM}}{2 E_{\overline{\nu}_e}}. \label{eq:radiative_NF_electromagnetic_energy_spectrum_coefficients}
\end{align}

The full electromagnetic energy spectrum, accurate to subpermille precision, for IBD with reactor antineutrinos is given by
\begin{equation} \label{eq:electromagnetic_energy_spectrum}
	\mathrm{d} \sigma_\mathrm{NLO} + \mathrm{d} \sigma^{1 \gamma}_\mathrm{LO} \to \left[ 1 + \frac{\alpha}{\pi} \left( \delta_v + \delta_s + \delta_\mathrm{I} + \delta_\mathrm{II} + \delta_\gamma \right) \right] \mathrm{d} \sigma_{\mathrm{LO}} + \mathrm{d} \sigma_{v, \mathrm{NF}} + \mathrm{d} \sigma^{1\gamma}_\mathrm{NF} + \mathrm{d} \sigma_\mathrm{recoil} + \mathrm{d} \sigma_\mathrm{wm} + \mathrm{d} \sigma_\mathrm{FF},
\end{equation}
and is manifestly independent of the unphysical parameters $\varepsilon_\gamma$ and $\lambda_\gamma$.

At the production threshold in forward kinematics, when $E_\mathrm{EM} \to \frac{\left( m_n + m_e \right)^2}{2 m_p} - m_p$, the electromagnetic energy spectrum coincides with the positron energy spectrum [see Eq.~(\ref{eq:threshold_result_positron})].

As for the positron energy spectrum above the lowest energy of the elastic IBD, individual corrections exhibit Sudakov-type double logarithms,
\begin{equation} \label{eq:sudakov2}
	\delta_v \underset{\beta \to 1}{\sim} -\frac{1}{4} \ln^2 \left( 1- \beta \right), \qquad \delta_s \underset{\beta \to 1}{\sim} -\frac{1}{4} \ln^2 \left( 1- \beta \right), \qquad \delta_\mathrm{II} \underset{\beta \to 1}{\sim} \frac{1}{2} \ln^2 \left( 1- \beta \right),
\end{equation}
but the complete electromagnetic energy spectrum is free from these logarithms~\cite{Sudakov:1954sw} and is free from positron-mass singularities as an infrared-safe observable~\cite{Bloch:1937pw,Nakanishi:1958ur,Yennie:1961ad,Kinoshita:1962ur}, after specifying the renormalization scale $\mu$ in Eq.~(\ref{eq:F1_FF}) independent of the positron mass. Note that the relativistic form factor $f_1$ has a twice as large coefficient in front of $\ln \frac{\mu^2}{m^2_e}$ compared to Eq.~(\ref{eq:F1_FF}), while both relativistic and nonrelativistic form factors have the same collinear behavior, i.e., the same $\ln m_e$ dependence in the limit $m_e \to 0$ in Eq.~(\ref{eq:F1_FF}), as it was observed at higher antineutrino energies in Refs.~\cite{Tomalak:2021hec,Tomalak:2022xup} and in the static limit in Refs.~\cite{Vogel:1983hi,Fayans:1985uej,Kurylov:2002vj,Fukugita:2004cq,Raha:2011aa,Ankowski:2016oyj,Tomalak:2021hec,Tomalak:2022xup}.

Finally, the electromagnetic energy spectrum has the following logarithmic behavior near endpoints of the elastic IBD kinematics:
\begin{equation} \label{eq:logarithm_near_end_point_positron_above}
	\frac{\mathrm{d} \sigma^{1 \gamma}_\mathrm{LO}}{\mathrm{d} \sigma_\mathrm{LO}} \approx \frac{\alpha}{\pi} 2 \left( 1 - \frac{1}{2 \beta} \ln \frac{1+\beta}{1-\beta} \right) \ln \frac{m^2_e}{E^2_\mathrm{EM} - \left( E_{\overline{\nu}_e} - \beta_n E_n \right)^2 - m^2_e},
\end{equation}
consistent with expectations from infrared logarithms in Eqs.~(\ref{eq:F1_FF}) and~(\ref{eq:soft_photon_correction}). This logarithmic enhancement is not captured by the ``static limit" approximation of the radiative phase space in Refs.~\cite{Vogel:1983hi,Fayans:1985uej,Kurylov:2002vj,Fukugita:2004cq,Raha:2011aa,Ankowski:2016oyj,Tomalak:2021hec,Tomalak:2022xup}.

\subsection{Electromagnetic energy spectrum in static limit}
\label{sec:static_limit}

In this section, we present analytic expressions for the electromagnetic energy spectrum from Refs.~\cite{Fukugita:2004cq} and~\cite{Raha:2011aa}, obtained by performing the phase-space integration in the static limit. The resulting electromagnetic energy spectrum in the unpolarized IBD can be expressed in terms of two functions of the velocity $\beta$ that depends on the total electromagnetic energy, cf. Eq.~(\ref{eq:beta_for_radiation}), as
\begin{equation} \label{eq:electromagnetic_energy_spectrum_static_limit}
	\frac{\mathrm{d} \sigma_\mathrm{NLO} + \mathrm{d} \sigma^{1 \gamma}_\mathrm{LO}}{\mathrm{d} E_\mathrm{EM}} \to \left[ 1 + \frac{\alpha}{\pi} \delta_1 \right] \frac{\mathrm{d} \sigma_{\mathrm{LO}}}{\mathrm{d} E_\mathrm{EM}} +  \frac{\alpha}{\pi} m_p \frac{G^2_F |V_{ud}|^2}{\pi} \frac{m^2_e + Q^2}{8 E^2_{\overline{\nu}_e}} \left( 3 \delta_1 - \delta_2  \right) \left( g^2_V - g^2_A \right),
\end{equation}
with the angle-independent ``outer" correction $\delta_1$:
\begin{equation} \label{eq:delta_1}
	\delta_1 = \frac{3}{4} \ln \frac{\mu^2}{m_e^2} + \frac{7}{2} + \frac{7 + 3 \beta^2}{8 \beta} \ln \frac{1 + \beta}{1 - \beta} + 2 \left( 1 - \frac{1}{2 \beta} \ln \frac{1 + \beta}{1-\beta} \right) \ln \frac{1 - \beta^2}{4 \beta^2} - \frac{1}{\beta} \ln^2 \frac{1 + \beta}{1-\beta} - \frac{4}{\beta} \mathrm{Li}_2 \frac{2 \beta}{1 + \beta},
\end{equation}
which modifies the $\Theta_e$-independent terms in Eq.~(\ref{eq:LO_static_diff_angle_full_1_over_m}), and the angle-dependent ``outer" correction $\delta_2$,
\begin{align} \label{eq: delta_2}
	\delta_2 &= \frac{9}{4} \ln \frac{\mu^2}{m_e^2} + 1 + 2 \frac{1 - \sqrt{1 - \beta^2}}{\beta^2} + \frac{1 - 4 \beta}{4 \beta}  \ln \frac{1 + \beta}{1-\beta} - \frac{ 1 - 4 \beta + 3 \beta^2}{16 \beta^2} \ln^2 \frac{1 + \beta}{1-\beta} - \frac{4}{\beta} \mathrm{Li}_2 \left[ 1 - \frac{\sqrt{1 - \beta}}{\sqrt{1 + \beta}} \right] \nonumber \\
	&+ 2 \left( 1 - \frac{1}{2 \beta} \ln \frac{1 + \beta}{1-\beta} \right) \ln \left[ \frac{ 1 + \beta}{2 \beta} \frac{\sqrt{1 + \beta} + \sqrt{1 - \beta}}{\sqrt{1 + \beta} - \sqrt{1 - \beta}} \right],
\end{align}
which modifies the terms with $\cos \Theta_e$ in Eq.~(\ref{eq:LO_static_diff_angle_full_1_over_m}). The angle-independent ``outer" correction can be equivalently written as~\cite{Tomalak:2022xup} after removing Coulomb corrections and accounting for the difference in renormalization schemes,
\begin{align} \label{eq: delta_1_short}
	\delta_1 = \frac{3}{4} \ln \frac{\mu^2}{m_e^2} + \frac{7}{2} + \frac{7 - 16 \beta + 3 \beta^2}{8 \beta} \ln \frac{1 + \beta}{1 - \beta} + 4 \left( 1 + \frac{1}{2 \beta} \ln \frac{1 + \beta}{1-\beta} \right)  \ln \frac{1 + \beta}{2 \beta} + \frac{4}{\beta} \mathrm{Li}_2 \frac{1 - \beta}{1 + \beta} - \frac{2}{3} \frac{\pi^2}{\beta}.
\end{align}

\section{Results and discussion}
\label{sec:results}

In this section, we present our main results and discuss their relevance for inverse beta decay, particularly in the context of reactor antineutrino experiments. In Section~\ref{sec:xsec}, we verify the consistency between the positron and neutron energy distributions and provide a detailed evaluation of the total IBD cross section as a function of the antineutrino energy, including a comprehensive uncertainty analysis. In Section~\ref{sec:averaged_energy}, we explore the effects of radiative corrections on the positron and electromagnetic energy spectra and present results for the corresponding averaged energies. Section~\ref{sec:averaged_angle} examines modifications to the positron scattering angle due to radiative effects. In Section~\ref{sec:energy_reconstuction_in_liquid_scintillator}, we quantify radiative effects on the reconstruction of the antineutrino energy in liquid scintillators. Beyond the Standard Model implications are considered in Section~\ref{sec:BSM}. In Section~\ref{sec:neutron_decay}, we extend our results to the case of charged-current neutrino-neutron elastic scattering, verify electron and neutrino energy spectra in the neutron decay, and update QED radiative corrections to the neutron lifetime and beta asymmetry. Finally, we compare our findings to previous results in the literature in Section~\ref{sec:literature_corrections}.

\subsection{Total cross section: Energy dependence and uncertainties}
\label{sec:xsec}

\begin{figure}[H]
	\centering
	\includegraphics[width=0.97\textwidth]{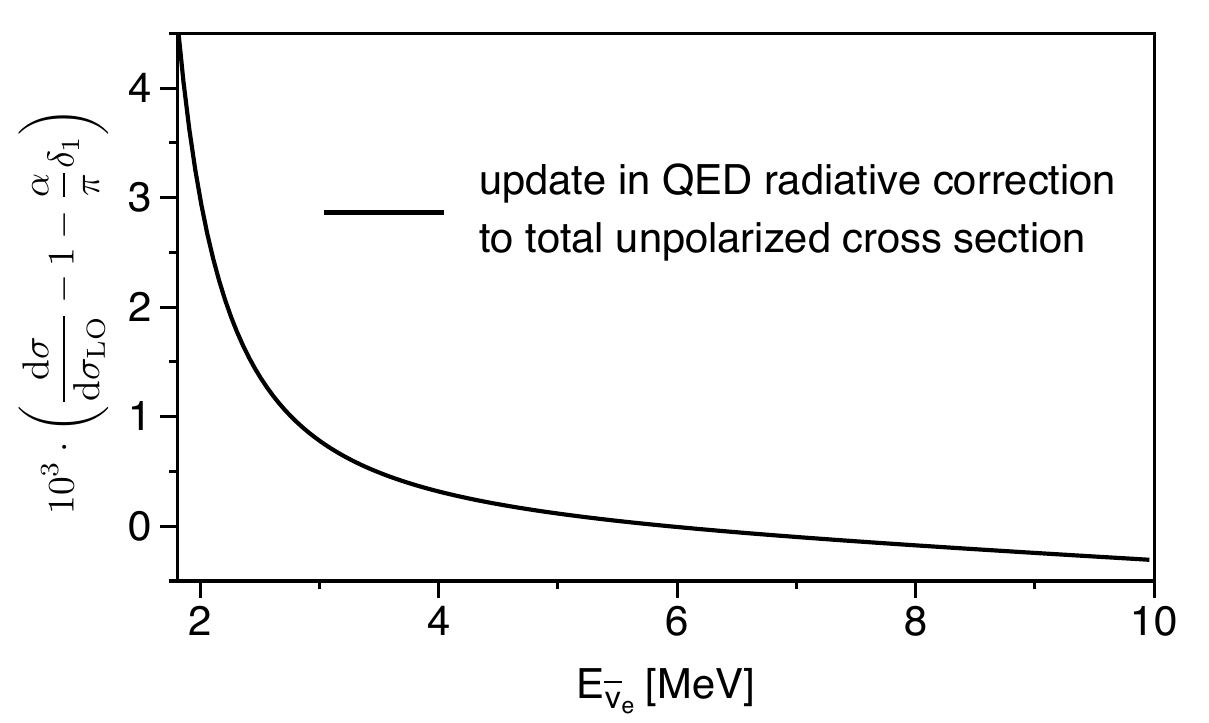}
	\caption{Difference in radiative correction to the total unpolarized cross section between results from this paper and the application of the multiplicative factor $\frac{\alpha}{\pi} \delta_1$ is shown as a function of the antineutrino beam energy $E_{\overline{\nu}_e}$. The width of the curve corresponds to the allowed range of the recoil positron energy in the radiation-free IBD. \label{fig:simple_expression}}
\end{figure}

The total cross section is obtained by integrating the electromagnetic energy spectrum presented in Section~\ref{sec:1xsec_electromagnetic_energy} over the full kinematic range permitted by the elastic inverse beta decay process. This result is fully consistent with the sum of two contributions: the integral of the positron energy spectrum from Section~\ref{sec:above} over the elastic IBD kinematics, and the integral of the pure radiative contribution from Section~\ref{sec:below}, spanning from the positron mass up to the lowest positron energy allowed in elastic IBD. Applying the radiative correction $\delta_1$ of Eq.~(\ref{eq:delta_1}) to the total cross section instead introduces differences at 1-2$\permil$ level compared to the integration based on Eq.~(\ref{eq:electromagnetic_energy_spectrum_static_limit}) and deviates from the result based on Sections~\ref{sec:1xsec_electromagnetic_energy},~\ref{sec:above},~and~\ref{sec:below} by less than 1$\permil$ at antineutrino beam energies above $E_{\overline{\nu}_e} \gtrsim 2.7~\mathrm{MeV}$. We illustrate the difference in the total unpolarized IBD cross section between the application of the multiplicative factor and the calculation of this paper as a function of the antineutrino energy $E_{\overline{\nu}_e}$ in Fig.~\ref{fig:simple_expression}.

To illustrate the contributions to the uncertainty of the total unpolarized IBD cross section, we provide benchmark values of the IBD cross section at antineutrino energies $E_{\overline{\nu}_e} = 3~\mathrm{MeV}$ and $E_{\overline{\nu}_e} = 5~\mathrm{MeV}$,\footnote{Contributions from the pseudoscalar interaction and unaccounted higher-order effects could shift the last significant digit of these values.}
\begin{align}
	\sigma \left( E_{\overline{\nu}_e} = 3~\mathrm{MeV} \right) &= \big[ 0.266664 \times 10^{-42} {\rm cm}^2 \big] \big[ 1 \pm {0.00171}_{\lambda} \pm {0.00064}_{V_{ud}} \pm {0.00004}_{g_V} \pm {0.00001}_{r^2_A} \big], \label{eq:sigmatot1} \\
	\sigma \left( E_{\overline{\nu}_e} = 5~\mathrm{MeV} \right) &= \big[ 1.28688 \times 10^{-42} {\rm cm}^2 \big] \big[ 1 \pm {0.00171}_{\lambda} \pm {0.00064}_{V_{ud}} \pm {0.00004}_{g_V} \pm {0.00004}_{r^2_A} \big]. \label{eq:sigmatot2}
\end{align}
The relative uncertainty from the CKM matrix element $V_{ud}$ is constant across all energies. Uncertainties from the axial-vector-to-vector coupling constant ratio $\lambda$ and the vector coupling constant $g_V$ are also nearly energy-independent. In contrast, the contribution from the squared nucleon axial-vector radius $r^2_A$ increases with energy. The results above are based on the PDG value for the axial-vector-to-vector coupling constant ratio $\lambda$~\cite{Beck:2019xye,ParticleDataGroup:2024cfk,Beringer:2024ady,Workman:2022ynf}. If instead we use the more precise measurement from the PERKEO-III experiment~\cite{Markisch:2018ndu,Dubbers:2018kgh} rather than the PDG-suggested value, the uncertainty from $\lambda$ is reduced by a factor of 2.3, without shifting the central value beyond the quoted $1\sigma$ band. The relative uncertainty from higher-order perturbative contributions for these two energies is negligibly small $\lesssim 10^{-5}$.

QED radiative corrections decrease the total cross section by $1.3\%$ and $1.6\%$ at energies $E_{\overline{\nu}_e} = 3~\mathrm{MeV}$ and $E_{\overline{\nu}_e} = 5~\mathrm{MeV}$, respectively. Including QCD and electroweak effects, estimated here as the difference to the $g_V = 1$ limit, the total suppression decreases to $0.4\%$ and $0.8\%$, respectively. The dominant sources of theoretical uncertainty are the axial-vector-to-vector coupling constant ratio $\lambda$ and the CKM matrix element $V_{ud}$.

\begin{figure}[H]
	\centering
	\includegraphics[width=0.97\textwidth]{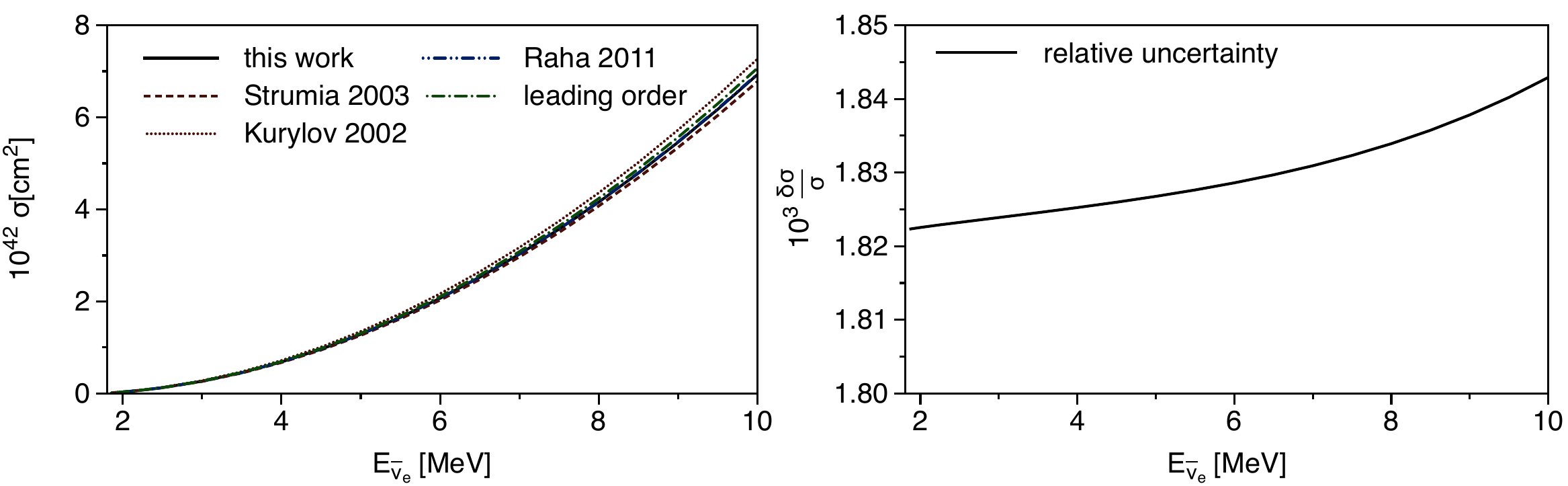}
	\caption{Total IBD cross section is shown as a function of the antineutrino beam energy $E_{\overline{\nu}_e}$ in the left panel. Our result is compared to the cross-section parameterization in Ref.~\cite{Strumia:2003zx}, labeled as ``Strumia 2003", the treatment of radiative corrections in Ref.~\cite{Kurylov:2002vj} as explained in Refs.~\cite{Strumia:2003zx,Hayes:2016qnu}, labeled as ``Kurylov 2002", and radiative corrections from Refs.~\cite{Fayans:1985uej,Vogel:1999zy,Fukugita:2004cq,Raha:2011aa}, described in Section~\ref{sec:static_limit} and labeled as ``Raha 2011". The relative uncertainty of the total cross section, dominated by $\lambda$ and $V_{ud}$, is illustrated in the right panel. \label{fig:total_cross_section}}
\end{figure}

In the following, we discuss cross sections as functions of antineutrino energy and compare our results with existing ones. We illustrate the total IBD cross section as a function of the incoming antineutrino energy $E_{\overline{\nu}_e}$ in the left panel in Fig.~\ref{fig:total_cross_section} and compare our result to the cross-section parameterization in Ref.~\cite{Strumia:2003zx} and previous calculations incorporating radiative corrections from Refs.~\cite{Kurylov:2002vj,Ricciardi:2022pru,Strumia:2003zx,Hayes:2016qnu} and Refs.~\cite{Fayans:1985uej,Vogel:1999zy,Fukugita:2004cq,Raha:2011aa}. Besides the latest compilation of radiative corrections from Refs.~\cite{Fayans:1985uej,Vogel:1999zy,Fukugita:2004cq,Raha:2011aa}, cf. Section~\ref{sec:static_limit} for details, at antineutrino energies $2.0~\mathrm{MeV} \lesssim E_{\overline{\nu}_e} \lesssim 7.5~\mathrm{MeV}$, none of the prior predictions fall within our estimated uncertainty band, primarily due to incomplete treatment of all contributions to radiative corrections. Our cross sections are above the results in Ref.~\cite{Strumia:2003zx} but remain below the values obtained using parameterizations of QED, QCD, and electroweak corrections in Refs.~\cite{Kurylov:2002vj,Ricciardi:2022pru,Strumia:2003zx,Hayes:2016qnu}, as well as results of Refs.~\cite{Fayans:1985uej,Vogel:1999zy,Fukugita:2004cq,Raha:2011aa}. We also present the relative uncertainty of the total cross section in the right panel in Fig.~\ref{fig:total_cross_section}. The relative cross-section uncertainty is nearly flat across the entire energy range and is well approximated by a constant value of $1.85\permil$.

\subsection{Positron, electromagnetic, and neutron energies}
\label{sec:averaged_energy}

\begin{figure}[htp]
	\centering
	\includegraphics[width=0.97\textwidth]{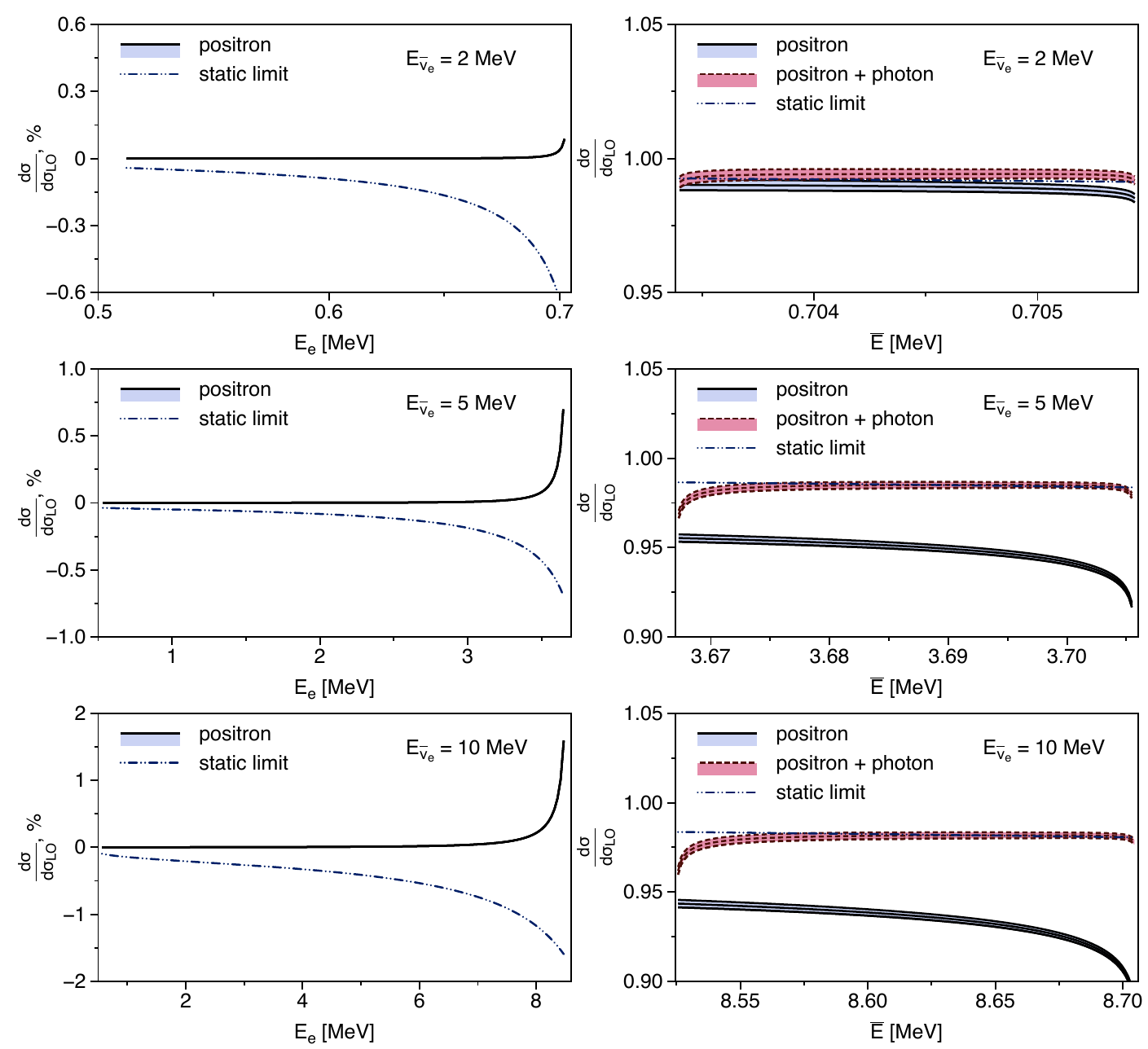}
	\caption{Effects from radiative corrections on the positron and electromagnetic energy spectra in IBD are presented as a function of the positron or electromagnetic energy for incoming antineutrino energies $E_{\overline{\nu}_e}= 2,~5,$ and $10~\mathrm{MeV}$. Ratio of our complete result to the leading-order prediction including recoil, weak magnetism, and nucleon radii corrections is compared to the electromagnetic energy spectrum in Refs.~\cite{Fayans:1985uej,Vogel:1999zy,Fukugita:2004cq,Raha:2011aa}, described in Section~\ref{sec:static_limit} and labeled as ``static limit". The left panel represents the pure radiative kinematic range, while the right panel illustrates the narrow kinematic range of the elastic IBD. $\overline{E}$ corresponds to the positron or electromagnetic energy in the right panel. Only uncertainties of the numerator are included. \label{fig:energy_spectra}}
\end{figure}

We investigate the impact of radiative corrections on the positron and electromagnetic energy spectra in inverse beta decay for fixed incoming antineutrino energies of $2$, $5$, and $10~\mathrm{MeV}$. In Fig.~\ref{fig:energy_spectra}, we present these effects as ratios of the corrected spectra to the corresponding spectra at leading order, which include recoil, weak magnetism, and nucleon radii corrections. For comparison, we also show the predictions for the electromagnetic energy spectrum available in the literature in Refs.~\cite{Fayans:1985uej,Vogel:1999zy,Fukugita:2004cq,Raha:2011aa}, described in Section~\ref{sec:static_limit} and denoted as the ``static limit".

We separate the contributions from the elastic IBD kinematics and the purely radiative region and illustrate them in the right and left panels, respectively. The positron spectra are normalized to the leading-order cross section evaluated at positron energies outside the elastic kinematic range. The spectral distortion in the electromagnetic, or neutron, energy is generally flat but displays logarithmic features near the endpoints of the elastic IBD kinematics. In contrast, the positron energy spectrum shows more variation across its energy range and exhibits logarithmic behavior below the kinematic endpoints.

Crucially, our calculation yields a positive radiative cross section, as expected from general principles. By comparison, the electromagnetic energy spectrum reported in Refs.~\cite{Fayans:1985uej,Vogel:1999zy,Fukugita:2004cq,Raha:2011aa} turns negative in the region of positron energies allowed only in the radiative process and, therefore, should not be applied in this region for precision analyses.

Additionally, we investigate the size of leading radiative-recoil corrections by numerically evaluating the sum of the bremsstrahlung contribution in Eq.~(\ref{eq:bremsstrahlung_in_terms_of_the_amplitude}) and the factorizable correction in Eq.~(\ref{eq:virtual_correction}), obtained from the leading-order IBD amplitudes with relativistic nucleon fields but keeping the HBChPT coupling to the photon field. For the positron energy spectrum, the corresponding corrections to the central results of this paper are negative and well below $\ll0.5\permil$ for the dominant spectral region in IBD with reactor antineutrinos. The differences increase with the energy of the incoming antineutrino. The largest deviations are observed at the lowest recoil positron energies in the elastic IBD kinematics and reach $\sim 1\permil$ at $E_{\overline{\nu}_e} = 10~\mathrm{MeV}$. In the pure radiative kinematic range, these corrections are negligibly small. The radiative-recoil corrections to the electromagnetic energy spectrum are also negative and increase in size with the reactor antineutrino energy. In contrast, the largest deviations are observed at the highest energy in elastic IBD kinematics. The deviations are well below $\ll 0.5\permil$ for the dominant spectral region in IBD and reach $\sim 1\permil$ at $E_{\overline{\nu}_e} = 10~\mathrm{MeV}$. The corresponding radiative-recoil corrections to the total IBD cross section increase in size with the antineutrino energy and are found to be below $1$-$2\times10^{-4}$ for the dominant spectral region in IBD, reaching $5.6\times10^{-4}$ at $E_{\overline{\nu}_e} = 10~\mathrm{MeV}$. The investigated enhanced radiative-recoil corrections originate from the factorizable terms correcting the leading-order cross section for the dominant contribution from soft and collinear regions and, therefore, can be well approximated with the product of the recoil correction in Eq.~(\ref{eq:recoil_differential}) and the factorizable correction to the energy spectrum, either the sum of Eqs.~(\ref{eq:vector_form_factor_correction}) and~(\ref{eq:soft_photon_correction}) with the replacement $\varepsilon_\gamma \to \frac{m_e}{2}$ or the total factorizable contribution. Besides small antineutrino energies just above the production threshold, the corresponding corrections in the charged-current neutrino-neutron elastic scattering are expected to be smaller in size due to compensation by additional factorizable positive Coulomb corrections that are absent in IBD with the electrically neutral neutron in the final state.

\begin{figure}[H]
	\centering
	\includegraphics[width=0.97\textwidth]{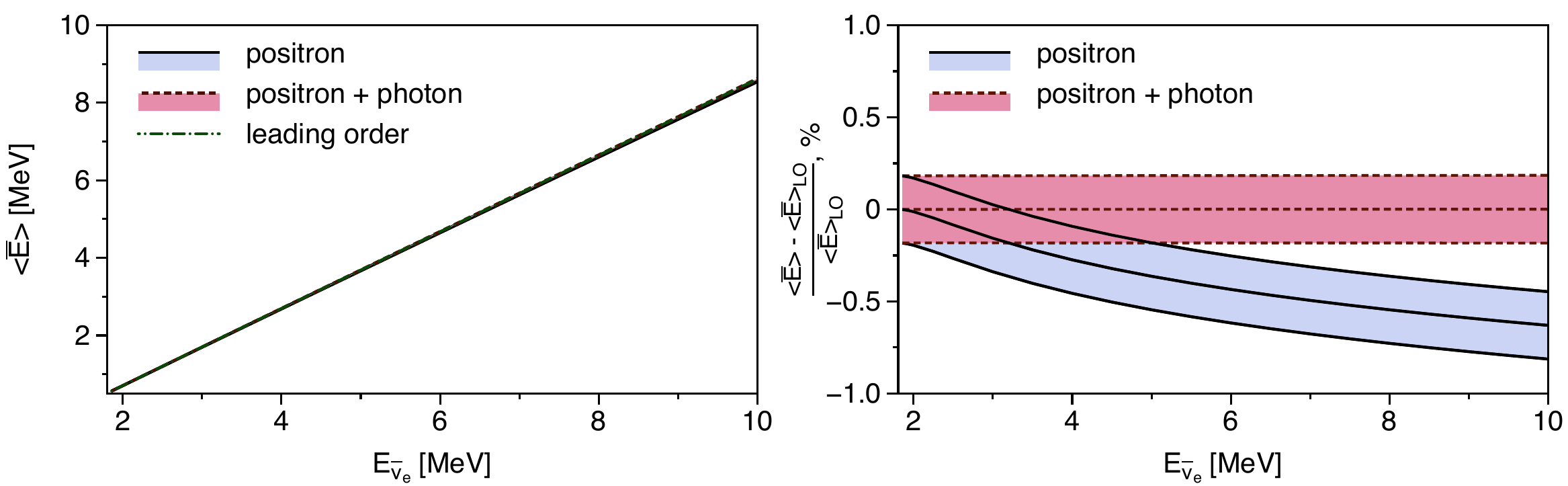}
	\caption{Averaged positron energy is compared to the averaged electromagnetic energy and averaging over the leading-order spectrum, including recoil, weak magnetism, and nucleon radii corrections, as a function of the antineutrino energy $E_{\overline{\nu}_e}$ in IBD. The left panel shows the absolute values, while the right panel presents the relative differences compared to averages over the leading-order spectrum with recoil, weak magnetism, and nucleon radii corrections. \label{fig:averaged_energy}}
\end{figure}

Next, we compute the averaged positron and electromagnetic energies at fixed antineutrino beam energy and present the results, accounting for uncertainties, in Fig.~\ref{fig:averaged_energy}. For comparison, we include averages based on the leading-order spectrum. Due to the narrow kinematic range of the elastic IBD, the leading-order and electromagnetic energy averages are nearly identical, driven by kinematic relations with minor recoil corrections. Radiative corrections do not alter the averaged positron energy at the reaction threshold and reduce it by $0.6\%$ at the antineutrino energy $E_{\overline{\nu}_e} = 10~\mathrm{MeV}$. This reduction increases monotonically with the antineutrino energy.

\subsection{Positron scattering angle}
\label{sec:averaged_angle}

\begin{figure}[H]
	\centering
	\includegraphics[width=0.8\textwidth]{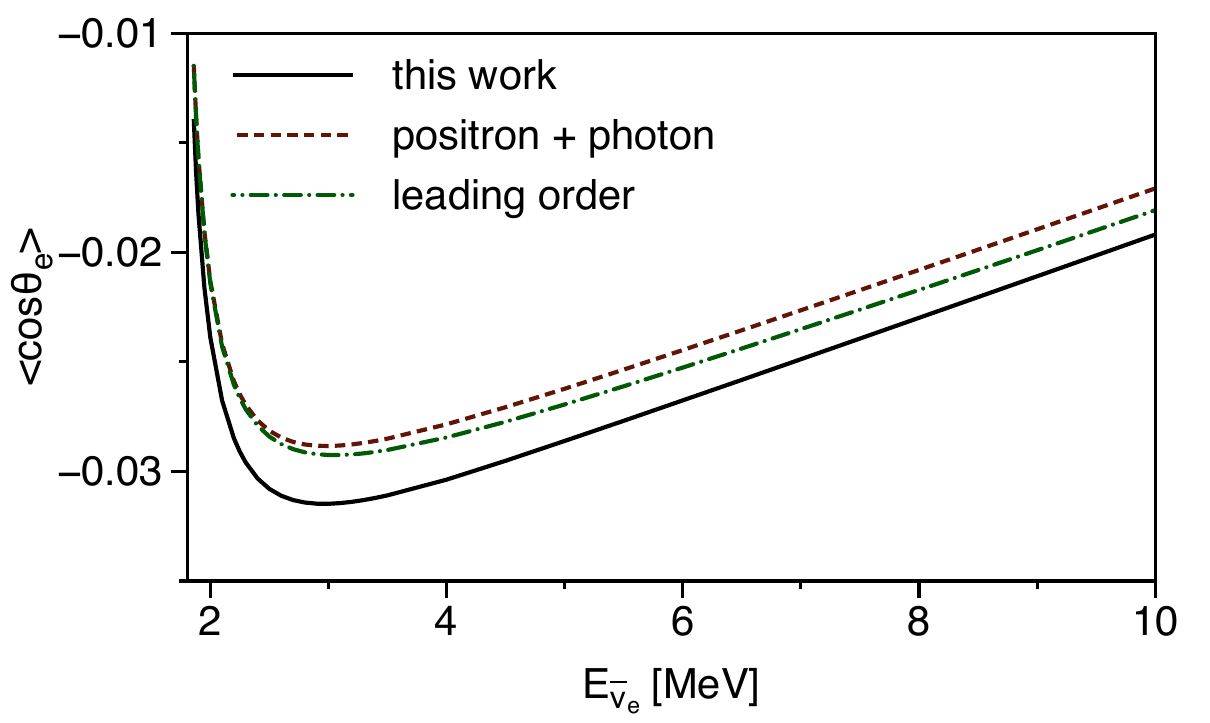}
	\caption{Averaged positron scattering angle is shown as a function of the antineutrino energy $E_{\overline{\nu}_e}$ in IBD. The result is compared to averaging over the leading-order spectrum, including recoil, weak magnetism, and nucleon radii corrections, and the electromagnetic energy spectrum under the assumption of elastic IBD kinematics. \label{fig:averaged_angle}}
\end{figure}
We evaluate the averaged positron scattering angle at fixed antineutrino beam energy by performing a weighted averaging over the two-dimensional distribution in Section~\ref{sec:2xsec_positron_energy_positron_angle} for radiative events with photons above the artificial energy cutoff $E_\gamma \ge \varepsilon_\gamma$. For photon energies below this cutoff $E_\gamma \le \varepsilon_\gamma$, we employ leading-order positron distributions, incorporating the virtual, soft, recoil, weak magnetism, and nucleon radii corrections described in Sections~\ref{sec:virtual_QED},~\ref{sec:soft_photons},~\ref{sec:recoil},~\ref{sec:weak_magnetism}, and~\ref{sec:form_factors}, respectively. We present the resulting averaged positron scattering angle as a function of the incoming antineutrino energy in Fig.~\ref{fig:averaged_angle}. The plot includes comparisons to the averaging over the electromagnetic energy spectrum, under the assumption of elastic IBD kinematics, and the leading-order spectrum with recoil, weak magnetism, and nucleon radii corrections included.

Our results demonstrate that the positron tends to scatter slightly backward on average. The inclusion of radiative corrections shifts the cosine of the averaged positron scattering angle by relative $\sim5$-$15\%$ underlining the importance of photon radiation effects in precision measurements. Furthermore, we confirm that the positron scattering angle is particularly sensitive to weak magnetism contributions, in agreement with findings in Ref.~\cite{Vogel:1999zy}.

\subsection{Antineutrino energy reconstruction in liquid scintillators}
\label{sec:energy_reconstuction_in_liquid_scintillator}

We quantitatively investigate the effects of radiative corrections in IBD on the reconstruction of the antineutrino energy in liquid scintillators, e.g., in the JUNO experiment. As the first approximation, the incoming antineutrino energy $E^\mathrm{rec}_{\overline{\nu}_e}$ can be reconstructed from the prompt energy deposit in the liquid scintillator $E_\mathrm{ls}$ corrected by the neutron-proton mass difference and the mass of annihilated atomic electron $m_n - m_p - m_e$, with elastic kinematic relation in the no-recoil approximation, as
\begin{equation} \label{eq:energy_reconstruction}
    E^\mathrm{rec}_{\overline{\nu}_e} = m_n - m_p + E_\mathrm{ls} - m_e.
\end{equation}
In the following, we investigate the quality of the antineutrino energy reconstruction for $4$ cases of interest,

1) Evaluating the reconstructed energy directly from Eq.~(\ref{eq:energy_reconstruction}) with the nonlinear positron energy response of JUNO~\cite{JUNO:2025fpc,JUNO:2025gmd} $R_e \left( E_e \right)$, i.e.,
\begin{align}
	E_\mathrm{ls} &=  R_e \left( E_{\overline{\nu}_e} - m_n + m_p + m_e \right) \left( E_{\overline{\nu}_e} - m_n + m_p + m_e \right). \label{eq:energy_reconstruction_first_approximation}
\end{align}

2) Accounting for nucleon recoil and nonlinear positron energy response $R_e \left( E_e \right)$. We provide a range of the deposited energy for the allowed kinematic range of the recoil positron energy,
\begin{equation}
    R_e \left( E^\mathrm{min}_e + m_e \right) \left( E_e^\mathrm{min} + m_e \right) \le E_\mathrm{ls} \le R_e \left( E^\mathrm{max}_e + m_e \right) \left(E_e^\mathrm{max} + m_e \right). \label{eq:energy_reconstruction_tree_level_range}
\end{equation}

3) Averaging tree-level expressions over the recoil positron energy and nonlinear positron energy response $R_e \left( E_e \right)$, we obtain for $E_\mathrm{ls}$
\begin{align} \label{eq:energy_reconstruction_tree_level_average}
	E_\mathrm{ls} &= N_\mathrm{LO}^{-1} \int R_e \left( E_e + m_e \right) \left(E_e + m_e \right) \left( \frac{\mathrm{d} \sigma_\mathrm{LO}}{\mathrm{d} E_e} + \frac{\mathrm{d} \sigma_\mathrm{recoil}}{\mathrm{d} E_e} + \frac{\mathrm{d} \sigma_\mathrm{wm}}{\mathrm{d} E_e} + \frac{\mathrm{d} \sigma_\mathrm{FF}}{\mathrm{d} E_e} \right) \mathrm{d} E_e,
\end{align}
with the normalization $N_\mathrm{LO}$,
\begin{align} \label{eq:normalization_tree_level}
	N_\mathrm{LO} &=  \int \left( \frac{\mathrm{d} \sigma_\mathrm{LO}}{\mathrm{d} E_e} + \frac{\mathrm{d} \sigma_\mathrm{recoil}}{\mathrm{d} E_e} + \frac{\mathrm{d} \sigma_\mathrm{wm}}{\mathrm{d} E_e} + \frac{\mathrm{d} \sigma_\mathrm{FF}}{\mathrm{d} E_e} \right) \mathrm{d} E_e.
\end{align}

4) Taking into account the radiation of one real photon and the nonlinear particle-dependent energy responses of JUNO~\cite{JUNO:2025fpc,JUNO:2025gmd} for positrons $R_e$ and for photons $R_\gamma$, we evaluate the corresponding energy deposit in the liquid scintillator $E_\mathrm{ls}$ as
\begin{align} \label{eq:real_scintillator}
	E_\mathrm{ls} &= N^{-1} \int R_e \left( E_e + m_e \right) \left( E_e + m_e \right) \left( \frac{\mathrm{d} \sigma_\mathrm{LO}}{\mathrm{d} E_e} + \frac{\mathrm{d} \sigma_\mathrm{recoil}}{\mathrm{d} E_e} + \frac{\mathrm{d} \sigma_\mathrm{wm}}{\mathrm{d} E_e} + \frac{\mathrm{d} \sigma_\mathrm{FF}}{\mathrm{d} E_e} \right) \mathrm{d} E_e \nonumber \\
	&+ N^{-1} \int R_e \left( E_e + m_e \right) \left( E_e + m_e \right) \left( \frac{\alpha}{\pi} \left( \delta_v + \delta_s + \delta_\mathrm{II} \right) \frac{\mathrm{d} \sigma_\mathrm{LO}}{\mathrm{d} E_e} + \frac{\mathrm{d} \sigma_{v, \mathrm{nf}}}{\mathrm{d} E_e} \right) \mathrm{d} E_e \nonumber \\
    &+ N^{-1} {\int } \left( R_e \left( E_e + m_e \right) \left( E_e + m_e \right) + R_\gamma \left( E_\gamma \right) E_\gamma \right) \frac{\mathrm{d} \sigma^{1 \gamma}}{\mathrm{d} E_e \mathrm{d} f \mathrm{d} E_\gamma} \mathrm{d} E_e \mathrm{d} f \mathrm{d} E_\gamma,
\end{align}
with the normalization $N$,
\begin{align} \label{eq:normalization}
	N &=  \int \left( \frac{\mathrm{d} \sigma_\mathrm{LO}}{\mathrm{d} E_e} + \frac{\mathrm{d} \sigma_\mathrm{recoil}}{\mathrm{d} E_e} + \frac{\mathrm{d} \sigma_\mathrm{wm}}{\mathrm{d} E_e} + \frac{\mathrm{d} \sigma_\mathrm{FF}}{\mathrm{d} E_e} \right) \mathrm{d} E_e \nonumber \\
    &+ \int \left( \frac{\alpha}{\pi} \left( \delta_v + \delta_s + \delta_\mathrm{II} \right) \frac{\mathrm{d} \sigma_\mathrm{LO}}{\mathrm{d} E_e}  + \frac{\mathrm{d} \sigma_{v, \mathrm{nf}}}{\mathrm{d} E_e} \right) \mathrm{d} E_e +  {\int } \frac{\mathrm{d} \sigma^{1 \gamma}}{\mathrm{d} E_e \mathrm{d} f \mathrm{d} E_\gamma} \mathrm{d} E_e \mathrm{d} f \mathrm{d} E_\gamma.
\end{align}

\begin{figure}[H]
\begin{center}
	\includegraphics[scale=0.435]{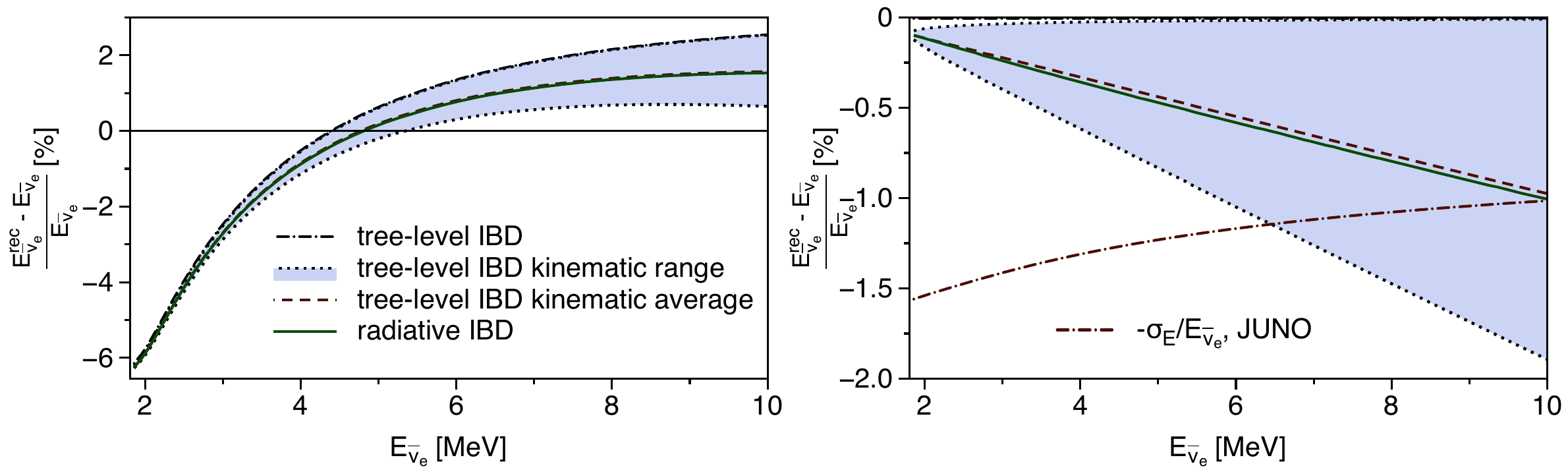}
	\caption{Ratio $\frac{E_{\overline{\nu}_e}^\mathrm{rec} - E_{\overline{\nu}_e}}{E_{\overline{\nu}_e}}$ is shown as a function of the antineutrino energy $E_{\overline{\nu}_e}$ in the left plot for $4$ scenarios of antineutrino energy reconstruction in this paper: (1) tree-level IBD, (2) tree-level IBD kinematic range, (3) tree-level IBD kinematic average, and (4) radiative IBD. Results after correcting for the positron non-linearity, subtracting the curve (1), are shown in the right plot. For reference, the expected JUNO energy resolution~\cite{JUNO:2024fdc}, with an opposite sign, is also presented by the red dash-dotted line.\label{fig:energy_reconstruction}}
\end{center}
\end{figure}

We evaluate the resulting antineutrino energy from $E_\mathrm{ls}$ with Eq.~(\ref{eq:energy_reconstruction}) for the above $4$ scenarios and illustrate our results in Fig.~\ref{fig:energy_reconstruction}. The four scenarios exhibit qualitatively similar behavior across the kinematic range relevant for reactor antineutrinos. At low antineutrino energy, $E_{\overline{\nu}_e} \lesssim 2$~MeV, all four curves approach $-6\%$, reflecting the strong positron nonlinearity $R_e$ near threshold. Accounting for the recoil nucleon kinematics at tree level\footnote{The recoil nucleon kinematics at tree level were considered in Ref.~\cite{JUNO:2024jaw}.} corrects the antineutrino energy reconstruction comparably to current experimental uncertainty, while simulations with vs without radiative corrections result in indistinguishable reconstruction. Including QED radiative corrections within scenario~(4) shifts the reconstructed energy by no more than $0.05\%$ relative to the tree-level average of scenario~(3). We therefore conclude that, for the purposes of antineutrino energy reconstruction in liquid scintillators, an accurate treatment of recoil kinematics at tree level is more important than the inclusion of NLO QED corrections.

\begin{figure}[H]
\begin{center}
	\includegraphics[scale=0.555]{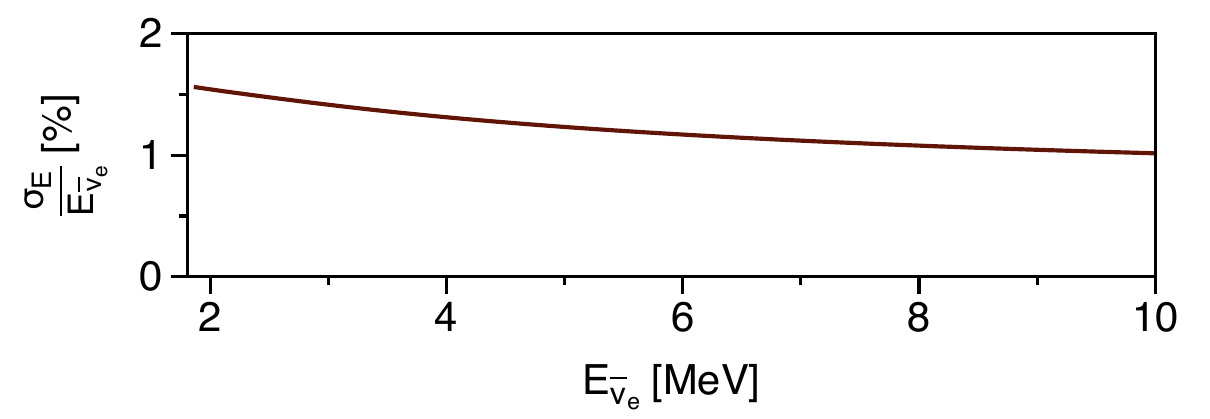}
	\caption{The experimental energy resolution $\frac{\sigma_E}{E_{\overline{\nu}_e}}$~\cite{JUNO:2024fdc} is shown as a function of the antineutrino energy $E_{\overline{\nu}_e}$ for antineutrino energy reconstruction in IBD, assuming only the positron energy deposit via Eq.~(\ref{eq:real_scintillator_reconstructed_positron_energy}). \label{fig:energy_resolution}}
\end{center}
\end{figure}

We also present the expected JUNO energy resolution in Fig.~\ref{fig:energy_resolution} from the study in~\cite{JUNO:2024fdc} resulting in the prediction
\begin{align} \label{eq:experimental_resolution_goal}
    \frac{\sigma_E}{E_{\overline{\nu}_e}} &= \frac{E_\mathrm{ls}}{E_{\overline{\nu}_e}} \sqrt{\frac{a^2}{E_\mathrm{ls}} + b^2 + \frac{c^2}{E_\mathrm{ls}^2} },
\end{align}
with the parameters $a = 0.02614~\sqrt{\mathrm{MeV}},$ $b=0.0064,$ and $c=0.01205~\mathrm{MeV}$, and $E_\mathrm{ls}$ from Eqs.~(\ref{eq:energy_reconstruction_first_approximation}),~(\ref{eq:energy_reconstruction_tree_level_average}), and~(\ref{eq:real_scintillator}). The results are indistinguishable for different choices of $E_\mathrm{ls}$, that is why we present only one curve.
\begin{figure}[H]
\begin{center}
	\includegraphics[scale=0.555]{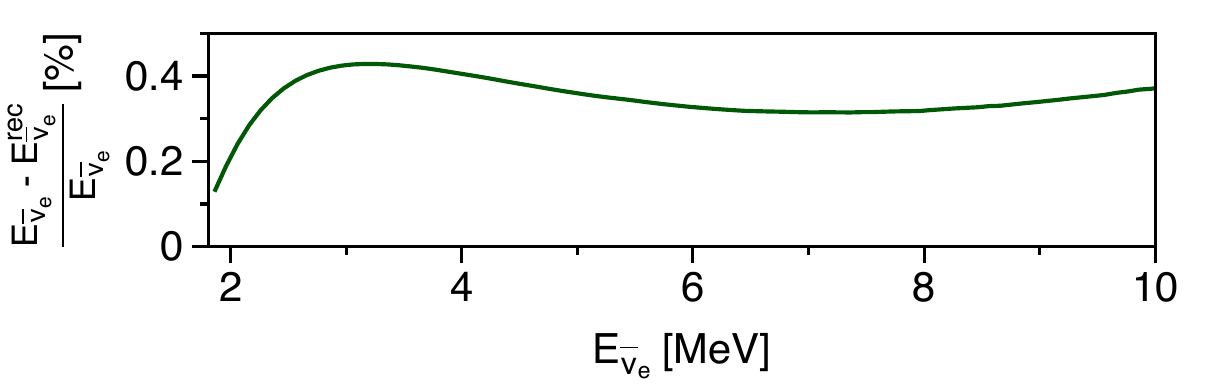}
	\caption{The ratio $\frac{E_{\overline{\nu}_e} - E_{\overline{\nu}_e}^\mathrm{rec}}{E_{\overline{\nu}_e}}$ is shown for antineutrino energy reconstruction in IBD, assuming only the positron energy deposit. \label{fig:energy_bias}}
\end{center}
\end{figure} 

Modern liquid scintillators cannot distinguish deposits of positrons, electrons, and photons. Therefore, we perform a more realistic simulation. For each occurrence of IBD, we replace the deposited energies and the sum of products of energies and responses under the integrals in scenarios (3) and (4) with the reconstructed positron energy $E^\mathrm{rec}_e + m_e$, assuming a radiation-free IBD process. We determine the energy $E^\mathrm{rec}_e$ from the solution of equality between the expected energy in radiation-free IBD and the real deposit
\begin{align} \label{eq:real_scintillator_reconstructed_positron_energy}
	   \left(E^\mathrm{rec}_e + m_e \right) R_e \left( E^\mathrm{rec}_e + m_e \right) = R_e \left( E_e + m_e \right) \left( E_e + m_e \right) + R_\gamma \left( E_\gamma \right) E_\gamma.
\end{align}
We obtain an excellent reconstruction of the antineutrino energy for tree-level kinematics after properly accounting for the nucleon recoil corrections. The reconstructed energy for the radiative process is determined from the following deposited energy:
\begin{align} \label{eq:real_scintillator_simulation}
    E_\mathrm{ls} &= N^{-1} \int \left( E_e + m_e \right) \left( \frac{\mathrm{d} \sigma_\mathrm{LO}}{\mathrm{d} E_e} + \frac{\mathrm{d} \sigma_\mathrm{recoil}}{\mathrm{d} E_e} + \frac{\mathrm{d} \sigma_\mathrm{wm}}{\mathrm{d} E_e} + \frac{\mathrm{d} \sigma_\mathrm{FF}}{\mathrm{d} E_e} \right) \mathrm{d} E_e \nonumber \\
	&+ N^{-1} \int \left( E_e + m_e \right) \left( \frac{\alpha}{\pi} \left( \delta_v + \delta_s + \delta_\mathrm{II} \right) \frac{\mathrm{d} \sigma_\mathrm{LO}}{\mathrm{d} E_e} + \frac{\mathrm{d} \sigma_{v, \mathrm{nf}}}{\mathrm{d} E_e} \right) \mathrm{d} E_e \nonumber \\
    &+ N^{-1} {\int } \left( E^\mathrm{rec}_e + m_e \right) \frac{\mathrm{d} \sigma^{1 \gamma}}{\mathrm{d} E_e \mathrm{d} f \mathrm{d} E_\gamma} \mathrm{d} E_e \mathrm{d} f \mathrm{d} E_\gamma,
\end{align}
where $E^\mathrm{rec}_e$ is given by Eq.~(\ref{eq:real_scintillator_reconstructed_positron_energy}). In Fig.~\ref{fig:energy_bias}, we present the quality of this more realistic energy reconstruction. Due to the differing energy responses to positrons and photons, the emission of real photons can introduce biases in the reconstruction of the antineutrino energy around $1/3$ of the expected experimental resolution. Such effects can be important for an accurate determination of the atmospheric oscillation parameters, neutrino squared mass differences, and consequently, the hierarchy of neutrino masses.

The current and future accuracy of the antineutrino energy reconstruction calls for the inclusion of nucleon recoil at tree-level kinematics~\cite{JUNO:2024jaw} and proper accounts for the radiation of real photons in the inverse beta decay reaction.

\subsection{New physics at leading order in nucleon recoil}
\label{sec:BSM}

We explore potential new physics contributions to the elastic IBD process at the level of invariant amplitudes. In general, the elastic IBD amplitude $T$ can be decomposed into eight invariant amplitudes with corresponding Lorentz structures as~\cite{Tomalak:2024yvq,Borah:2024hvo}
\begin{align}
	T &= -\sqrt{2} G_F V^\star_{ud} \overline{\bar{\nu}}_e \gamma^\rho \mathrm{P}_\mathrm{L} \overline{e} \bar{N} \left[ \left( \bar{g}_M - \bar{f}_2 + \bar{f}_A \gamma_5 \right) \gamma_\rho - 2 \fAtbar \gamma_5 \frac{{P}_\rho}{m_n} + \bar{f}_2 \frac{i \sigma_{\rho \nu} q^\nu}{2 m_n} \right] \tau^- N \nonumber \\
&- \sqrt{2} G_F V^\star_{ud} \frac{m_e}{m_n} \left[ \frac{\bar{f}_{T}}{4} \overline{\bar{\nu}}_e \sigma^{\mu \nu} \mathrm{P}_\mathrm{R} \overline{e} \bar{N} \sigma_{\mu \nu} \tau^- N - \overline{\bar{\nu}}_e \mathrm{P}_\mathrm{R} \overline{e} \bar{N} \left( \bar{f}_3 + \bar{f}_P \gamma_5 - \frac{\bar{f}_R}{4} \frac{\gamma^\mu K_\mu}{m_n} \gamma_5 \right) \tau^- N \right], \label{eq:amplitude_decomposition}
\end{align}
with the averaged lepton and nucleon momenta $K_\mu = \frac{ k_\mu + k^\prime_\mu }{2}$ and $P_\mu = \frac{ p_\mu + p^\prime_\mu }{2}$, respectively, and $\bar{g}_E = \bar{g}_M - \frac{Q^2}{4m_n^2} \bar{f}_2$. At zero momentum transfer, the invariant amplitudes can be expressed in terms of the low-energy coupling constants as
\begin{align}
	\bar{g}_M &= g_V + \mu_p - \mu_n - 1, \qquad \bar{f}_A = - g_A, \qquad \bar{f}_2 = \mu_p - \mu_n - 1, \nonumber \\
	\bar{f}_{3} &= - \frac{m_n}{m_e} C_S, \qquad \bar{f}_P = - \frac{m_n}{m_e} C_P, \qquad \bar{f}_T = \frac{2 m_n}{m_e} C_T. \label{eq:amplitude_normalization}
\end{align}
The amplitude $\fAtbar$ does not contribute in the limit of isospin symmetry. The amplitude $\bar{f}_R$ corresponds to a higher-order effective operator, $\frac{m_e S \cdot k}{2 m^2_N}$, analogous to spin-dependent contact lepton-nucleon interactions in the electromagnetic sector, generated via two-photon exchange~\cite{Drell:1966kk,Bernabeu:1973uf,Caswell:1985ui,Pineda:1998kj,Hill:2012rh,Miller:2012ne,Tomalak:2017owk,Peset:2021iul}. The normalization conventions for the amplitudes and their relation to the Lee-Yang coefficients $C_i$ are provided in Refs.~\cite{Lee:1956qn,Gonzalez-Alonso:2018omy,Tomalak:2024yvq}, while a further connection to the quark level can be performed with nucleon matrix elements following Refs.~\cite{Lee:1956qn,Jackson:1957zz,Severijns:2006dr,Cirigliano:2009wk,Bhattacharya:2011qm,Cirigliano:2012ab,Naviliat-Cuncic:2013ylu,Gonzalez-Alonso:2016etj,Falkowski:2017pss,Gonzalez-Alonso:2018omy,Falkowski:2019xoe,Falkowski:2020pma,Falkowski:2021bkq,Cirigliano:2023nol,Dawid:2024wmp}, with lattice-QCD inputs from Refs.~\cite{Gonzalez-Alonso:2013ura,BMW:2014pzb,Gonzalez-Alonso:2018omy,Gupta:2018qil,FlavourLatticeAveragingGroupFLAG:2021npn,Wang:2025nsd}.
 
The general expression for the unpolarized elastic scattering cross section and polarization observables in the isospin limit has been derived in Refs.~\cite{Tomalak:2024yvq} and~\cite{Borah:2024hvo}. In this paper, we consider the unpolarized cross section
\begin{equation} \label{eq:cross_section_E0_terms}
\frac{\mathrm{d} \sigma \left( E_{\overline{\nu}_e}, Q^2 \right) }{\mathrm{d} Q^2}= \frac{G^2_F |V_{ud}|^2}{2\pi} \frac{m_n^2}{E^2_{\overline{\nu}_e}} \left[ \left( \tau + r_e^2 \right)A \left(\nu,~Q^2 \right) + \frac{\nu}{m_n^2} B \left(\nu,~Q^2 \right) + \frac{\nu^2}{m_n^4} \frac{C \left(\nu,~Q^2 \right)}{1+ \tau} \right],
\end{equation}
with kinematic variables
\begin{equation} \label{eq:cross_section_kinematic_notations}
	\nu = \frac{E_{\overline{\nu}_e}}{m_n} - \tau - r_e^2, \qquad \tau = \frac{Q^2}{4m_n^2}, \qquad r_e = \frac{m_e}{2 m_n}.
\end{equation}
The structure-dependent quantities $A$, $B$, and $C$ are functions of $E_{\overline{\nu}_e}$ and $Q^2$. They are expressed in terms of the invariant amplitudes as
\begin{align}
	A &= \tau | \bar{g}_M |^2 - | \bar{g}_E |^2 + \left( 1+ \tau \right) | \bar{f}_A |^2 - r_e^2 \left( | \bar{g}_M |^2 + | \bar{f}_A + 2 \bar{f}_P |^2 - 4 \left( 1 + \tau \right) \left( |\bar{f}_P|^2 + |\bar{f}_3|^2\right)\right) \nonumber \\
& - 4 \tau \left( 1+ \tau \right) |\fAtbar|^2 + \frac{r_e^2}{4} \left( \nu^2 + 1 + \tau - \left(1 + \tau + r_e^2 \right)^2 \right) |\bar{f}_R|^2 - r_e^2 \left( 1 + 2 r_e^2 \right) |\bar{f}_T|^2 \nonumber \\
&- 2 r_e^2 \mathfrak{Re} \Big[ \left( \bar{g}_E + 2 \bar{g}_M - 2 \left( 1 + \tau \right) \fAtbar \right) \bar{f}_T^\star \Big] + r_e^2 \left( 1 + \tau + r_e^2 \right) \mathfrak{Re}\Big [ \bar{f}_A \bar{f}_R^\star \Big] + 2 r_e^4 \mathfrak{Re}\Big [ \bar{f}_P \bar{f}_R^\star \Big], \\
	B &= \mathfrak{Re} \Big[ 4 \tau \bar{f}^*_A \bar{g}_M - 4 r_e^2 \left( \bar{f}_A - 2 \tau \bar{f}_P \right)^* \fAtbar - 4 r_e^2 \bar{g}_E \bar{f}_3^\star - 2 r_e^2 \left( 3 \bar{f}_A - 2 \tau \left( \bar{f}_P + \bar{f}_3\right) \right) \bar{f}_T^\star + r_e^4 \left( \bar{f}_T + 2 \fAtbar \right) \bar{f}_R^\star \Big], \\
	C &= \tau |\bar{g}_M|^2 + |\bar{g}_E|^2 + \left( 1 + \tau \right) |\bar{f}_A|^2 + 4\tau \left( 1 + \tau \right) |\fAtbar|^2 + 2 r_e^2 \left( 1 + \tau \right) |\bar{f}_T|^2 - r_e^2 \left( 1 + \tau \right) \mathfrak{Re} \Big[ \bar{f}_A \bar{f}_R^\star \Big]. \label{eq:ABC_parameters}
\end{align}
New physics scenarios can be studied by allowing the values of these amplitudes to deviate from the Standard Model. At leading order in $1/m_n$ expansion, only vector and axial-vector amplitudes contribute, reproducing the results in Eq.~(\ref{eq:LO_static_diff}). At next-to-leading order $\mathcal{O} \left( \frac{1}{m_n} \right)$, scalar and tensor amplitudes $\bar{f}_3$ and $\bar{f}_T$, as well as the amplitude $\fAtbar$, become relevant. The corresponding recoil corrections at reactor antineutrino energies can be consistently incorporated using the following substitutions:\footnote{The corresponding substitutions for charged-current neutrino-neutron elastic scattering, with invariant amplitudes for this channel, are given by
\begin{align}
	A &\to \left( 1 - \frac{E_0}{m_n} \right) \left( A + \frac{4 E_0}{m_n} \mathfrak{Re} \left[ \left( 1 - \frac{r^2_e}{2 \tau} \right) \left( {g}_M - {f}_2 \right) {f}^\star_3 + \frac{3}{4} \frac{r^2_e}{\tau} {f}_A {f}^\star_T + \frac{1}{2} \left( 1 - \frac{r^2_e}{\tau} \right) {f}_A \fAt^\star \right] \right), \\
	B &\to \left( 1 + \left( 1 - \frac{r^2_e}{\tau} \right) \frac{E_0}{2 E_{{\nu}_e}} \right) B, \\
	C &\to \left( 1 + \frac{E_0}{E_{{\nu}_e}} + \frac{E_0}{m_n} \frac{Q^2 + m^2_e}{4 E^2_{{\nu}_e}} - \frac{m^2_e}{2 m_n E_{{\nu}_e}} \right) \left( C + \frac{2 E_0}{m_n} \mathfrak{Re} \left[ {f}_A \fAt^\star \right] \right). \label{eq:ABC_parameters_recoil_neutron}
\end{align}}
\begin{align}
	A &\to \left( 1 - \frac{E_0}{m_n} \right) \left( A - \frac{4 E_0}{m_n} \mathfrak{Re} \left[ \left( 1 - \frac{r^2_e}{2 \tau} \right) \left( \bar{g}_M - \bar{f}_2 \right) \bar{f}^\star_3 - \frac{3}{4} \frac{r^2_e}{\tau} \bar{f}_A \bar{f}^\star_T + \frac{1}{2} \left( 1 - \frac{r^2_e}{\tau} \right) \bar{f}_A \fAtbar^\star \right] \right), \\
	B &\to \left( 1 - \left( 1 - \frac{r^2_e}{\tau} \right) \frac{E_0}{2 E_{\overline{\nu}_e}} \right) B, \\
	C &\to \left( 1 - \frac{E_0}{E_{\overline{\nu}_e}} - \frac{2 E_0}{m_n} - \frac{E_0}{m_n} \frac{Q^2 + m^2_e}{4 E^2_{\overline{\nu}_e}} + \frac{2 E^2_0 + m^2_e}{2 m_n E_{\overline{\nu}_e}} \right) \left( C - \frac{2 E_0}{m_n} \mathfrak{Re} \left[ \bar{f}_A \fAtbar^\star \right] \right). \label{eq:ABC_parameters_recoil}
\end{align}
Notably, at reactor antineutrino energies, the relative contribution from the amplitude $\bar{f}_P$ in the structure-dependent function $A$ remains below $2\times10^{-5}$ and can thus be safely neglected.

To provide robust constraints on effective interactions and new-physics scenarios, the QED radiative corrections from this work should be applied to ensure a reliable interpretation of modern and future high-precision IBD experiments.

\subsection{Charged-current neutrino-nucleon elastic scattering and neutron decay}
\label{sec:neutron_decay}

The framework developed in this work is readily extendable to charged-current neutrino-neutron elastic scattering and to neutron decay. In this case, the result is obtained by including the Coulomb corrections to virtual QED contributions in Section~\ref{sec:virtual_QED}\footnote{Contributions from the hydrogen bound states in reaction with a ``free" neutron in the initial state start to enter at $\mathcal{O} \left( \alpha^3 \right)$~\cite{refId0,Sherk:1949xx,Galzenati:1960ug,Bahcall:1961zz,Kabir:1967usj,Nemenov:1979fh,Wilkinson:1982hu,Song:1987ug,Byrne:2001sj,Kouzakov:2004dw,Schott:2006pt,Faber:2009ts,Melnikov:2014lwa,McAndrew:2014iia,Gupta:2018pll,Cao:2025lrw} and, therefore, can be neglected.} and interchanging the proton and neutron masses, $m_p \leftrightarrow m_n$, implying the no-recoil approximation for the final-state proton $p \cdot k_\gamma = m_p E_\gamma$, and flipping the sign of the weak magnetism term. Notably, the charged-current neutrino-neutron elastic scattering process has no kinematic threshold and shares the same phase-space structure and integration regions as the IBD process for incident antineutrino energies above the photon-emission threshold, $E_{\overline{\nu}_e} \ge E_{\overline{\nu}_e}^\mathrm{\gamma}$. 

We also reproduced numerically the known results for the electron and neutrino energy spectra in neutron decay from Refs.~\cite{Sirlin:1967zza} and~\cite{Sirlin:2011wg} (cf. also Refs.~\cite{Batkin:1995vp} and~\cite{Gluck:2022ogz} for the neutrino energy spectrum), by applying the integration technique from Ref.~\cite{Ram:1967zza}. Accounting for full kinematic range in the radiative phase space results in permille-level spectral distortions, above the level of the modern experimental precision~\cite{Gonzalez:2025und}, and increases the size of the radiative correction $\Delta_{\rm TOT} - \left( g^2_V - 1\right) $ in Eq.~(\ref{eq:LO_neutron_lifetime_with_corrections}) by $\sim0.25\permil$, which is $3$-$4$ times above the theoretical uncertainty from QED radiative corrections and around the size of the total uncertainty from the radiative correction $\Delta_R$~\cite{Cirigliano:2023fnz}. For a $\sim0.25\permil$ shift, we averaged the increase in the neutron lifetime, with numerically equal shifts for the superallowed transitions, obtained from the neutrino energy spectrum $0.20\permil$ and the electron energy spectrum $0.30\permil$, finding a perfect consistency between integrated neutrino and electron energy spectra after accounting for the full kinematic range in the radiative phase space.\footnote{Evaluating the neutron lifetime with the neutrino energy spectrum from Ref.~\cite{Sirlin:2011wg} or with the electron energy spectrum from Ref.~\cite{Sirlin:1967zza} gives exactly the same result after integrating over the ranges that are leading in the nucleon recoil. The neutron lifetime evaluated with the neutrino energy spectrum is above the result based on the electron spectrum by $2.7\times10^{-5}$, when integrated over the exact kinematics, while exploiting the relativistic expressions for the leading-order energy spectra as part of QED radiative corrections increases this difference by the uncertainty of QED radiative corrections from Ref.~\cite{Cirigliano:2023fnz} up to $9.5\times10^{-5}$.} This update would shift the extraction of $V_{ud}$ from superallowed nuclear $\beta$ decays down by around the size of the uncertainty from $\Delta_R$ from $V_{ud} =0.97373(31)$~\cite{Hardy:2020qwl} to $V_{ud} =0.97361(31)$. Accounting also for the update in short-distance radiative corrections from Ref.~\cite{Cirigliano:2023fnz}, $V_{ud}$ reduces down to $V_{ud} = 0.97348(31)$.

\begin{figure}[H]
	\centering
	\includegraphics[width=0.97\textwidth]{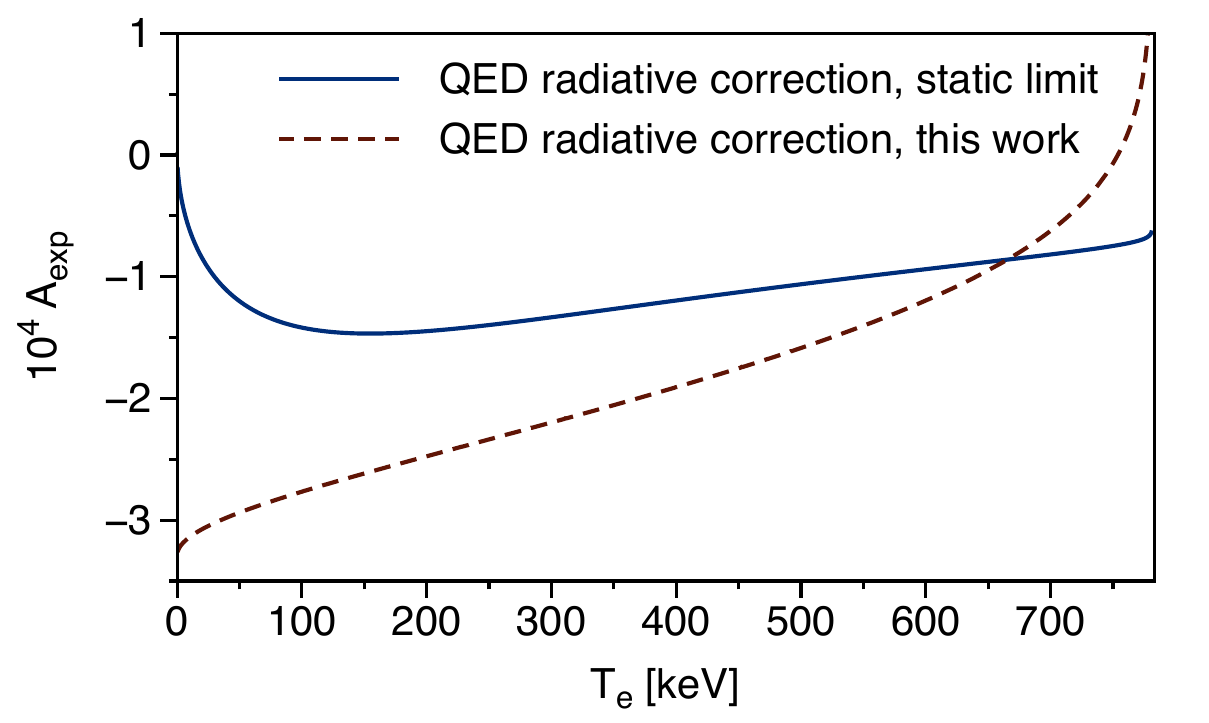}
	\caption{QED radiative correction to the beta asymmetry $A_\mathrm{exp}$ in the polarized neutron decay~\cite{Shann:1971fz,Garcia:1978bq,Ando:2004rk,Cirigliano:2022hob} is shown as a function of the electron kinetic energy $T_e$. The total correction from Refs.~\cite{Shann:1971fz,Garcia:1978bq,Ando:2004rk,Cirigliano:2022hob}, shown as a blue solid line, is compared to the result from this work that accounts for the full kinematic range in the radiative phase-space integration, shown as a red dotted line. \label{fig:beta_asymmetry}}
\end{figure}

To verify QED radiative corrections in extracting the nucleon axial-vector coupling constant from the beta asymmetry in polarized neutron decay $A_\mathrm{exp}$, we study the corrections to this observable within the no-recoil approximation for the final-state proton and without kinematic approximations. The asymmetry is defined as a relative difference between the number of electrons emitted in the hemisphere of the neutron spin direction $N^\uparrow$ vs the hemisphere in the opposite direction to the neutron spin $N^\downarrow$,
\begin{equation}
A_\mathrm{exp} = \frac{N^\uparrow - N^\downarrow}{N^\uparrow + N^\downarrow}.
\end{equation}
We present the resulting QED radiative correction from Refs.~\cite{Shann:1971fz,Garcia:1978bq,Ando:2004rk,Cirigliano:2022hob} and results from this work that account for the full kinematic range in the radiative phase space for allowed positron kinematics in Fig.~\ref{fig:beta_asymmetry} and for energy ranges of UCNA~\cite{UCNA:2012fhw,UCNA:2017obv,UCNA:2019dlk} and PERKEO-III~\cite{Markisch:2018ndu,Saul:2019qnp} experiments in Table~\ref{tab:beta_asymmetry}. As input parameters, we take the vector coupling constant from Ref.~\cite{Cirigliano:2023fnz} and the axial-vector-to-vector coupling constant ratio at low energy from the PDG $\lambda^{\mathrm{I}} = 1.2754(13)$~\cite{Beck:2019xye,ParticleDataGroup:2024cfk,Beringer:2024ady,Workman:2022ynf}. We confirm the magnitude of QED radiative corrections from Refs.~\cite{Shann:1971fz,Garcia:1978bq,Ando:2004rk,Cirigliano:2022hob} implemented in measurements of the beta asymmetry in the neutron decay~\cite{UCNA:2012fhw,UCNA:2017obv,UCNA:2019dlk,Markisch:2018ndu,Saul:2019qnp} and assume the correct sign in the analysis of Refs.~\cite{UCNA:2012fhw,UCNA:2017obv,UCNA:2019dlk,Markisch:2018ndu,Saul:2019qnp}. However, accounting for the full kinematic range in the radiative phase space modifies the resulting radiative corrections by an energy-dependent factor of the order of the correction itself. This update shifts the ratio $\lambda$ in~\cite{Markisch:2018ndu,Saul:2019qnp} upwards by $1.9\times10^{-4}$, $3$ times smaller than the size of the experimental uncertainty in the PERKEO-III measurement~\cite{Markisch:2018ndu,Saul:2019qnp}.\footnote{Should the Collaboration misinterpret the sign of the QED radiative correction, the value of $\lambda$ would shift downwards by $8.4\times10^{-4}$ or approximately $1.5\sigma$ of the experimental uncertainty.}
\begin{table}
	\centering
	\caption{The beta asymmetry $A_\mathrm{exp}$ in the polarized neutron decay, the QED radiative correction to it from Refs.~\cite{Shann:1971fz,Garcia:1978bq,Ando:2004rk,Cirigliano:2022hob}, labeled as ``static limit", and the result for the full kinematic range in the radiative phase-space integration are presented for typical recoil electron energy ranges.}
	\label{tab:beta_asymmetry}
	\begin{tabular}{|c|c|c|c|}
	\hline          
	Electron energy range & Asymmetry & QED correction &QED correction \\
	$[\mathrm{keV}]$ & $A_\mathrm{exp}$ & static limit, $0.1\permil$ & this work, $0.1\permil$ \\
	\hline
	$220$-$670$~\cite{UCNA:2012fhw}     & -0.10196 & -1.00 & -1.60 \\
	$190$-$740$~\cite{UCNA:2017obv}     & -0.09798 & -0.92 & -1.61 \\
	$300$-$700$~\cite{Markisch:2018ndu} & -0.09885 & -0.97 & -1.65 \\
	$197$-$694$~\cite{Saul:2019qnp}        & -0.09777 & -0.90 & -1.59 \\
	\hline
	\end{tabular}
\end{table}

\subsection{Comparison to literature}
\label{sec:literature_corrections}

Our results for recoil and weak magnetism corrections to the differential cross section are in full agreement with the general expressions presented in Refs.~\cite{Seckel:1993dc,Lopez:1997ki,Strumia:2003zx,Ankowski:2016oyj,Ricciardi:2022pru}. We agree with Refs.~\cite{Vogel:1999zy,Raha:2011aa,Hayes:2016qnu} in the limit of vanishing electron mass. In Ref.~\cite{Fayans:1985uej}, we note a sign error in front of the squared magnetic moment term proportional to $\left(\mu_p - \mu_n \right)^2$. Our results are in agreement with Ref.~\cite{Horowitz:2001xf} up to corrections proportional to $E_0$ that are not included in the analysis in Ref.~\cite{Horowitz:2001xf}. For the total cross section, our recoil correction agrees with the results of Refs.~\cite{Fayans:1985uej} and~\cite{Seckel:1993dc}, although we do not observe agreement with Ref.~\cite{Vogel:1983hi} besides the absolute value of the $g_V g_A$ interference term. We confirm the weak magnetism correction to the total cross section as reported in Refs.~\cite{Vogel:1983hi,Fayans:1985uej,Seckel:1993dc}, and we reproduce the correct analytic threshold behavior as given in Ref.~\cite{Fayans:1985uej}.

All previous evaluations of QED radiative corrections to IBD were performed using relativistic fields for both the positron and proton. In contrast, our calculation consistently employs a heavy baryon framework, in which the low-energy constants are valid only when nucleons are treated as heavy fields. This allows us to control the constant term of order $\mathcal{O} \left( \frac{\alpha}{\pi} \right)$ in a theoretically consistent manner. While Refs.~\cite{Kurylov:2002vj,Strumia:2003zx,Hayes:2016qnu} provide parameterizations that include electroweak, QCD, and QED corrections, these contributions are separated in our effective field theory framework. Prior evaluations of the energy spectrum were limited to the electromagnetic spectrum in the static limit, whereas our work avoids such a kinematic approximation. We provide radiative corrections not only to the total cross section and electromagnetic energy spectrum but also to the positron energy spectrum, as well as the double- and triple-differential distributions in radiative IBD. Our results are based on a fully consistent phase-space integration and should thus be considered the most reliable for use in high-precision experimental analyses.

In the soft-photon limit, our results for both virtual and real radiative corrections agree with Refs.~\cite{Towner:1998bh,Kurylov:2001av,Kurylov:2002vj,Fukugita:2004cq,Fukugita:2005hs}, and we reproduce the correct behavior in the collinear regime as discussed in Refs.~\cite{Tomalak:2021hec} and~\cite{Tomalak:2022xup}. Importantly, our result for the infrared-safe observable electromagnetic energy spectrum is free of positron-mass singularities, in agreement with the Kinoshita-Lee-Nauenberg theorem~\cite{Bloch:1937pw,Nakanishi:1958ur,Yennie:1961ad,Kinoshita:1962ur}. This feature has been demonstrated at higher beam energies in charged-current (anti)neutrino-nucleon elastic scattering~\cite{Tomalak:2021hec,Tomalak:2022xup}, as well as in the static limit by previous works~\cite{Vogel:1983hi,Fayans:1985uej,Kurylov:2002vj,Fukugita:2004cq,Raha:2011aa,Ankowski:2016oyj,Tomalak:2021hec,Tomalak:2022xup}. Finally, we confirm that the positron energy spectrum contains the expected single-logarithmic behavior in the limit of small electron mass, in agreement with the results in Ref.~\cite{Dicus:1982bz}, based on Refs.~\cite{Abers:1968zz,Dicus:1970mk,Appelquist:1972wau,Beg:1972ajq,Sirlin:1977sv}.

\section{Conclusions and outlook}
\label{sec:summary}

In this work, we performed a comprehensive analysis of inverse beta decay cross sections for reactor antineutrino energies. Our calculation starts with heavy baryon chiral perturbation theory, incorporating the latest precise determination of the vector coupling constant $g_V$ as a fundamental input for the most accurate predictions. For the first time, we present fully analytic results for QED radiative corrections to the positron energy spectrum, double- and triple-differential distributions in the radiative IBD process, as well as provide the electromagnetic energy spectrum without the no-recoil approximation in the evaluation of the radiative phase-space integrals.

Our findings show that QED radiative corrections affect observables at the few-percent level, with a pronounced dependence on the specific experimental setup. We have illustrated these effects for energy spectra, averaged energies, and the positron scattering angle. A thorough uncertainty analysis confirms that the nucleon axial-vector coupling constant $g_A$ and the CKM matrix element $V_{ud}$ remain the dominant sources of cross-section uncertainty for reactor antineutrino energies, contributing relative errors of $1.71\permil$ and $0.64\permil$, respectively. Reducing these uncertainties through improved external experimental inputs would enable subpermille precision in IBD cross-section predictions. We have quantified non-negligible effects from radiative corrections in inverse beta decay on the reconstruction of the antineutrino energy with modern and future liquid scintillator detectors. Furthermore, we have explored the potential of reactor antineutrino measurements to constrain physics beyond the Standard Model.

The results reported here are directly applicable to studies of reactor antineutrino fluxes, neutrino oscillation experiments, and searches for new particles with reactor antineutrino sources, as well as to any precise antineutrino measurements with energy below $10~\mathrm{MeV}$ where IBD is an important detection channel.

To extend the precision description of IBD with reactor antineutrinos to the energy range from 10 to 60~MeV, relevant for supernova neutrinos and pion-decay-at-rest sources, we account for QED contributions involving pions and deltas in Refs.~\cite{Tomalak:2026btz,Tomalak:2026ido}.

\section*{Acknowledgments}
FeynCalc~\cite{Mertig:1990an,Shtabovenko:2016sxi}, LoopTools~\cite{Hahn:1998yk}, JaxoDraw~\cite{Binosi:2003yf}, Mathematica~\cite{Mathematica}, and DataGraph~\cite{JSSv047s02} were extremely useful in this work. This work is supported by the National Science Foundation of China under Grants No. 12347105 and No. 12447101. This work is supported in part by the National Natural Science Foundation of China under Grant No. 12342502.

\newpage
\bibliography{IBD}{}

\end{document}